\documentstyle[prb,aps,multicol,epsf,eqsecnum]{revtex}

\begin{document}



\title{Spin-1 effective Hamiltonian with three degenerate orbitals:\\
 An application to the case of V$_2$O$_3$. }

\author{S. Di Matteo,$^{1,4}$ N.B. Perkins,$^{2,3}$ C.R. Natoli$^{1,2}$}
\address {$^1$ESRF, B.P. 220, F-38043 Grenoble Cedex 9 (France)\\
$^2$Laboratori Nazionali di Frascati, INFN, 
Casella Postale 13, I-00044 Frascati (Italy)\\
$^3$BLTP, JINR, Dubna, 141980,(Russia)\\
$^4$Dipartimento di Scienze Fisiche ``E.R. Caianiello'', 
Universit\`a di Salerno, 84081, Baronissi (SA), (Italy)}
\date{\today}

\maketitle

\begin{abstract}

Motivated by recent neutron and x-ray observations in V$_2$O$_3$, we derive 
 the effective Hamiltonian in the strong coupling limit of an Hubbard model with three degenerate 
$t_{2g}$ states containing two electrons coupled to spin $S = 1$, and use 
it to re-examine the low-temperature ground-state properties of this compound. 
An axial trigonal distortion of the cubic states is also taken into account.
Since there are no assumptions about the symmetry properties of the hopping 
integrals involved, the resulting spin-orbital Hamiltonian can be generally 
applied to any crystallographic configuration of the transition metal ion 
giving rise to degenerate $t_{2g}$ orbitals. 

Specializing to the case of V$_2$O$_3$ we consider the low temperature 
antiferromagnetic insulating phase. We find two variational regimes, depending on the relative size of the correlation energy of the vertical pairs and the in-plane interaction energy. The former favors the formation of stable molecules throughout the crystal, while the latter tends to break this correlated state.
Using the appropriate variational wave functions
we determine in both cases the minimizing orbital solutions for various spin 
configurations, compare their energies and draw the corresponding phase 
diagrams in the space of the relevant parameters of the problem. We find that 
none of the symmetry-breaking stable phases with the real spin structure 
presents an orbital ordering compatible with the magnetic space group 
indicated by very recent observations of non-reciprocal x-ray gyrotropy in 
V$_2$O$_3$. 
We do however find a compatible solution with very small excitation 
energy in two distinct regions of the phase space, which might turn into
the true ground state of V$_2$O$_3$ due to the favorable coupling with the 
lattice.
We illustrate merits and drawbacks of the various solutions and discuss them 
in relation to the present experimental evidence.

\vspace{0.5cm}


\end{abstract}

\vspace{0.3cm}

\protect\begin{multicols}{2} 

\section{Introduction}

The crystal and electronic structure of V$_2$O$_3$ (vanadium sesquioxide) has 
been the subject of intensive theoretical and experimental studies over the 
past three decades.\cite{rice69,mcwhan70,dernier70,moon70,nebenzahl71}
As is well known, this compound is considered to be
the prototype of the Mott-Hubbard systems, showing metal-insulator transitions
from a paramagnetic metallic (PM) phase to an antiferromagnetic insulating
(AFI) phase at low temperatures
($\approx 150 K$) and from a PM phase to a paramagnetic insulating (PI) 
phase at higher temperatures ($\approx 500 K$), due to the interplay between 
band formation and electron Coulomb correlation.\cite{mcwhan70} 
Actually it is the only known example among transition-metal 
oxides to show a PM to PI transition.\cite{bao98} 
In the paramagnetic phases
the crystal has the corundum structure, in which the V ions are
arranged in V-V pairs along the $c$-hexagonal axis and form a honeycomb 
lattice in the basal $ab$ plane. Each V$^{3+}-$ion has $3d^2$ configuration and
is surrounded by an oxygen octahedron with a small trigonal 
distortion which lifts the three-fold degeneracy of  $t_{2g}$ orbitals into a
nondegenerate $a_{1g}$ and doubly degenerate $e_g$  orbitals separated 
by the distortion energy $\Delta_t$, with the $e_g$ states lying lower. 
Cooling down to the AFI phase, a strongly destructive 
first order transition takes place and the system becomes monoclinic. At
the same time a peculiar antiferromagnetic (AF) spin structure that breaks the 
original trigonal symmetry of the corundum in the basal plane sets in. The magnetic order consists of ferromagnetic planes perpendicular to 
the monoclinic $b_m$ axis that are stacked antiferromagnetically. This is rather 
surprising, since the corundum lattice is known to be non frustrated.

Different theoretical models have been proposed to explain the anomalous 
properties of V$_2$O$_3$. In the late seventies Castellani, Natoli and 
Ranninger \cite{cnr1,cnr2,cnr3}  (CNR) attempted 
in a series of papers a realistic description of the complex magnetic
properties and phase diagram of V$_2$O$_3$. They 
realized that the peculiar structure observed in the AFI 
phase could not be explained in terms of a single band Hubbard model and 
that the introduction of additional degrees of freedom in the model, 
in terms of orbital degeneracy of the atomic $3d$ states involved, was
necessary to explain the experimental findings. They argued 
that there was only one magnetic electron in the doubly degenerate $e_g$ 
band, the other electron being involved in a strong covalent diamagnetic 
bond formed by the $a_{1g}$ states along the vertical pair.
In this way a phase diagram showing the regions of stability of the 
various spin-orbital orderings was obtained as a function of the in-plane
hopping integrals and the $J/U_2$ ratio, $J$ being the intra-atomic exchange
integral and $U_2$ the Coulomb integral between different orbitals.
Implicit in the model was the assumption that $J$ was not too strong, so 
that a spin $S=1/2$ for the $V^{3+}-$ion was favored instead of $S=1$.
The importance of the orbital degrees of freedom in the physics of V$_2$O$_3$
was later confirmed by several experimental facts.
A series of neutron scattering experiments \cite{bao98,bao97} 
showed evidence that the wave vector of the short range antiferromagnetic 
correlations in both paramagnetic phases was rotated by $30^°$ degrees with 
respect to that observed in the AFI phase. In all three phases the 
atoms along the vertical pairs are ferromagnetically coupled; however one 
passes from a complete antiferromagnetic coupling along the three in-plane 
bonds in the high temperature paramagnetic phases to one ferromagnetic and two 
antiferromagnetic in the low temperature phase, with consequent breaking of the 
corundum trigonal symmetry. Moreover the magnetic fluctuations in the PM 
and PI phases remain short range down to the transition temperature. 
Finally NMR studies by Takigawa {\it et al.} \cite{taki} in the PM phase 
showed clear evidence of the role of the orbital degrees of freedom in the 
relaxation mechanism of the nuclear spins. These findings strongly point 
to an interpretative scheme in which the orbital degrees of freedom are 
frozen in the AFI phase to give rise to the peculiar AF spin structure 
and are responsible for the short range magnetic fluctuations in the 
high temperature phases.

However a direct experimental evidence of the orbital ordering (OO) and 
of the spin state of the V-ions was still lacking and was made possible only 
by 
a series of x-ray spectroscopies carried out with the very bright beam of a 
third generation 
synchrotron radiation source. For the first time one could  
subject to experimental test the assumptions and the predictions of
the CNR\cite{cnr1} model. It all started when Fabrizio {\it et al.} 
\cite{fabrizio98} 
suggested that the existence of an OO of the kind suggested by 
CNR\cite{cnr1} could be revealed by means of x-ray resonant 
diffraction at the Vanadium K-edge. Following this suggestion 
Paolasini {\it et al.}\cite{paolasini99} carried out such measurements.
On the basis of a classification of forbidden reflections into pure 
magnetic and orbital, they interpreted the (111) 
monoclinic reflection as evidence of an orbital ordering in V$_2$O$_3$, 
in keeping with the prediction of the CNR model. However in the same paper
a measurement of the ratio between orbital and spin moments by non resonant 
magnetic scattering provided a value $\langle L\rangle / \langle S\rangle\, =  -0.3$ which,
together with the value of the magnetic moment $ \langle L\rangle  +\, 2 \langle S\rangle \,=  1.2\, \mu_B$ 
seen by neutrons,\cite{moon70} gives $2 \langle S\rangle\, =  1.7\,\mu_B$, compatible more with a 
spin $S=1$ than with a spin $S=1/2$ state of the V atoms. This finding was 
pointing to an inconsistency of the interpretative framework.
Additional evidence for a spin $S=1$ state of the Vanadium atoms came 
from the interpretation of linear dichroism in the Vanadium 
L-edge absorption spectra,\cite{park00} where evidence 
of a reduced occupation of $a_{1g}$ orbitals was also found (25\% in
the PM phase, 20\% in the PI phase, 17\% in the AFI phase, instead of the 50\% postulated by CNR).
These are strong indication that in the AFI phase of V$_2$O$_3$ intra-atomic 
correlations prevail over band delocalization, contrary to the assumption
made by CNR.\cite{cnr1}

Actually the transition between the two regimes had been examined in the 
second paper of the series CNR,\cite{cnr2} where a realistic calculation 
was performed in the Hartree-Fock (HF) approximation,
using a bare tight binding band structure for the $a_{1g}$ and $e_g$
orbitals in the Hubbard Hamiltonian. This method is very akin to the more 
modern LDA + U method.\cite{aza}
There it was shown that the value $J/U_1 \approx 0.2$, where $U_1$ is the 
intra-atomic Coulomb parameter within the same orbital, marks the transition
between a $V$ spin $S=1/2$ regime described above and a spin $S=1$ regime, 
where a stable self-consistent solution was found with the real (i.e., 
observed) spin structure (RS), no orbital ordering, 1.5 electrons in the 
$e_g$ and 0.5 in the $a_{1g}$ bands aligned by intra-atomic exchange, with 
a spin moment of $1.7 \mu_B$. 
This solution was metallic but was stabilized  by the monoclinic distortion  
with the concomitant opening of a gap for convenient, reasonable values of 
U and J. At the time this possibility was discarded in favor of the spin 
$S=1/2$ solution showing an orbital ordering, mainly on the basis of general 
considerations of the phase diagram.\cite{mcwhan70} 

More recently Ezhov {\it et al.}\cite{ezhov99}  using the 
LDA +U method,\cite{aza} proposed substantially the same solution with the 
same stabilization mechanism 
(with no surprise since the two methods are conceptually identical and 
the range of parameters and the initial bare density of states rather similar).
Relying on an estimate of the effective electron-electron interaction 
parameters performed by Solovyev {\it et al.}\cite{solov96} on the basis 
of the LDA + U method in perovskite LaVO$_3$, they noted that $J \approx 1$ 
eV, not screened in the solid, and $U_2 \approx 3$ eV. Moreover 
their LDA calculations indicate that the $a_{1g}-e_g$ splitting due to 
the trigonal field is $\Delta_t \approx 0.4$ eV with the $e_g$ band 
center below that of the $a_{1g}$ band. They then argued that with these 
values of $J$ and $\Delta_t$ the $|e_g^1 e_g^2\rangle$ configuration for the 
AFI is more favorable than the $|a_{1g}e_g^1\rangle$ configuration, implying 
that the AFI ground state is not degenerate (no OO) with spin $S=1$ on 
each V atom.

However the fact that the spin $S=1$ solution seems to point to a lack of
orbital ordering must be an artifact of the Hartree-Fock approximation,
implicit in the LDA + U scheme, together with the high $J$ value, since
an examination of the electronic states of the $V^{3+}-$ion leads again to
an orbital degeneracy. Indeed out of the three one electron states 
$|e_g^1\rangle$, $|e_g^2\rangle$ and $|a_{1g}\rangle$ in octahedral symmetry 
one can form three degenerate two electron states: 
$|e_g^1 e_g^2\rangle \: \equiv |0\rangle$, 
$|a_{1g}e_g^1\rangle \: \equiv |-1\rangle$, 
$|a_{1g}e_g^2\rangle \: \equiv |1\rangle$ 
(they constitute the spin and orbital 
triplet ground state of a two electron system in a strong cubic crystal 
field \cite{ball}). In the presence of a trigonal distortion $\Delta_t > 0$,
the singlet state $|0\rangle$ would lie lowest, followed
by the doublet $|+1\rangle$, $|-1\rangle$. Putting the two atoms belonging
to a vertical pair in the same $|e_g^1 e_g^2\rangle$ state to gain 
$\Delta_t$, as suggested by Ezhov {\it et al.},\cite{ezhov99} would loose 
the much bigger gain coming from the allowed hopping processes in a 
configuration in which one atom is in the state $|e_g^1 e_g^2\rangle$ and 
the other in $|a_{1g}e_g^{1,2}\rangle$ and vice-versa.

This situation was realized by Mila {\it et al.} in their paper \cite{mila00} 
where they took up an old suggestion by J. Allen \cite{allen76} that ``the 
magnetic and optical properties of all the phases of V$_2$O$_3$ show a loss 
of $V^{3+}$-ion identity''. These findings, together with the results by 
inelastic neutron scattering quoted above \cite{bao98,bao97}
indicate that the vertical bond is  quite stable and coupled to 
total spin $S_{tot}=2$ with the non polar part of the wavefunction  
given by $|\psi_{ \pm 1}\rangle =  (|0\rangle_a |\pm 1\rangle_b + |\pm 1\rangle_a |0\rangle_b) / \sqrt{2}$ 
(where $a$ and $b$ indicate the two V centers). This state is
clearly doubly degenerate, due to the freedom in the choice of the two 
degenerate states $|\pm 1\rangle$. As a consequence Mila {\it et al.} proposed a simple 
spin-orbital Hamiltonian in which the vertical pairs are described by a 
spin $S_{tot}=2$ and a pseudospin $\tau = 1/2$ for the orbital degeneracy, 
this last being lifted by the in-plane interaction of the pairs. The result 
was a phase diagram in which the spin structures with three in-plane AF bonds,
with two AF and one ferromagnetic (F) bond, with one AF and two F bonds and finally with three F bonds 
(respectively called G, C, A, and F in their paper) follow each other as being the most stable solutions as a function of the 
increasing parameter $J/U_1$ (Fig. 2 of Mila {\it et al.}\cite{mila00}). 
Moreover spin structures C and A present a ferro-orbital configuration, in the 
sense that the vertical pairs are found in the same orbital configuration 
throughout the crystal, $|\psi_{-1}\rangle $ for C and $|\psi_{+1}\rangle$ for A. 
Structure C is indeed the one actually realized in the AFI phase of 
V$_2$O$_3$. 
This model seems to reconcile the existence of orbital degeneracy with the
spin $S=1$ state of the V-ions. However, as will be shown below, there 
still remain problems. The stability region of the C phase 
in the parameter space is very small, like its stabilization energy and 
the percentage of occupancy of the $|a_{1g} \rangle$
with respect to the $|e_g \rangle$ states in the molecular state $|\psi_{ \pm 1}\rangle$ of the AFI phase
is 25\%, to be compared with the value of $17\%$ found by Park {\it et al.}\cite{park00}
Furthermore, the magnetic group corresponding to the ferro-orbital ordering of the C 
phase is not compatible with that suggested by the very recent observation 
of non-reciprocal x-ray gyrotropy by Goulon {\it et al.} \cite{goulon00} Finally, 
this solution cannot give rise to mixed orbital and magnetic reflections, as claimed by Mila {\it et al.}\cite{mila00}, but only 
to pure magnetic, so that it is also in conflict  with Paolasini {\it et al.} 
\cite{paolasini99,paolasini00}  data, as will be discussed in section VII. 

All these considerations led us to re-examine the microscopic Hubbard 
Hamiltonian used to describe the ground state properties of
V$_2$O$_3$ and to study the strong 
coupling limit of this Hamiltonian with three bands and two electrons 
per site coupled to spin $S = 1$, along the patterns developed in 
CNR\cite{cnr1} for spin $S = 1/2$, in the hope 
that starting  from the fundamental Hamiltonian one could cure the 
problems still present in the model proposed by Mila {\it et al.} 
\cite{mila00}  This model is indeed based on interacting  
vertical V-pairs coupled to spin $S_{tot} = 2$ and it is not clear 
at present what is the relation of their spin-orbital Hamiltonian to 
that derived from the atomic limit. 

The paper is organized as follows.
In sections II and III we introduce the three-band Hubbard Hamiltonian 
and perform a second order perturbation expansion 
in the hopping parameters in order
to  derive an effective $S=1$ Hamiltonian ($H_{\rm eff}$) for a system with two 
particles in three degenerate orbitals.
Introducing a pseudospin representation for the orbital degrees of freedom,
we rewrite this Hamiltonian in terms of spin and pseudospin operators.
The resulting spin-orbital Hamiltonian, whose novelty
consists in the fact that both spin and pseudospin degrees of freedom are 
3-components vector operators ($\vec{S}=1$, $\vec{\tau}=1$), complements 
those introduced for cuprates,\cite{kugel73,fei97} for V$_2$O$_3$ 
\cite{cnr1} and more recently for
manganites.\cite{ish96,shiina97,fei99,oles99} 
Since no assumptions are made about the crystal symmetry for the hopping 
parameters, this Hamiltonian can be generally 
applied to any crystallographic configuration of the transition metal ion 
giving rise to degenerate $t_{2g}$ orbitals.
We also discuss some general 
facts about $H_{\rm eff}$, prove a theorem regarding 
the spin ground state for each bond and generalize it to the quasi-degenerate case, where a crystal field splitting is present. 

In view of an application to V$_2$O$_3$, Section IV presents the 
crystallographic and magnetic structure of this compound and discusses
the magnetic space group, both in relation to neutron measurements and 
to recent observation of non-reciprocal x-ray gyrotropy by Goulon 
{\it et al.}\cite{goulon00} From these experiments we are able to determine 
the true magnetic space group of the AFI phase, which should coincide 
with the invariance group of the broken symmetry phase derived from the 
minimization procedure of $H_{\rm eff}$ over the appropriate variational state. 
This section also fixes the parameters of the Hubbard Hamiltonian which 
are thought to be appropriate to the case of V$_2$O$_3$.

Preliminary to the actual minimization for the entire crystal, 
Section V is devoted to 
an in depth study of the energetics of the vertical pairs, both from the 
point of view of the the original Hubbard Hamiltonian and the effective 
spin-orbital Hamiltonian. Apart from checking that they 
give the same result in the atomic limit, we also establish the limit 
of validity of $H_{\rm eff}$. Finally, this study will be used to determine the 
regions of stability of the various competing molecular states in 
the parameter space described by the hopping integrals, $J/U_2$ and the 
trigonal distortion $\Delta_t$, in order to assess the  best variational 
wave function for the entire crystal.

We find two variational regimes and they are considered in Section VI: in the first one, 
the correlation energy of the vertical pairs is big compared to the in-plane 
interaction energy and this favors the formation of stable vertical molecules throughout the crystal (molecular regime). In the other situation, the bigger in-plane interaction energy favors  uncorrelated atomic sites, due to a larger variational Hilbert space, thus breaking the molecules (atomic regime).
In this Section we specialize the effective Hamiltonian derived in 
Section III to the case of V$_2$O$_3$ in the AFI phase, to determine its 
orbital and spin ground state phase diagram in the two regimes. 
In the molecular regime, 
besides the spin-orbital configurations already found by Mila {\it et al.} 
\cite{mila00} (phases A, C, G, F), a new stable phase (RS') appears 
with the same real spin structure as the C phase and an anti-ferro in-plane orbital ordering, but  
the magnetic groups associated to the C and RS' phases are 
incompatible with the one suggested by the non-reciprocal x-ray gyrotropy 
measurements.\cite{goulon00} We do however find a compatible solution in 
the same region of the phase space as the RS' solution with an excitation 
energy of less than $1~meV$.
In the atomic regime we still find a solution with the real spin structure,
however with an orbital order again incompatible with the findings by 
Goulon {\it et al}.\cite{goulon00}

Finally Section VII reviews the implications of the results obtained in the 
previous sections in relation to the present experimental evidence and 
provides an outlook on still open problems.

Since the derivation and the form of the effective Hamiltonian is rather 
cumbersome, for convenience of the reader we have tried to use, whenever 
possible, a pictorial representation of the relevant states, deferring the 
actual calculations and the final expressions to Appendices A to C. 
Appendix D contains useful formulas to calculate the average 
of $H_{\rm eff}$ over molecular variational wave functions for 
various spin orderings.

\section{The model}

To be as general as possible, in the present and in the following sections we 
shall ignore the specific crystallographic symmetries of V$_2$O$_3$  and 
simply deal with the strong coupling limit of a Hubbard Hamiltonian with 
three degenerate $t_{2g}$ states containing two electrons coupled to spin 
$S = 1$. This means that we shall treat the intra-atomic exchange $J$ as an 
high energy parameter with respect to the hopping integrals.
The quasi-degenerate case, in which the three degenerate $t_{2g}$ levels 
are split by a trigonal field of the same order magnitude as the 
hopping term, can be easily accommodated in the formalism.

Following CNR\cite{cnr1} for the notations, we work with the same trigonal basis $e_g^{(1)}$, $e_g^{(2)}$ and $a_{1g}$, to be referred in the following as orbitals 
1, 2 and 3. The total Hamiltonian can then be written as: 

\begin{equation}
H=H_0+H' ~,
\label{acca}
\end{equation}

\noindent where $H_0$ consists of the sum over the whole lattice of 
the atomic interaction terms $H_{0j}^{(n)}$ (where $j$ is the site-index 
which for simplicity of notation will be dropped from the fermion 
operators):

\begin{eqnarray}
\begin{array}{lll}

\vspace{0.3cm}

H_{0j}^{(1)}&=&U_{11}n_{1\uparrow}n_{1\downarrow}+U_{22}
n_{2\uparrow}n_{2\downarrow}
+U_{33}n_{3\uparrow}n_{3\downarrow}\\[0.3cm]

H_{0j}^{(2)}&=&\sum_{\sigma} {\big [}
(U_{12}-J_{12})n_{1\sigma}n_{2\sigma}\\
 & &+(U_{13}-J_{13})n_{1\sigma}n_{3\sigma}\\
 & &+(U_{23}-J_{23})n_{2\sigma}n_{3\sigma}
{\big ]}\\[0.3cm]

H_{0j}^{(3)}&=&
+U_{12}(n_{1\uparrow}n_{2\downarrow}+n_{1\downarrow}n_{2\uparrow}) \\
& & +U_{13}(n_{1\uparrow}n_{3\downarrow}+n_{1\downarrow}n_{3\uparrow}) \\
& &+U_{23}(n_{2\uparrow}n_{3\downarrow}+n_{2\downarrow}n_{3\uparrow})\\[0.3cm]

H_{0j}^{(4)}&=&
+J_{12}(c_{1\uparrow}^{+}c_{1\downarrow}^{+}c_{2\downarrow}
c_{2\uparrow} + c_{2\uparrow}^{+}c_{2\downarrow}^{+}
c_{1\downarrow}c_{1\uparrow})  \\
 & &+J_{13}(c_{1\uparrow}^{+}c_{1\downarrow}^{+}c_{3\downarrow}
c_{3\uparrow} + c_{3\uparrow}^{+}c_{3\downarrow}^{+}
c_{1\downarrow}c_{1\uparrow})\\
 & &+J_{23}(c_{2\uparrow}^{+}c_{2\downarrow}^{+}c_{3\downarrow}
c_{3\uparrow}+c_{3\uparrow}^{+}c_{3\downarrow}^{+}c_{2\downarrow}
c_{2\uparrow}) \\[0.3cm]

H_{0j}^{(5)}&=&
-J_{12}(c_{1\uparrow}^{+}c_{1\downarrow}c_{2\downarrow}^{+}
c_{2\uparrow}+c_{1\downarrow}^{+}c_{1\uparrow}c_{2\uparrow}^{+}
c_{2\downarrow})\\
 & &-J_{13}(c_{1\uparrow}^{+}c_{1\downarrow}c_{3\downarrow}^{+}
c_{3\uparrow}+c_{1\downarrow}^{+}c_{1\uparrow}c_{3\uparrow}^{+}
c_{3\downarrow})\\
 & &-J_{23}(c_{2\uparrow}^{+}c_{2\downarrow}c_{3\downarrow}^{+}
c_{3\uparrow}+c_{2\downarrow}^{+}c_{2\uparrow}c_{3\uparrow}^{+}
c_{3\downarrow}) ~.
\end{array}
\label{h2}
\end{eqnarray}

The meaning of these five on-site terms is the following:
\begin{itemize}
\item $H_{0j}^{(1)}$ describes the Coulomb repulsion between two electrons 
on the same orbital;
\item $H_{0j}^{(2)}$ represents the repulsion between electrons with the 
same spin on different orbitals, given by the Coulomb minus the exchange 
energy;
\item $H_{0j}^{(3)}$ describes the Coulomb repulsion between two electrons 
with opposite spin on different orbitals;
\item $H_{0j}^{(4)}$ represents the energy due to the jump of a pair of 
electrons with opposite spins from one orbital to another;
\item $H_{0j}^{(5)}$ is the exchange term  of the 
process described in $H_{0j}^{(3)}$;
\end{itemize}

The kinetic $(H'_t)$ and crystal field $(H'_{cf})$ terms are given by: 
\begin{eqnarray}
H' & = & \sum_{jj'}\sum_{mm' \sigma }t_{jj'}^{mm'}
c^{+}_{jm \sigma }c_{j'm' \sigma } + \sum_{j m \sigma } 
\Delta_{m} n_{jm\sigma} \nonumber \\
   & = & H'_t + H'_{cf}~,
\label{h3}
\end{eqnarray}
where the summation is over all sites $j$ and  all possible
spin ($\sigma=\uparrow,\downarrow$) and  orbital ($m,m'=1,2,3$) 
configurations. Moreover $\Delta_{1} = \Delta_{2} = 0$ and 
$\Delta_{3} = \Delta_t > 0$.

Since the trigonal field splits the degeneracy of the two electron states 
by an amount comparable with the hopping integrals, 
in performing the atomic limit we shall apply quasi degenerate perturbation 
theory,\cite{Messiah} whereby $H_0$ is the unperturbed Hamiltonian 
with the reference energy of all three levels equal to zero. 
The perturbation term $H'$ will then lift the spin and 
orbital  degeneracy of the ground state of $H_0$.
Our aim will be to find out a representation of this perturbation Hamiltonian 
in terms of the spin and orbital degrees of freedom, to describe the 
insulating phase of a spin-1 Mott-Hubbard system.

As shown in CNR\cite{cnr1} in the trigonal basis we still have: 
$U_{11}=U_{22}=U_{33}\equiv U_{1}$; 
$U_{12}=U_{13}=U_{23}\equiv U_{2}$; 
$J_{12}=J_{23}=J_{13}\equiv J$, together with the relation
\begin{equation}
U_1=U_2+2J~.
\label{trigon}
\end{equation}
Consider first the zeroth-order ground state of $H$.
The strong Hund's coupling $J$ favors the triplet states with energy
$E_t=U_2-J$ with respect to the singlet with energy  $E_s=U_2+J$,
and to the states with two electrons on a single orbital,
with even higher energy $E_d=U_1+J=U_2+3J$.
Hence, in keeping with the above assumptions,
we can construct our zeroth-order Hilbert subspace  
using only triplet states with energy $E_t$,  dropping out all the others.

The total ground state of $H_0$ can be written as a tensorial product of 
the atomic states over the entire crystal:
$$|\Psi_0 \rangle\equiv\prod_{j=1}^{N}|\alpha_{j} \rangle~,$$
where $|\alpha_{j} \rangle$ denotes the 9-fold degenerate atomic state on site 
$j$ with spin $S=1$ and N is the total number of sites. 
Therefore $|\Psi_0 \rangle$ is $9^N$-fold degenerate.
The atomic subspace can be pictorially represented as follows:

\begin{figure}
\begin{picture}(0,0)
\put (30,0){\makebox(10,0){ $|\alpha  \rangle_{1}$ =}}
\put (80,0){\circle{24}}
\put (80,0){\line (0,1){12}}
\put (80,0){\line(1,0){12}}
\put (80,0){\line(-1,0){12}}
\put (80,0){\makebox(-12,12){$\uparrow$}}
\put (80,0){\makebox(12,12){$\uparrow$}}
\end{picture}
\end{figure}

\begin{figure}
\begin{picture}(0,0)
\put (30,0){\makebox(12,0){ $|\alpha  \rangle_{2}$ =}}
\put (80,0){\circle{24}}
\put (80,0){\line (0,1){12}}
\put (80,0){\line(1,0){12}}
\put (80,0){\line(-1,0){12}}
\put (80,0){\makebox(-12,12){$\uparrow$}}
\put (68,0){\makebox(24,-12){$\uparrow$}}
\end{picture}
\end{figure}

\begin{figure}
\begin{picture}(0,0)
\put (30,0){\makebox(12,0){ $|\alpha  \rangle_{3}$ =}}
\put (80,0){\circle{24}}
\put (80,0){\line (0,1){12}}
\put (80,0){\line(1,0){12}}
\put (80,0){\line(-1,0){12}}
\put (80,0){\makebox(12,12){$\uparrow$}}
\put (68,0){\makebox(24,-12){$\uparrow$}}
\end{picture}
\end{figure}

\begin{figure}
~~~\begin{picture}(0,0)
\put (30,0){\makebox(12,0){ $|\alpha  \rangle_{4} =\frac{1}{\sqrt{2}} {\bigg (}$ }}
\put (80,0){\circle{24}}
\put (80,0){\line (0,1){12}}
\put (80,0){\line(1,0){12}}
\put (80,0){\line(-1,0){12}}
\put (80,0){\makebox(-12,12){$\uparrow$}}
\put (80,0){\makebox(12,12){$\downarrow$}}
\put (98,0){\makebox(3,0){$ + $}}
\put (120,0){\circle{24}}
\put (120,0){\line (0,1){12}}
\put (120,0){\line(1,0){12}}
\put (120,0){\line(-1,0){12}}
\put (120,0){\makebox(-12,12){$\downarrow$}}
\put (120,0){\makebox(12,12){$\uparrow$}}
\put (140,0){\makebox(3,0){${\bigg )}$}}
\end{picture}
\end{figure}

\begin{figure}
~~~\begin{picture}(0,0)
\put (30,0){\makebox(12,0){ $|\alpha  \rangle_{5} =\frac{1}{\sqrt{2}} {\bigg (}$ }}
\put (80,0){\circle{24}}
\put (80,0){\line (0,1){12}}
\put (80,0){\line(1,0){12}}
\put (80,0){\line(-1,0){12}}
\put (80,0){\makebox(-12,12){$\downarrow$}}
\put (68,0){\makebox(24,-12){$\uparrow$}}
\put (98,0){\makebox(3,0){$ + $}}
\put (120,0){\circle{24}}
\put (120,0){\line (0,1){12}}
\put (120,0){\line(1,0){12}}
\put (120,0){\line(-1,0){12}}
\put (120,0){\makebox(-12,12){$\uparrow$}}
\put (108,0){\makebox(24,-12){$\downarrow$}}
\put (140,0){\makebox(3,0){${\bigg )}$}}
\end{picture}
\end{figure}

\begin{figure}
~~~\begin{picture}(0,0)
\put (30,0){\makebox(12,0){ $|\alpha  \rangle_{6} =\frac{1}{\sqrt{2}} {\bigg (}$ }}
\put (80,0){\circle{24}}
\put (80,0){\line (0,1){12}}
\put (80,0){\line(1,0){12}}
\put (80,0){\line(-1,0){12}}
\put (80,0){\makebox(12,12){$\uparrow$}}
\put (68,0){\makebox(24,-12){$\downarrow$}}
\put (98,0){\makebox(3,0){$ + $}}
\put (120,0){\circle{24}}
\put (120,0){\line (0,1){12}}
\put (120,0){\line(1,0){12}}
\put (120,0){\line(-1,0){12}}
\put (120,0){\makebox(12,12){$\downarrow$}}
\put (108,0){\makebox(24,-12){$\uparrow$}}
\put (140,0){\makebox(3,0){${\bigg )}$}}
\end{picture}
\end{figure}

\begin{figure}
\begin{picture}(0,0)
\put (30,0){\makebox(12,0){ $|\alpha  \rangle_{7}$ =}}
\put (80,0){\circle{24}}
\put (80,0){\line (0,1){12}}
\put (80,0){\line(1,0){12}}
\put (80,0){\line(-1,0){12}}
\put (80,0){\makebox(-12,12){$\downarrow$}}
\put (80,0){\makebox(12,12){$\downarrow$}}
\end{picture}
\end{figure}

\begin{figure}
\begin{picture}(0,0)
\put (30,0){\makebox(12,0){ $|\alpha  \rangle_{8}$ =}}
\put (80,0){\circle{24}}
\put (80,0){\line (0,1){12}}
\put (80,0){\line(1,0){12}}
\put (80,0){\line(-1,0){12}}
\put (80,0){\makebox(-12,12){$\downarrow$}}
\put (68,0){\makebox(24,-12){$\downarrow$}}
\end{picture}
\end{figure}

\begin{figure}
\begin{picture}(0,0)
\put (30,0){\makebox(12,0){ $|\alpha  \rangle_{9}$ =}}
\put (80,0){\circle{24}}
\put (80,0){\line (0,1){12}}
\put (80,0){\line(1,0){12}}
\put (80,0){\line(-1,0){12}}
\put (80,0){\makebox(12,12){$\downarrow$}}
\put (68,0){\makebox(24,-12){$\downarrow$}}
\end{picture}
\end{figure}

\vspace{0.5cm}

In this notation, the circle represents a site and each sector in a given circle denotes an orbital, according to the following prescription:

\vspace{0.3cm}

\begin{figure}
\begin{picture}(0,30)
\put (100,15){\circle{40}}
\put (100,15){\line (0,1){20}}
\put (100,15){\line(1,0){20}}
\put (100,15){\line(-1,0){20}}
\put (100,15){\makebox(-18,15){$e_g^{(1)}$}}
\put (100,15){\makebox(18,15){$e_g^{(2)}$}}
\put (85,15){\makebox(30,-20){$a_{1g}$}}
\end{picture}
\end{figure}


For example, $|\alpha  \rangle_{1} \equiv c^+_{2\uparrow}c^+_{1\uparrow} |0\rangle $.

We can now introduce the perturbation $H'$ that partially
removes the degeneracy of 
the $|\alpha_{j} \rangle$.
Then the quasi-degenerate perturbation theory up to 
second order gives the 
following eigenvalue equation:
\begin{eqnarray}
{\big |}\sum_{\beta}\frac{\langle \alpha|H'_t|\beta \rangle\langle \beta|H'_t|\alpha' \rangle}
{E_{\alpha} - E_{\beta}}+\delta_{\alpha\alpha'}
\langle \alpha|H'_{cf}|\alpha' \rangle - E {\big |}=0 ~,\nonumber
\end{eqnarray}

\vspace{-1.2cm}

\begin{eqnarray}
\label{h5}
\end{eqnarray}

\vspace{-0.6cm}

\noindent since $H'_{cf}$ is site diagonal and $H'_t$ changes the site occupation.
Here $|\alpha \rangle$ and $|\alpha' \rangle$ are two particular states belonging 
to the $9^N$ degenerate ground state manifold and $|\beta \rangle$ is one of 
the intermediate states with one site singly and another site 
triply occupied. The first order term is trivial and only partially lifts 
the degeneracy of the $|\alpha \rangle$ states according to their orbital 
population. It will be taken into account at the end.
We shall therefore concentrate on the second 
order term. Since $H'_t$ involves only two sites in the hopping, the 
difference between the excited state $|\beta \rangle$ and the 
zeroth-order ground state $|\alpha \rangle$ is only in these two sites. For both
sites the atomic $|\alpha_{j} \rangle$  is a  two-electron state, while one of
the atomic
 $|\beta \rangle$ state is one-electron and the other is a three-electrons
 state. This implies that the 
denominator $E_{\alpha}-E_{\beta}$, which is the energy difference between
initial ($|\alpha \rangle$) 
and intermediate ($|\beta \rangle$) states for the whole crystal, is actually
given only by the contribution
 of the two sites involved in the hopping process.
 Thus in the eigenvalue equation (\ref{h5}), 
  the energy of  the ground state  should be taken as
 the energy of two sites: $E_\alpha=2E_t=2(U_2-J)$.
On the other hand, the energy $E_{\beta}$ consists only of the three 
electrons-site contribution,
 as the one electron atom does not contribute to $H_0$.

We  consider the full multiplet structure of the intermediate states,
$|\beta_{\lambda} \rangle$, i.e., the eigenstates of $H_0$ with three 
electrons. They  are twenty, ten of which are shown below 
(only the site with three electrons) and
the other ten are obtained simply by reversing the spin.

\vspace{-0.2cm}

\begin{figure}
\begin{picture}(0,0)
\put (30,0){\makebox(10,0){ $|\beta_0 \rangle$ =}}
\put (80,0){\circle{24}}
\put (80,0){\line (0,1){12}}
\put (80,0){\line(1,0){12}}
\put (80,0){\line(-1,0){12}}
\put (80,0){\makebox(-12,12){$\uparrow$}}
\put (80,0){\makebox(12,12){$\uparrow$}}
\put (68,0){\makebox(24,-12){$\uparrow$}}
\end{picture}
\end{figure}
\begin{figure}
~~~~\begin{picture}(0,0)
\put (30,0){\makebox(12,0){ $|\beta_1 \rangle =\frac{1}{\sqrt{2}} {\bigg (}$ }}
\put (80,0){\circle{24}}
\put (80,0){\line (0,1){12}}
\put (80,0){\line(1,0){12}}
\put (80,0){\line(-1,0){12}}
\put (80,0){\makebox(-12,12){$\downarrow\uparrow$}}
\put (68,0){\makebox(24,-12){$\uparrow$}}
\put (98,0){\makebox(3,0){$ - $}}
\put (120,0){\circle{24}}
\put (120,0){\line (0,1){12}}
\put (120,0){\line(1,0){12}}
\put (120,0){\line(-1,0){12}}
\put (120,0){\makebox(12,12){$\downarrow\uparrow$}}
\put (108,0){\makebox(24,-12){$\uparrow$}}
\put (140,0){\makebox(5,0){${\bigg )}$}}
\end{picture}
\end{figure}
\begin{figure}
~~~~\begin{picture}(0,0)
\put (30,0){\makebox(12,0){ $|\beta_2 \rangle =\frac{1}{\sqrt{2}} {\bigg (}$ }}
\put (80,0){\circle{24}}
\put (80,0){\line (0,1){12}}
\put (80,0){\line(1,0){12}}
\put (80,0){\line(-1,0){12}}
\put (80,0){\makebox(-12,12){$\downarrow\uparrow$}}
\put (68,0){\makebox(24,-12){$\uparrow$}}
\put (98,0){\makebox(3,0){$ + $}}
\put (120,0){\circle{24}}
\put (120,0){\line (0,1){12}}
\put (120,0){\line(1,0){12}}
\put (120,0){\line(-1,0){12}}
\put (120,0){\makebox(12,12){$\downarrow\uparrow$}}
\put (108,0){\makebox(24,-12){$\uparrow$}}
\put (140,0){\makebox(5,0){${\bigg )}$}}
\end{picture}
\end{figure}
\begin{figure}
~~~~\begin{picture}(0,0)
\put (30,0){\makebox(12,0){ $|\beta_3 \rangle=\frac{1}{\sqrt{2}}{\bigg (}$ }}
\put (80,0){\circle{24}}
\put (80,0){\line (0,1){12}}
\put (80,0){\line(1,0){12}}
\put (80,0){\line(-1,0){12}}
\put (80,0){\makebox(-12,12){$\uparrow$}}
\put (80,0){\makebox(12,12){$\downarrow\uparrow$}}
\put (98,0){\makebox(3,0){$ - $}}
\put (120,0){\circle{24}}
\put (120,0){\line (0,1){12}}
\put (120,0){\line(1,0){12}}
\put (120,0){\line(-1,0){12}}
\put (120,0){\makebox(-12,12){$\uparrow$}}
\put (108,0){\makebox(24,-12){$\downarrow\uparrow$}}
\put (140,0){\makebox(5,0){${\bigg )}$}}
\end{picture}
\end{figure}
\begin{figure}
~~~~~\begin{picture}(0,0)
\put (30,0){\makebox(12,0){ $|\beta_4 \rangle = \frac{1}{\sqrt{2}} {\bigg (}$ }}
\put (80,0){\circle{24}}
\put (80,0){\line (0,1){12}}
\put (80,0){\line(1,0){12}}
\put (80,0){\line(-1,0){12}}
\put (80,0){\makebox(-12,12){$\uparrow$}}
\put (80,0){\makebox(12,12){$\downarrow\uparrow$}}
\put (98,0){\makebox(3,0){$ + $}}
\put (120,0){\circle{24}}
\put (120,0){\line (0,1){12}}
\put (120,0){\line(1,0){12}}
\put (120,0){\line(-1,0){12}}
\put (120,0){\makebox(-12,12){$\uparrow$}}
\put (108,0){\makebox(24,-12){$\downarrow\uparrow$}}
\put (140,0){\makebox(5,0){${\bigg )}$}}
\end{picture}
\end{figure}
\begin{figure}
~~~~~\begin{picture}(0,0)
\put (30,0){\makebox(12,0){ $|\beta_5 \rangle =\frac{1}{\sqrt{2}} {\bigg (}$ }}
\put (80,0){\circle{24}}
\put (80,0){\line (0,1){12}}
\put (80,0){\line(1,0){12}}
\put (80,0){\line(-1,0){12}}
\put (80,0){\makebox(-12,12){$\downarrow\uparrow$}}
\put (80,0){\makebox(12,12){$\uparrow$}}
\put (98,0){\makebox(3,0){$ - $}}
\put (120,0){\circle{24}}
\put (120,0){\line (0,1){12}}
\put (120,0){\line(1,0){12}}
\put (120,0){\line(-1,0){12}}
\put (120,0){\makebox(12,12){$\uparrow$}}
\put (108,0){\makebox(24,-12){$\downarrow\uparrow$}}
\put (140,0){\makebox(5,0){${\bigg )}$}}
\end{picture}
\end{figure}
\begin{figure}
~~~~~\begin{picture}(0,0)
\put (30,0){\makebox(12,0){ $|\beta_6 \rangle = \frac{1}{\sqrt{2}} {\bigg (}$ }}
\put (80,0){\circle{24}}
\put (80,0){\line (0,1){12}}
\put (80,0){\line(1,0){12}}
\put (80,0){\line(-1,0){12}}
\put (80,0){\makebox(-12,12){$\downarrow\uparrow$}}
\put (80,0){\makebox(12,12){$\uparrow$}}
\put (98,0){\makebox(3,0){$ + $}}
\put (120,0){\circle{24}}
\put (120,0){\line (0,1){12}}
\put (120,0){\line(1,0){12}}
\put (120,0){\line(-1,0){12}}
\put (120,0){\makebox(12,12){$\uparrow$}}
\put (108,0){\makebox(24,-12){$\downarrow\uparrow$}}
\put (140,0){\makebox(5,0){${\bigg )}$}}
\end{picture}
\end{figure}
\begin{figure}
~~~~~\begin{picture}(0,0)
\put (30,0){\makebox(12,0){ $|\beta_7 \rangle =\frac{1}{\sqrt{3}} {\bigg (}  $ }}
\put (80,0){\circle{24}}
\put (80,0){\line (0,1){12}}
\put (80,0){\line(1,0){12}}
\put (80,0){\line(-1,0){12}}
\put (80,0){\makebox(-12,12){$\uparrow$}}
\put (80,0){\makebox(12,12){$\uparrow$}}
\put (68,0){\makebox(24,-12){$\downarrow$}}
\put (98,0){\makebox(3,0){$ + $}}
\put (120,0){\circle{24}}
\put (120,0){\line (0,1){12}}
\put (120,0){\line(1,0){12}}
\put (120,0){\line(-1,0){12}} 
\put (120,0){\makebox(-12,12){$\uparrow$}}
\put (120,0){\makebox(12,12){$\downarrow$}}
\put (108,0){\makebox(24,-12){$\uparrow$}}
\put (138,0){\makebox(3,0){$ + $}}
\put (160,0){\circle{24}}
\put (160,0){\line (0,1){12}}
\put (160,0){\line(1,0){12}}
\put (160,0){\line(-1,0){12}}
\put (160,0){\makebox(-12,12){$\downarrow$}}
\put (160,0){\makebox(12,12){$\uparrow$}}
\put (148,0){\makebox(24,-12){$\uparrow$}}
\put (180,0){\makebox(5,0){$ {\bigg )} $}}
\end{picture}
\end{figure}
\begin{figure}
~~~~~\begin{picture}(0,0)
\put (30,0){\makebox(12,0){ $|\beta_8 \rangle =\frac{1}{\sqrt{2}} {\bigg (}$ }}
\put (80,0){\circle{24}}
\put (80,0){\line (0,1){12}}
\put (80,0){\line(1,0){12}}
\put (80,0){\line(-1,0){12}}
\put (80,0){\makebox(-12,12){$\uparrow$}}
\put (80,0){\makebox(12,12){$\downarrow$}}
\put (68,0){\makebox(24,-12){$\uparrow$}}
\put (98,0){\makebox(3,0){$ - $}}
\put (120,0){\circle{24}}
\put (120,0){\line (0,1){12}}
\put (120,0){\line(1,0){12}}
\put (120,0){\line(-1,0){12}}
\put (120,0){\makebox(-12,12){$\downarrow$}}
\put (120,0){\makebox(12,12){$\uparrow$}}
\put (108,0){\makebox(24,-12){$\uparrow$}}
\put (140,0){\makebox(5,0){${\bigg )}$}}
\end{picture}
\end{figure}
\begin{figure}
~~~~~~~\begin{picture}(0,0)
\put (30,0){\makebox(12,0){ $|\beta_9 \rangle=\frac{1}{\sqrt{6}} {\bigg (}2$ }}
\put (80,0){\circle{24}}
\put (80,0){\line (0,1){12}}
\put (80,0){\line(1,0){12}}
\put (80,0){\line(-1,0){12}}
\put (80,0){\makebox(-12,12){$\uparrow$}}
\put (80,0){\makebox(12,12){$\uparrow$}}
\put (68,0){\makebox(24,-12){$\downarrow$}}
\put (98,0){\makebox(3,0){$ - $}}
\put (120,0){\circle{24}}
\put (120,0){\line (0,1){12}}
\put (120,0){\line(1,0){12}}
\put (120,0){\line(-1,0){12}}
\put (120,0){\makebox(-12,12){$\uparrow$}}
\put (120,0){\makebox(12,12){$\downarrow$}}
\put (108,0){\makebox(24,-12){$\uparrow$}}
\put (138,0){\makebox(3,0){$ - $}}
\put (160,0){\circle{24}}
\put (160,0){\line (0,1){12}}
\put (160,0){\line(1,0){12}}
\put (160,0){\line(-1,0){12}}
\put (160,0){\makebox(-12,12){$\downarrow $}}
\put (160,0){\makebox(12,12){$\uparrow $}}
\put (148,0){\makebox(24,-12){$\uparrow $}}
\put (180,0){\makebox(5,0){$ {\bigg )}~~. $}}
\end{picture}
\end{figure}
\vspace{0.5cm}

Due to the cubic symmetry 
 some of  the intermediate states are degenerate, so that there are 
only three different excited levels with energies
\begin{eqnarray}
\begin{array}{l}
E_{\beta_0}=E_{\beta_7}=3(U_2-J)~,\\
E_{\beta_1}=E_{\beta_3}=E_{\beta_5}=E_{\beta_8}=E_{\beta_9}=3U_2~,\\
E_{\beta_2}=E_{\beta_4}=E_{\beta_6}=3U_2+2J~.
\end{array}
\label{h4}
\end{eqnarray}
Accordingly, this implies that we have only three different 
 values for the denominator:
\begin{eqnarray}
E_{\alpha}-E_{\beta_\lambda}=\left \{ 
\begin{array}{l}
-(U_2-J)\\
-(U_2+2J)\\
-(U_2+4J)~.
\end{array}
\right. 
\label{h4bis}
\end{eqnarray}

Another classification of the states can be done according to their total
spin: 
it is easy to check, by summing three spin $\frac{1}{2}$, that $|\beta_0 \rangle$
has $|S=\frac{3}{2}, 
S_z=\frac{3}{2} \rangle$, $|\beta_7 \rangle$ has $|S=\frac{3}{2}, S_z=\frac{1}{2} \rangle$
and all the others
 have  $|S=\frac{1}{2}, S_z=\frac{1}{2} \rangle$. 
The same criteria apply to the other 10 states with opposite spin, by 
reversing the sign of $S_z$.

To proceed further, we  introduce the following operator: 
\begin{eqnarray}
X_j=\sum_{\lambda }
\frac{|\beta_{j\lambda} \rangle\langle \beta_{j\lambda}|}
{E_{\alpha}-E_{\beta_{\lambda}}}~,
\label{h6}
\end{eqnarray}
which can also be written more explicitly:

\begin{eqnarray}
\begin{array}{l}
X_j=-\frac{1}{U_2-J}X_j^{(1)}-\frac{1}{U_2+2J}X_j^{(2)}-
\frac{1}{U_2+4J}X_j^{(3)} ~, \\
\end{array} 
\label{h7}
\end{eqnarray}

\noindent where $X_j^{(1)}$ collects the subspace spanned by the eigenvectors 
$(|\beta_0 \rangle,~ |\beta_7 \rangle)$ and 
those with the same orbital occupancy, but with opposite spin,
$X_j^{(2)}$  corresponds to the eigenvectors 
($ |\beta_1 \rangle,~ |\beta_3 \rangle,~ |\beta_5 \rangle,~ |\beta_8 \rangle,~ |\beta_9 \rangle$) and $X_j^{(3)}$ to ($ |\beta_2 \rangle,~ |\beta_4 \rangle,~ |\beta_6 \rangle $), plus those with opposite spin, respectively.
Explicit  expressions of the operators $X_j^{(i)}$ are presented in Appendix A.

In the subspace spanned by the ground states $|\alpha \rangle$ only terms of the kind
\begin{eqnarray}
\begin{array}{c}
H_{\rm eff}=\sum_{ij}\sum_{m m' n n'}
\sum_{\sigma \sigma '}\\[0.2cm]
t_{ij}^{nm}t_{ji}^{m'n'} 
c_{in\sigma}^{+}c_{jm\sigma}X_j
c_{jm'\sigma '}^{+}c_{in'\sigma '}
\end{array}
\end{eqnarray}
can contribute. In this same subspace, the operator $c_{jm\sigma}X_jc_{jm'\sigma '}^{+}$ is equivalent to:
\begin{eqnarray}
\begin{array}{c}
c_{jm\sigma}X_jc_{jm'\sigma '}^{+}=c_{jm\sigma}[X_j,c_{jm'\sigma '}^{+}]=\\[0.2cm]
-[X_j,c_{jm'\sigma '}^{+}]c_{jm\sigma}+\{ c_{jm\sigma},[X_j,c_{jm'\sigma '}^{+}] \}= \\[0.2cm]
\{ c_{jm\sigma},[X_j,c_{jm'\sigma '}^{+}] \}~.
\end{array}
\label{commut}
\end{eqnarray}

The last expression is used to reduce the number of fermion operators on site $j$. 
The commutators are evaluated  in  Appendix B.

The explicit form obtained for the effective Hamiltonian $H_{\rm eff}$, expressed 
in terms of the fermion operators, is reported in Appendix A. It is composed of three terms:
$H_{\rm eff}=H_{\rm eff}^{(1)}+H_{\rm eff}^{(2)}+H_{\rm eff}^{(3)}$, corresponding to those in 
Eq. (\ref{h7}).

Our next task is to find a representation which allows to rewrite $H_{\rm eff}$ in terms of two operators describing, respectively,  the spin  and the
orbital degrees of freedom.

\section{The spin-orbital representation of the effective Hamiltonian.}

We can characterize each atomic state $|\alpha_{j}\rangle$ 
by two quantum numbers: the spin $S$ and  pseudospin $\tau$.
The pseudospin operator describes the orbital occupation and has an algebra 
which is exactly analogous to that of the usual spin operator.
Due to the 9-fold degeneracy of each $|\alpha_{j} \rangle$ state,  we  need a 
representation with both total $S_j=1$ and $\tau_j=1$.

Consider the pseudospin representation, first.
The orbital quantization axis can be selected arbitrarily, and  we
choose the following convention:    

\begin{figure}
\begin{picture}(0,30)
\put (25,0){\circle{30}}
\put (25,0){\line (0,1){15}}
\put (25,0){\line(1,0){15}}
\put (25,0){\line(-1,0){15}}
\put (25,0){\makebox(-15,15){$\bullet$}}
\put (10,0){\makebox(30,-15){$\bullet$}}
\put (100,0){\makebox(100,0){$|\tau=1,\tau_z=-1\rangle$}}
\put (30,0){\makebox(150,0){$\Longrightarrow$}}
\put (20,0){\makebox(100,0){$=c^{+}_3c^{+}_1|0 \rangle$}}
\end{picture}
\end{figure}

\vspace{-0.3cm}

\begin{figure}
\begin{picture}(0,30)
\put (25,0){\circle{30}}
\put (25,0){\line (0,1){15}}
\put (25,0){\line(1,0){15}}
\put (25,0){\line(-1,0){15}}
\put (25,0){\makebox(15,15){$\bullet$}}
\put (25,0){\makebox(-15,15){$\bullet$}}
\put (100,0){\makebox(100,0){$|\tau=1,\tau_z=0\rangle$}}
\put (30,0){\makebox(150,0){$\Longrightarrow$}}
\put (20,0){\makebox(100,0){$=c^{+}_2c^{+}_1|0 \rangle$}}
\end{picture}
\end{figure}

\vspace{-0.3cm}

\begin{figure}
\begin{picture}(0,30)
\put (25,0){\circle{30}}
\put (25,0){\line (0,1){15}}
\put (25,0){\line(1,0){15}}
\put (25,0){\line(-1,0){15}}
\put (25,0){\makebox(15,15){$\bullet$}}
\put (10,0){\makebox(30,-15){$\bullet$}}
\put (100,0){\makebox(100,0){$|\tau=1,\tau_z=1\rangle$}}
\put (30,0){\makebox(150,0){$\Longrightarrow$}}
\put (20,0){\makebox(100,0){$=c^{+}_3c^{+}_2|0 \rangle$}}
\end{picture}
\end{figure}

\vspace{-2.8cm}
\begin{equation}
\label{rappres}
\end{equation}

\vspace{2.4cm}

Note that this representation is valid for both spin directions, 
so that we can omit the spin indices, and consider only the orbital ones: $c_m^{+}$ ($m=1,2,3$).

With this choice we have the following relations between the fermion
and the pseudospin operators:

\vspace{1.0cm}

\begin{eqnarray}
\begin{array}{lll}
\tau^{+} & = & \sqrt{2}(c_2^{+}c_3-c_3^{+}c_1)~,
\end{array}
\label{htx}
\end{eqnarray}
\begin{eqnarray}
\tau^- & = & \sqrt{2}(c_3^{+}c_2-c_1^{+}c_3)~,
\label{htxx}
\end{eqnarray}
\begin{eqnarray}
\tau_z & = & c_2^{+}c_2-c_1^{+}c_1~.
\label{ht14}
\end{eqnarray}

\indent The factor $\sqrt{2}$ is necessary in order to have the correct commutator, $[\tau^{+},\tau^-]=2\tau_z$, and the correct value of the matrix element for a spin one operator, namely:
\begin{equation}
\tau^{\pm}|1,\tau_z \rangle=\sqrt{(1\mp\tau_z)(2\pm\tau_z)}|1, \;
\tau_z\pm1 \rangle~.
\label{tau+-}
\end{equation}
All possible orbital transitions can be described in the following way with the
pseudospin operators:
\begin{eqnarray}
\begin{array}{rl}
&c_1^{+}c_1c_2^{+}c_2\equiv
(1+\tau_z)(1-\tau_z) : |1,0 \rangle\rightarrow |1,0 \rangle\\[0.2cm]
&c_1^{+}c_1c_3^{+}c_3\equiv
-\frac{1}{2}\tau_z(1-\tau_z) : |1,-1 \rangle\rightarrow |1,-1 \rangle\\[0.2cm]
&c_2^{+}c_2c_3^{+}c_3\equiv
+\frac{1}{2}\tau_z(1+\tau_z) : |1,1 \rangle\rightarrow |1,1 \rangle\\[0.2cm]
&c_1^{+}c_1c_2^{+}c_3\equiv
-\frac{1}{\sqrt{2}}\tau^{+}\tau_z : |1,-1 \rangle\rightarrow |1,0 \rangle\\[0.2cm]
&c_1^{+}c_1c_3^{+}c_2\equiv
-\frac{1}{\sqrt{2}}\tau_z\tau^- : |1,0 \rangle\rightarrow |1,-1 \rangle\\[0.2cm]
&c_2^{+}c_2c_1^{+}c_3\equiv
-\frac{1}{\sqrt{2}}\tau^-\tau_z : |1,1 \rangle\rightarrow |1,0 \rangle\\[0.2cm]
&c_2^{+}c_2c_3^{+}c_1\equiv
-\frac{1}{\sqrt{2}}\tau_z\tau^{+} : |1,0 \rangle\rightarrow |1,1 \rangle\\[0.2cm]
&c_3^{+}c_3c_1^{+}c_2\equiv
+\frac{1}{2}\tau^-\tau^- : |1,1 \rangle\rightarrow |1,-1 \rangle\\[0.2cm]
&c_3^{+}c_3c_2^{+}c_1\equiv
+\frac{1}{2}\tau^{+}\tau^{+} : |1,-1 \rangle\rightarrow |1,1 \rangle~.
\end{array}
\label{t}
\end{eqnarray}

Equivalence in Eq. (\ref{t}) between fermion and pseudospin operators must be interpreted as equality of matrix elements between corresponding states according to the representation (\ref{rappres}). The states after colons indicate the only matrix element different from zero. For example, the fourth line means that:

\begin{equation}
\langle \alpha|c_1^{+}c_1c_2^{+}c_3|\alpha' \rangle\equiv\langle 1,\tau_z|-\frac{1}{\sqrt{2}}\tau^{+}\tau_z|1,\tau_z' \rangle ~,\\
\label{example}
\end{equation}

\noindent and  the only allowed transition is from the state $|1,-1 \rangle$ to the state $|1,0 \rangle$. Note that, in Eq. (\ref{example}), we consider only the orbital part of the $|\alpha\rangle$ states.

Similarly we can introduce an analogous representation for the spin variable 
in the triplet spin states. This case is less straightforward, as 
we have to deal with the sum of two spins $\frac{1}{2}$. This leads to a 
space whose dimensionality is four and  we must project  out the singlet subspace.

The most direct way to introduce the spin representation is to show how it is possible to write the correspondence between the matrix elements of the fermion and the spin operators, in analogy with Eq. (\ref{t}). 
Considering all the possible spin transitions that leave us in the
 triplet subspace, $|S=1,S_z \rangle$, spanned by the $|\alpha\rangle_i$,  we have:
\begin{eqnarray}
\begin{array}{l}
c_{m\sigma}^{+}c_{m\sigma}
c_{m'\overline{\sigma}}^{+}c_{m''\overline{\sigma}} 
\equiv \frac{1}{2}(1-S_z^2) : |1,0 \rangle\rightarrow |1,0 \rangle \\[0.2cm]
c_{m\sigma}^{+}c_{m\overline{\sigma}}
c_{m'\overline{\sigma}}^{+}c_{m''\sigma} 
\equiv \frac{1}{2}(1-S_z^2) : |1,0 \rangle\rightarrow |1,0 \rangle \\[0.2cm]
c_{m\downarrow}^{+}c_{m\downarrow}
c_{m'\downarrow}^{+}c_{m''\downarrow} \equiv
\frac{S_z}{2}(S_z-1) : |1,-1 \rangle\rightarrow |1,-1 \rangle \\[0.2cm]
c_{m\uparrow}^{+}c_{m\uparrow}c_{m''\uparrow}^{+}
c_{m'\uparrow} \equiv \frac{S_z}{2}(S_z+1) : |1,1 \rangle\rightarrow |1,1 \rangle 
\\[0.2cm]
c_{m\downarrow}^{+}c_{m\downarrow}c_{m'\downarrow}^{+}c_{m''\uparrow}
\equiv-\frac{1}{2}S_zS^- : |1,0 \rangle\rightarrow |1,-1 \rangle \\[0.2cm]
c_{m\downarrow}^{+}c_{m\downarrow}c_{m'\uparrow}^{+}c_{m''\downarrow}
\equiv-\frac{1}{2}S^{+}S_z : |1,-1 \rangle\rightarrow |1,0 \rangle \\[0.2cm]
c_{m\uparrow}^{+}c_{m\uparrow}
c_{m'\downarrow}^{+}c_{m''\uparrow} \equiv
\frac{1}{2}S^-S_z : |1,1 \rangle\rightarrow |1,0 \rangle \\[0.2cm]
c_{m\uparrow}^{+}c_{m\uparrow}c_{m'\uparrow}^{+}c_{m''\downarrow}
\equiv\frac{1}{2}S_zS^{+} : |1,0 \rangle\rightarrow |1,1 \rangle 
\end{array}
\label{s}
\end{eqnarray}

\vspace{0.4cm}

 Due to the properties of the Hubbard Hamiltonian, we have no transitions 
$|1,1 \rangle\leftrightarrow |1,-1 \rangle$. The labels $m$, $m'$ and 
$m''$ ($=1,2,3$) refer to orbital occupancy, according to Eq. (\ref{t}). This implies that $m \neq m'$ and $m\neq m''$, but there is  no constraint on $m'$ and $m''$.  
The spin label $\sigma$ ($\overline{\sigma}$) can be either 
$\uparrow$ ($\downarrow$) or $\downarrow$ ($\uparrow$). Again, as in Eq. (\ref{t}), such equivalence must be interpreted as equality of matrix elements between corresponding states in the $|\alpha \rangle$ subspace. The only difference with the previous case is due to the factor $\frac{1}{\sqrt{2}}$ that appears each time a state involved in the transition has a component in the singlet state $S=0$. For example, the second line of Eq. (\ref{s}) means:

\begin{eqnarray}
\begin{array}{c}
\langle \alpha|c_{m\uparrow}^{+}c_{m\downarrow}c_{m'\downarrow}^{+}c_{m''\uparrow}|\alpha' \rangle \equiv \\[0.3cm]
\langle S=1,S_z|\frac{1}{2}(1-S_z^2)|S=1,S_z' \rangle ~, 
\end{array}
\label{example2}
\end{eqnarray}

\noindent and the only allowed transition is between triplet states with $S_z=0$.
Note that this time, due to the Pauli principle, we indicate explicitly the orbital labels $m$, $m'$, $m''$, in contrast to  Eqs. (\ref{t}) and (\ref{example}), where we did not indicate the spin labels. Nonetheless, Eq. (\ref{example2}) expresses an equivalence between matrix elements in the spin space, only, the orbital degrees of freedom having already been taken care of. The global spin-orbit representation is obtained through the direct product of Eqs. (\ref{example}) and (\ref{example2}).
For example, considering both the orbital and the spin degrees of freedom, we have the equality:

\begin{eqnarray}
\begin{array}{c}
\langle \alpha|c_{1\uparrow}^{+}c_{1\downarrow}c_{2\downarrow}^{+}c_{3\uparrow}|\alpha' \rangle \equiv  
\langle \tau =1,\tau_z|-\frac{1}{\sqrt{2}}\tau^{+}\tau_z|\tau=1,\tau_z' \rangle \\[0.3cm]
 \times   \langle S=1,S_z|\frac{1}{2}(1-S_z^2)|S=1,S_z' \rangle ~.\\
\end{array}
\nonumber
\end{eqnarray}

\vspace{-0.8cm}

\begin{eqnarray}
\label{example3}
\end{eqnarray}

\vspace{0.4cm}

In this way it becomes possible to rewrite the effective Hamiltonian 
$H_{\rm eff}$ in terms 
of the spin and pseudospin operators.
A straightforward but nevertheless tedious algebra leads to the
expression for the spin-orbital $H_{\rm eff}$ which we report in 
Appendix C.
Note that the main difficulty of this Hamiltonian consists in the great 
number of terms to deal with ($3^4=81$) which is due to 
the fact that there is no conservation law for the pseudospin quantum 
number $\tau_z$ in the hopping process. This  prevents from having more 
symmetrical expressions for the orbital part of $H_{\rm eff}$, in contrast with the spin part that retains the usual spherical symmetry.

We can now make some general considerations about $H_{\rm eff}$.
First it will be expedient to write it in the following form:
\begin{eqnarray}
\begin{array}{lll}
H_{\rm eff} & = & -\frac{1}{3}\frac{1}{U_2-J}\sum_{ij}{\big [}
2+\vec S_i\cdot \vec S_j {\big ]}O^{(1)}_{ij}\\[0.3cm]
 & & -\frac{1}{4}\frac{1}{U_2+4J}\sum_{ij}{\big [}
1-\vec S_i\cdot \vec S_j {\big ]}O^{(2)}_{ij}\\[0.3cm]
 & & -\frac{1}{12}\frac{1}{U_2+2J}\sum_{ij}{\big [}
1-\vec S_i\cdot \vec S_j {\big ]}O^{(3)}_{ij}~,
\end{array}
\label{simpform}
\end{eqnarray}
where $O^{(k)}_{ij}$ are the orbital contributions to the energy 
corresponding to three different terms of the effective Hamiltonian 
$H_{\rm eff}$ given in Appendix C.
It is clear that the magnetic behavior of the system described by 
$H_{\rm eff}$ is strongly affected by the orbital degrees of freedom, 
that determine the sign and order of magnitude of the exchange constants. 
Nonetheless, even without considering the explicit form of the
$O^{(k)}_{ij}$, we can easily demonstrate the following statement about 
$H_{\rm eff}$ starting from its form (\ref{simpform}): ``It is impossible to 
get the ground state of each single bond in a configuration of total spin 
$\vec{S}_i+\vec{S}_j=1$''.\\
The proof proceeds as follows. The spin scalar product can take the 
following values, depending on the kind of coupling between the two 
nearest neighbor spins:
\begin{equation}
\left\{
\begin{array}{lll}
\vec{S}_i+\vec{S}_j=2 & \Longrightarrow  & \vec{S}_i\cdot\vec{S}_j=+1 \\
\vec{S}_i+\vec{S}_j=1 & \Longrightarrow  & \vec{S}_i\cdot\vec{S}_j=-1 \\
\vec{S}_i+\vec{S}_j=0 & \Longrightarrow  & \vec{S}_i\cdot\vec{S}_j=-2 ~.\\
\end{array}
\right.
\end{equation}

If we consider a single bond, $ij$, this implies the following form for 
the Hamiltonian in the three different cases:

\begin{equation}
\left\{
\begin{array}{lll}
\vec{S}_i+\vec{S}_j=2 & \Longrightarrow  & 
H_{\rm eff}(ij)=H_{\rm eff}^{F}(ij)  \\
\vec{S}_i+\vec{S}_j=1 & \Longrightarrow  & 
H_{\rm eff}(ij)=\frac{1}{3}H_{\rm eff}^{F}(ij)+
\frac{2}{3}H_{\rm eff}^{A}(ij)\\
\vec{S}_i+\vec{S}_j=0 & \Longrightarrow  & 
H_{\rm eff}(ij)=H_{\rm eff}^{A}(ij) ~,\\
\end{array}
\right.
\label{eq32}
\end{equation}

where we defined:
\begin{equation}
H_{\rm eff}^{F}(ij)=-\frac{1}{U_2-J}O^{(1)}_{ij}
\label{hf}
\end{equation}

and
\begin{equation}
H_{\rm eff}^{A}(ij)=-\frac{3}{4}\frac{1}{U_2+4J}O^{(2)}_{ij}-
\frac{1}{4}\frac{1}{U_2+2J}O^{(3)}_{ij}~.
\label{haf}
\end{equation}

It is easy to see 
that if $H_{\rm eff}^{A}(ij)<\!H_{\rm eff}^{F}(ij)$, then the minimum 
of energy is achieved for 
the antiferromagnetic  configuration  $\vec{S}_i+\vec{S}_j=0$, while, if, 
on the contrary,
$H_{\rm eff}^{F}(ij)<\!H_{\rm eff}^{A}(ij)$ then the minimum of energy 
is for the
ferromagnetic one $\vec{S}_i+\vec{S}_j=2$.
Thus for any value of the parameters 
the minimum of the energy is never achieved in the configuration 
$\vec{S}_i+\vec{S}_j=1$, except for points in parameters space where 
$H_{\rm eff}^{A}=H_{\rm eff}^{F}$ and all three configurations 
are degenerate.
This gives us the important result that we can simply study the two spin 
configurations $\vec{S}_i+\vec{S}_j=2$ and $\vec{S}_i+\vec{S}_j=0$, corresponding to the ferromagnetic and the antiferromagnetic bonds, respectively. Note that the two Hamiltonians $H_{\rm eff}^{F}$ and $H_{\rm eff}^{A}$ act as projectors on the subspaces of two spins coupled to $\vec{S}_{ij}=2$ and $\vec{S}_{ij}=0$. In the following we shall call them ferromagnetic and antiferromagnetic parts of $H_{\rm eff}$.

Another important consequence is the 
impossibility to describe by $H_{\rm eff}$ the old 
CNR\cite{cnr1} solutions, where the vertical molecule 
(see section V) was bound in a molecular spin $\vec{S}_{ij}=1$, as a 
particular 
limit in the parameters space. 
The two models differ drastically from the beginning (the older being an 
$S=1/2$ atomic-limit description), and 
 there is no solution  of the new model
where the molecule is bound in a total spin $\vec{S}_{ij}=1$ state, 
as long as the effective Hamiltonian $H_{\rm eff}$ is a good 
representation of the problem. 

Finally we should remember to add to the second order $H_{\rm eff}$ the 
first order contribution coming from the crystal field, whose form is 
easily seen to be
\begin{equation}
H_{\rm eff}^{cf} = \sum_{j } \Delta_{z} \tau_{jz}^2 ~,
\label{hcf}
\end{equation}
where $\Delta_{1} = \Delta_{-1} = \Delta_t$ and $\Delta_{0} = 0$.

Before ending this section, it might be instructive to present the result 
of a preliminary study that led us to the spin-orbital representation of
$H_{\rm eff}$: the case of two electrons coupled to spin $S = 1$
and two orbitals per site. The usefulness of this solution is that one 
could derive the answer on physical grounds. In fact, 
due to the high $J$ value, the crystal ground state is composed by a 
collection of atomic states, each with one electron
per orbital ferromagnetically coupled, in order to maximize Hund's 
exchange, as in the more general 3-orbital case. The main difference 
is that now there is no orbital degeneracy. Due to  the ``freezing'' of 
configurations the only way to decrease the atomic energy with 2$^{nd}$ 
order hoppings is to have an antiferromagnetic coupling between nearest 
neighbors with exchange constants given by an average over all possible 
atomic hoppings among the four orbitals involved. 
As expected, we obtained:
\begin{equation}
H_{\rm eff}=-\frac{1}{(U_2+3J)}\sum_{ij}\langle t_{ij}^2 \rangle{\big [}
1-\vec S_i\cdot \vec S_j {\big ]} ~,
\label{twoorb}
\end{equation}
where $\langle t_{ij}^2 \rangle\equiv\sum_{m,m'=1}^{2}(t_{ij}^{mm'})^2$ and $U_2+3J$ 
is the energy difference between the atomic ground state and the polar
intermediate state. This derivation was also useful to find out the spin-1 
representation in Eqs. (\ref{s}), (\ref{example2}) in terms of fermion operators.

The result in Eq. (\ref{twoorb}) confirms the expectation that with no 
orbital degeneracy it is impossible to break the trigonal symmetry in 
the corundum structure of V$_2$O$_3$, since $\langle t_{ij}^2 \rangle$ is invariant
under a rotation of $2\pi/3$ around the hexagonal $c_H$ axis (see Table II 
of next Section).  


\section{The crystal and magnetic structure of V$_2$O$_3$}

Before embarking in the study of V$_2$O$_3$, we shall try to establish 
the magnetic symmetry group of the AFI phase from the analysis of the 
present experimental evidence, since this will turn out to be important 
for deciding between different ground state solutions with the 
real spin structure obtained by the minimization procedure of $H_{\rm eff}$.

The crystal structure of V$_2$O$_3$, the choice of the Wannier basis 
functions, their symmetry properties and the role of the oxygens, have 
been discussed in detail in the papers by CNR,\cite{cnr1,cnr2} 
to which we refer for what not explicitly
repeated in this paper. We adopt here the same conventions and definitions. 
For convenience we shall remind here some basic facts about this compound.

\begin{figure}
      \epsfysize=105mm
      \centerline{\epsffile {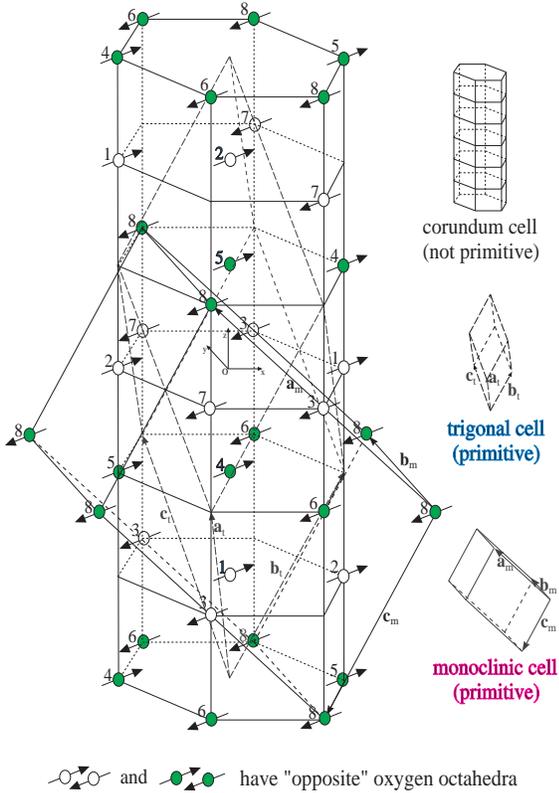}}

\vspace{0.5cm}

\caption{Corundum structure together with the unit cells 
for the trigonal and monoclinic phase. Only Vanadium ions 
are shown: filled and empty  circles correspond to ions with
Oxygen octahedra differently oriented in  the space.
Arrows indicate the direction of spins.}
\label{boh}
\end{figure}

Figure \ref{boh} shows the non primitive hexagonal unit cell of V$_2$O$_3$, 
together with the primitive trigonal one, in the corundum paramagnetic phase.
Only Vanadium atoms are shown. There are two formula units per cell 
(four V atoms) and each metal ion is surrounded  
by a slightly distorted oxygen octahedron with the
three-fold symmetry axis directed along the c$_H$-hexagonal vertical axis.
The distortion corresponds to a compression of the octahedron along c$_H$.
The octahedra around the V-ions represented by the filled circles 
in Fig. \ref{boh} are rotated by $180^o$ about the c$_H$
axis with respect to those around the V-ions represented by the empty circles, this 
orientation varying from plane to plane. Only 2/3 of the octahedra centers 
are occupied by the metallic ion. The space group of the corundum phase is $R\overline{3}c$ 
with the following generators (written in the conventional notation\cite{tinkham}): 

\begin{eqnarray}
\begin{array}{lll}
\{{\hat E,0}\} & = & {\rm Identity} \\
\{{\hat I,0}\} & = & {\rm Inversion ~around ~the ~midpoint ~of ~V_4} \\
           &   & {\rm and ~V_5~(point ~O ~in ~Fig. ~\ref{boh})} \\
\{{\hat C_3,0}\} & = & {\rm Rotation ~of~ } \pi/3 {\rm~ around ~the~ c_H~ axis} \\
\{{\hat C_2,\frac{1}{2}\vec{a}_m }\} &= & {\rm Rotation ~of~ }\pi {\rm~ 
                around ~the~} \vec{b}_m {\rm~ axis} \\
              &  & {\rm followed ~by ~a ~translation ~of 
                      ~\frac{1}{2}} \vec{a}_m ~, \\
\end{array}
\nonumber
\end{eqnarray}

\noindent where $\vec{a}_m$ and $\vec{b}_m$ are two of the 
basis vectors of the monoclinic unit cell shown in Fig. \ref{boh}. The origin O has been chosen as the inversion point between atom V$_4$ and V$_5$ in the trigonal cell. The translation associated with the $\hat C_2$ rotation can be expressed also in terms of the trigonal basis vectors defined in Fig. \ref{boh} as $\frac{1}{2}(\vec{a}_t+\vec{b}_t+\vec{c}_t)$.
The corresponding crystal point group, obtained by setting all translations 
to zero, is $D_{3d}$.\cite{tinkham}
Note that among the symmetry operations of $R\overline{3}c$ there is a glide plane, obtained through the combination of $\hat I$ and $\hat C_2$:

\begin{eqnarray}
\begin{array}{lll}
\{\hat \sigma_b,\frac{1}{2}\vec{a}_m\}  &= &{\rm Reflection~ through~ a~ 
               plane~ orthogonal} \\ 
         &  & {\rm to ~the~} \vec{b}_m {\rm ~axis ~followed ~by ~a 
               ~translation} \\
         &  & {\rm of ~\frac{1}{2}} \vec{a}_m ~.\\
\end{array}
\nonumber
\end{eqnarray}

By lowering the temperature, the system makes a disruptive first order 
transition to a monoclinic phase, with further distortion of the octahedra 
to accommodate a rotation by about $1.8^o$ of the vertical Vanadium pairs 
(e.g., V$_1$ and V$_4$) in the $a_m - c_m$ plane towards the adjacent 
octahedral voids.\cite{dernier70}  As a consequence one bond in the basal 
plane becomes longer than the other two by about 0.1 \AA, the trigonal 
symmetry is lost and the lattice space group lowers to $I2/a$ with the 
same generators except for $(\hat C_3,0)$. Its crystal point group is $C_{2h}$. The monoclinic cell is 
body-centered and, containing four formula units, it is not primitive, from the point of view of the bare crystal lattice. However 
concomitant to the structural transition a magnetic order sets in, 
with ferromagnetic $a_m - c_m$ planes stacked antiferromagnetically and with an AF wave vector given by $\frac{1}{2}\vec{b}_m$.\cite{moon70} Because of this magnetic order, the monoclinic 
cell becomes primitive, due to the AF coupling of the magnetic moments on the V-ions connected by the body-centered translation.
The magnetic moments of the V-ions, indicated by arrows in the 
figure, lie in the $a_m - c_m$ plane, at an angle of $71^o$ away from the 
c$_H$ axis.\cite{moon70,bao98} Notice 
that the in-plane longer bond corresponds to the ferromagnetic coupling and is orthogonal to $\vec{b}_m$.

Under these conditions one can easily check that the time-reversal 
operator $\hat T$ followed by the non primitive translation 
$\frac{1}{2}(\vec{a}_m+\vec{b}_m+\vec{c}_m)$ is a symmetry operation 
of the magnetic structure, so that the magnetic point group is 
$C_{2h}\otimes\hat T$. Each operation should be followed by the appropriate 
translation as indicated here: 

\begin{eqnarray}
\begin{array}{lll}
1){\rm ~} {\hat E}, {\rm ~} {\hat I} &\rightarrow & {\rm No ~translations} \\
2){\rm ~} {\hat C_2}, {\rm ~} {\hat \sigma_b} &\rightarrow & 
\frac{1}{2}(\vec{b}_m+\vec{c}_m) \\
3){\rm ~} {\hat T}, {\rm ~} {\hat T \hat I} &\rightarrow & 
\frac{1}{2}(\vec{a}_m+\vec{b}_m+\vec{c}_m) \\
4){\rm ~} {\hat T \hat  C_2}, {\rm ~} {\hat T \hat\sigma_b } &
\rightarrow & \frac{1}{2}\vec{a}_m ~. 
\end{array}
\nonumber
\end{eqnarray}

Notice that the translation associated to $\hat C_2$ and $\hat \sigma_b$ 
has changed. In fact, since now the application of these two operations 
changes the direction of the magnetic moment,\cite{tinkham} 
the total translation must be 
$$\frac{1}{2}\vec{a}_m + \frac{1}{2}(\vec{a}_m+\vec{b}_m+\vec{c}_m) \equiv 
\frac{1}{2}(\vec{b}_m+\vec{c}_m)$$
and the role of $\hat C_2$ and $\hat \sigma_b$ in the paramagnetic lattice 
is taken, in the AFI phase, by $\hat T \hat C_2$ and $\hat T \hat \sigma_b$.

Under these operations the correspondence between the charge and 
magnetic states of the various metal sites with their oxygen environment 
is given in Table I, with reference to the numbering of Fig. \ref{boh}:

\vspace{5mm}
\noindent
Table I. Correspondence table between magnetic sites in V$_2$O$_3$

\begin{center}
$
\begin{array}{||l|c|c|c|c|c|c|c|c||}  \hline
\hat E:               & 1 & 2 & 3 & 4 & 5 & 6 & 7 & 8  \\ \hline
\hat I:               & 2 & 1 & 7 & 5 & 4 & 8 & 3 & 6  \\ \hline
\hat C_2:             & 8 & 6 & 5 & 7 & 3 & 2 & 4 & 1  \\ \hline
\hat \sigma_b:        & 6 & 8 & 4 & 3 & 7 & 1 & 5 & 2  \\ \hline
\hat T:               & 7 & 3 & 2 & 8 & 6 & 5 & 1 & 4  \\ \hline
\hat T \hat I:        & 3 & 7 & 1 & 6 & 8 & 4 & 2 & 5  \\ \hline
\hat T \hat C_2:      & 4 & 5 & 6 & 1 & 2 & 3 & 8 & 7  \\ \hline
\hat T \hat \sigma_b: & 5 & 4 & 8 & 2 & 1 & 7 & 6 & 3  \\ \hline
\end{array}
$
\end{center}

Now the recent observation of non reciprocal x-ray gyrotropy by Goulon 
{\it et al.} \cite{goulon00} in the AFI phase of V$_2$O$_3$ points to a 
reduction of magnetic symmetry. 
In this experiment a transverse x-ray linear dichroism at the 
Vanadium K-edge is observed and interpreted as due to a dipole-quadrupole interference effect. 
This signal changes sign according to whether the externally applied magnetic field 
is parallel or antiparallel to the direction of the incident x-ray beam along 
the $c_H$ axis. Therefore neither $\hat T$ nor $\hat I$ can be separately 
symmetry operations, but their product $\hat T \hat I$ is. There are seven 
subgroups of four elements of the magnetic point group $C_{2h}\otimes\hat T$.
They are listed below:

\begin{eqnarray*}
1.~~~~~ && \hat E, \hat I, \hat C_2, \hat \sigma_b  \\
2.~~~~~ && \hat E, \hat I, \hat T, \hat T \hat I   \\
3.~~~~~ && \hat E, \hat I, \hat T\hat C_2, \hat T\hat \sigma_b  \\
4.~~~~~ && \hat E, \hat T, \hat C_2, \hat T \hat  C_2  \\
5.~~~~~ && \hat E, \hat T, \hat \sigma_b, \hat T\hat \sigma_b  \\
6.~~~~~ && \hat E, \hat C_2, \hat T \hat I,\hat T \hat \sigma_b  \\
7.~~~~~ && \hat E,  \hat \sigma_b, \hat T\hat I, \hat T  \hat C_2 ~.  
\end{eqnarray*}
It is immediately clear that only groups 6 and 7 are eligible candidates, 
i.e., group $C_{2h}(C_s)$ and $C_{2h}(C_2)$, respectively 
$\underline{2}/m$ and $2/\underline{m}$ in international notation. Both are
magnetoelectric (ME); however the first one gives rise to an off-diagonal ME 
tensor whereas in the second one this tensor presents diagonal components. 
\cite{birss}  It is possible to discriminate between them by noting that the existence of
an off-diagonal ME tensor explains why Astrov and Al'shin failed\cite{astrov} 
to find a ME effect in V$_2$O$_3$, since their experiment was set up to look 
for diagonal components. This is a strong indication that 
$C_{2h}(C_s)$ is the correct magnetic group for V$_2$O$_3$. 

The origin of the reduction of magnetic symmetry from $C_{2h}\otimes\hat T$ 
to $C_{2h}(C_s)$ can reasonably be ascribed to an orbital ordering in the 
magnetic and charge density of V$_2$O$_3$ due to electron correlations. 
However the ferro-orbital C phase found by Mila {\it et al.}\cite{mila00}
does not provide the correct answer, since the corresponding magnetic group 
is easily seen to be $C_{2h}\otimes\hat T$, due to the fact that all sites 
are occupied by the same orbital. The same can be said for the other stable 
phases with the real spin structure found in this work.

We speculate that 
the presence of an excited configuration with the correct magnetic group 
$C_{2h}(C_s)$, very near the ground state and with the favorable coupling 
with the lattice, can provide the solution to this puzzle.

\begin{figure}
\epsfysize=40mm
\centerline{\epsffile{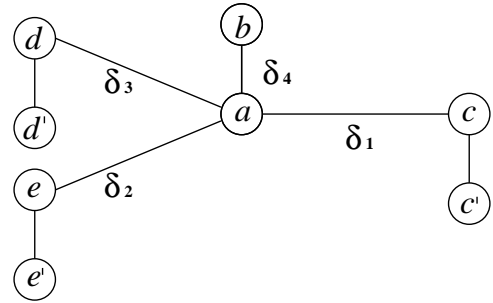}}
\vspace{0.5cm}
\caption{The neighbor structure of the vertical bonds. 
Referring to Fig.\ref{boh}, if the atom $a$ coincides with V$_1$, 
then we have the correspondences $b\rightarrow$ V$_4$, $c\rightarrow$ V$_2$, $c'\rightarrow$ V$_5$,
 $d\equiv e\rightarrow$ V$_3$, and $d'\equiv e'\rightarrow$ V$_6$.}
\label{bonds}
\end{figure}

In order to proceed in the following sections with the minimization of 
$H_{\rm eff}$ we need to have a reasonable guess at the various parameters 
appearing in it, namely the hopping integrals and the Coulomb and exchange 
atomic parameters. In Fig. \ref{bonds} we show half of the 
cluster of nearest neighbors to a given molecule (the other half can be 
deduced with the help of Fig. \ref{boh}) to  illustrate the notation that 
will be used later on.

Following CNR\cite{cnr1,cnr2}
we present in Table II    the transfer integrals
evaluated by exploiting the symmetry properties 
of the corundum structure without taking into account 
the monoclinic distortion of the bonds, since again our aim is 
to show how electronic correlations can break the initial 
trigonal symmetry of the lattice. 
The deviations from this symmetry will be considered later on to 
illustrate if and how the monoclinic distortion can 
stabilize the orbitally ordered state with the correct magnetic 
spin structure.

\vspace{5mm}

\noindent
Table II. Transfer integrals along different  
bonds in the corundum phase.

\begin{center}
\begin{tabular}{|c|c|c|c|c|} \hline
direction & $\delta_1$ & $\delta_{2}$ & $\delta_3$ & $\delta_4$ \\ \tableline
$t_{11}$ & $-\alpha$ & $-\frac{1}{4}\alpha+\frac{3}{4}\beta$ &  
$-\frac{1}{4}\alpha+\frac{3}{4}\beta $ & $\mu$ \\ \tableline
$t_{22} $ & $ \beta $ & $-\frac{3}{4}\alpha+\frac{1}{4}\beta $&
$  -\frac{3}{4}\alpha+\frac{1}{4}\beta $ & $\mu$ \\ \tableline
$t_{33} $ & $\sigma $& $\sigma $& $\sigma$ & $\rho $ \\ \tableline
$t_{12}$ & 0 & $\frac{\sqrt{3}}{4}(\alpha+\beta)$ &
$ -\frac{\sqrt{3}}{4}(\alpha+\beta) $ & 0 \\ \tableline
$t_{13} $ & 0 & $\frac{\sqrt{3}}{2}\tau $& $-\frac{\sqrt{3}}{2}\tau $ & 0 \\
\tableline
$t_{23} $ & $-\tau $& $\frac{1}{2}\tau $& $\frac{1}{2}\tau $ & 0 \\ \tableline
$t_{21}$ & 0 & $\frac{\sqrt{3}}{4}(\alpha+\beta)$ &
$ -\frac{\sqrt{3}}{4}(\alpha+\beta)$ & 0 \\ \tableline
$t_{31}$ & 0 & $\frac{\sqrt{3}}{2}\tau$ & $-\frac{\sqrt{3}}{2}\tau $ & 0 \\
\tableline
$t_{32}$ & $-\tau$ &$ \frac{1}{2}\tau $& $\frac{1}{2}\tau$ & 0 \\ \hline  
\end{tabular}
\end{center}

\vspace{5mm}
\noindent
Table III. Transfer integrals (eV) from tight binding  
calculations used by CNR\cite{cnr2} and from
LAPW-calculations by Mattheiss.\cite{mattheiss94}

\begin{center}
\begin{tabular}{|c|c|c|}\hline 
 & Castellani {\em et al.}\cite{cnr2} & Mattheiss\cite{mattheiss94}
\\\tableline 
$\mu$ & $0.2$ & $0.2$ \\\tableline
$\rho$ & $-0.72$ & $-0.82$ \\\tableline
$-\alpha$ & $-0.13$ & $-0.14$ \\\tableline
$\beta$ & $-0.04$ & $-0.05$ \\\tableline
$\sigma$ & $0.05$ & $0.05$ \\\tableline
$-\tau$ & $-0.23$ & $-0.27$  \\\hline
\end{tabular}
\end{center}
\vspace{0.5cm}

In the following we shall assume for the Coulomb 
and exchange parameters the values suggested by Ezhov 
{\it et al.}\cite{ezhov99} and Mila {\it et al.}\cite{mila00}  
i.e., $J\simeq 0.7 \div 1.0~ eV $, $U_2\sim 2.5~ eV$, 
and for the hopping parameters those derived by Mattheiss\cite{mattheiss94} and shown in Table III. 
This set will be referred as the standard set.
By fitting the LAPW band structure of V$_2$O$_3$ to a tight-band 
calculation,  Mattheiss \cite{mattheiss94} has provided the relevant 
Slater-Koster integrals that have been used to calculate the appropriate 
hopping integrals. 
As on can see from Table III they are quite close to those estimated by 
CNR. \cite{cnr1,cnr2}

\section{The energetics of the vertical molecule. }

As realized by CNR \cite{cnr1,cnr2} and later by Mila {\it et al.}, 
\cite{mila00} the formation of the vertical molecular bond ($\delta_4$ in Fig. \ref{bonds}) is the 
key to the understanding of the physics of V$_2$O$_3$ in all three 
phases. This fact is indeed supported by the experimental 
evidence both from optical spectra \cite{allen76} and inelastic 
neutron scattering.\cite{bao98,bao97}  However the solution proposed
by CNR was appropriate to low values of $J$ ($\leq$ 0.2 eV) and 
values of the trigonal distortion which where supposed to be quite 
small, as suggested by Rubinstein. \cite{rubin70} 
The new solution proposed by Mila {\it et al.} 
\cite{mila00} reconciles the present evidence for an high value of
$J$ ($\geq$ 0.7 eV) and the consequent spin $S=1$ state of the V-ions 
\cite{paolasini99,park00} with the existence of an orbitally degenerate 
molecular state, while being rather stable against a sizable value of 
the trigonal distortion. It is interesting to study how this can come 
about, since this investigation can provide a clue to the kind of 
variational wave function to be used in the minimization of $H_{\rm eff}$, 
will delimit the regions of stability of the solution in the parameter 
space and indicate competing states that might be relevant for the 
phenomenon of the metal-insulator transition in V$_2$O$_3$.

\subsection {The approximate solution using $H_{\rm eff}$ }

In considering the vertical pair it is convenient, as shown in Section V-B, to introduce the 
following molecular quantum numbers: the
total spin $S^M=S_a+S_b$, its $z$-component $S^M_z=S_{az}+S_{bz}$ and total $z$-component of pseudospin
$\tau_z^M=\tau_{az}+\tau_{bz}$.
Along the vertical bond $\delta_4$ only $t_{11} = t_{22} = \mu$ and 
$t_{33} = \rho$ are different from zero (as seen from Table II) and 
their values are given in Table III. Specializing to this case the
effective Hamiltonian given in Appendix C we obtain for the ferromagnetic state
($S^M=2$) the following expression:

\begin{eqnarray}
H_{\rm eff}^{F} &=& -\frac{1}{3}\frac{1}{U_2-J} 
{\big [}2+\vec S_a\cdot \vec S_b {\big ]}
(\mu^2 f_{\mu}+\rho^2 f_{\rho} -\mu\rho f_{\mu\rho}) ~, \nonumber \\
    & &
\label{fb}
\end{eqnarray}

\noindent and for the antiferromagnetic bond ($S_M=0$):
 
\begin{eqnarray}
\label{afb}
H_{\rm eff}^{A} &=& -\frac{1}{2}\frac{1}{U_2+4J} 
{\big [} 1- \vec S_a\cdot \vec S_b {\big ]} \nonumber \\
   & & \times [\mu^2 (a_{\mu}+a'_{\mu})
+\rho^2  a_{\rho}
+\frac{1}{2}\mu\rho  a_{\mu\rho}] \nonumber \\
   & & -\frac{1}{2}\frac{1}{U_2+2J} 
{\big [} 1- \vec S_a\cdot \vec S_b {\big ]}  \\
   & & \times [\mu^2 (a_{\mu}-a'_{\mu})
+\rho^2 a_{\rho}
-\frac{1}{2}\mu\rho a_{\mu\rho}] \nonumber \\
   & & -\frac{1}{6}\frac{1}{U_2+2J} 
{\big [} 1- \vec S_a\cdot \vec S_b {\big ]} \nonumber \\
   & & \times [2\mu^2 a''_{\mu}
+2\rho^2 a''_{\rho}
+\mu\rho a''_{\mu\rho}] ~, \nonumber 
\end{eqnarray}

\noindent where, respectively, we have defined: 

\begin{eqnarray}
\label{fb1}
f_{\mu} & = &\tau_{az}^2+\tau_{bz}^2-
\tau_{az}\tau_{bz}-\tau_{az}^2\tau_{bz}^2 \nonumber \\
& & -\frac{1}{2}\tau_a^-\tau_a^-\tau_b^{+}
\tau_b^{+}-\frac{1}{2}\tau_a^{+}
\tau_a^{+}\tau_b^-\tau_b^-  \nonumber \\
f_{\rho} & = &\tau_{az}^2+\tau_{bz}^2-2\tau_{az}^2\tau_{bz}^2 \\
f_{\mu\rho } & = & \tau_a^-\tau_{az}\tau_{bz}\tau_b^{+}+
\tau_{az} \tau_a^-\tau_b^{+}\tau_{bz} \nonumber \\
& & +\tau_a^{+}\tau_{az}\tau_{bz}\tau_b^-+\tau_{az}\tau_a^{+}
\tau_b^-\tau_{bz} \nonumber
\end{eqnarray}

\noindent and

\begin{eqnarray}
\label{afb1}
a_{\mu} & = & 2-\tau_{az}^2-\tau_{bz}^2+
\frac{1}{2}(\tau_{az}\tau_{bz}
+\tau_{az}^2\tau_{bz}^2) \nonumber \\
a'_{\mu}  & = & +\frac{1}{4}(\tau_a^-\tau_a^-\tau_b^-\tau_b^-+\tau_a^{+}
\tau_a^{+}
\tau_b^{+}\tau_b^{+}) \nonumber \\
a_{\mu\rho} & = & \tau_a^-\tau_{az}\tau_b^-\tau_{bz}+
\tau_{az}\tau_a^-\tau_{bz}\tau_b^- \nonumber \\
& & +\tau_a^{+}\tau_{az}\tau_b^{+}\tau_{bz}+\tau_{az}\tau_a^{+}
\tau_{bz}\tau_b^{+} \nonumber \\
a_{\rho} & = & \tau_{az}^2\tau_{bz}^2  \\
a''_{\mu} & = & \tau_{az}^2+\tau_{bz}^2-\tau_{az}\tau_{bz}-
\tau_{az}^2\tau_{bz}^2 \nonumber \\
& & +\frac{1}{4}(\tau_a^-\tau_a^-\tau_b^{+}
\tau_b^{+}+\tau_a^{+}\tau_a^{+}\tau_b^-\tau_b^-) \nonumber \\
a''_{\rho} & = & \tau_{az}^2+\tau_{bz}^2-2\tau_{az}^2\tau_{bz}^2 \nonumber \\
a''_{\mu\rho} & = &
\tau_a^-\tau_{az}\tau_{bz}\tau_b^{+}+
\tau_{az}\tau_a^-\tau_b^{+}\tau_{bz} \nonumber \\
& & +\tau_a^{+}\tau_{az}\tau_{bz}\tau_b^-+
\tau_{az}\tau_a^{+}\tau_b^-\tau_{bz} ~. \nonumber 
\end{eqnarray}

We assume for the moment $\Delta_t = 0$.
Based on Eq. (\ref{fb})  and with the 
definitions of Eq. (\ref{fb1}) we can easily evaluate eigenvalues and eigenstates of $H_{\rm eff}^{F}$. Neglecting the 5-fold spin degeneracy and taking into account only the orbital one, we find the following doubly degenerate ground state with $\tau^M_z=\pm1$:
 
\begin{equation}
|\psi^o_{\pm} \rangle_{ab}=\frac{1}{\sqrt{2}}(|\pm1 \rangle_a|0 \rangle_b+|\pm1 \rangle_b|0 \rangle_a)
\label{psidim}
\end{equation}

Equation (\ref{psidim}) represents only the orbital part of the ground state. 
The whole state (e.g., with $S^M_z=2$) can be pictured as: 

\begin{figure}
\begin{picture}(0,20)
\put (30,0){\makebox(10,0){$|\gamma_{-} \rangle_{ab}=\frac{1}{\sqrt{2}}
{\Bigg (} $ }}
\put (80,18){\circle{24}}
\put (80,18){\line (0,1){12}}
\put (80,18){\line(1,0){12}}
\put (80,18){\line(-1,0){12}}
\put (80,6){\line(0,-1){12}}
\put (80,18){\makebox(-12,12){$\uparrow$}}
\put (68,18){\makebox(24,-12){$\uparrow$}}
\put (80,-18){\circle{24}}
\put (80,-18){\line (0,1){12}}
\put (80,-18){\line(1,0){12}}
\put (80,-18){\line(-1,0){12}}
\put (80,-18){\makebox(-12,12){$\uparrow$}}
\put (80,-18){\makebox(12,12){$\uparrow$}}
\put (100,0){\makebox(5,0){$+ $ }}
\put (120,18){\circle{24}}
\put (120,18){\line (0,1){12}}
\put (120,18){\line(1,0){12}}
\put (120,18){\line(-1,0){12}}
\put (120,6){\line(0,-1){12}}
\put (120,18){\makebox(-12,12){$\uparrow$}}
\put (120,18){\makebox(12,12){$\uparrow$}}
\put (120,-18){\circle{24}}
\put (120,-18){\line (0,1){12}}
\put (120,-18){\line(1,0){12}}
\put (120,-18){\line(-1,0){12}}
\put (120,-18){\makebox(-12,12){$\uparrow$}}
\put (108,-18){\makebox(24,-12){$\uparrow$}}
\put (142,0){\makebox(3,0){${\Bigg )} $ }}
\end{picture}
\end{figure}
\vspace{0.5cm}

\noindent or as 

\begin{figure}
\begin{picture}(0,20)
\put (30,0){\makebox(10,0){$|\gamma_{+} \rangle_{ab}=\frac{1}{\sqrt{2}}
{\Bigg (} $ }}
\put (80,18){\circle{24}}
\put (80,18){\line (0,1){12}}
\put (80,18){\line(1,0){12}}
\put (80,18){\line(-1,0){12}}
\put (80,6){\line(0,-1){12}}
\put (80,18){\makebox(12,12){$\uparrow$}}
\put (68,18){\makebox(24,-12){$\uparrow$}}
\put (80,-18){\circle{24}}
\put (80,-18){\line (0,1){12}}
\put (80,-18){\line(1,0){12}}
\put (80,-18){\line(-1,0){12}}
\put (80,-18){\makebox(-12,12){$\uparrow$}}
\put (80,-18){\makebox(12,12){$\uparrow$}}
\put (100,0){\makebox(5,0){$+ $ }}
\put (120,18){\circle{24}}
\put (120,18){\line (0,1){12}}
\put (120,18){\line(1,0){12}}
\put (120,18){\line(-1,0){12}}
\put (120,6){\line(0,-1){12}}
\put (120,18){\makebox(-12,12){$\uparrow$}}
\put (120,18){\makebox(12,12){$\uparrow$}}
\put (120,-18){\circle{24}}
\put (120,-18){\line (0,1){12}}
\put (120,-18){\line(1,0){12}}
\put (120,-18){\line(-1,0){12}}
\put (120,-18){\makebox(12,12){$\uparrow$}}
\put (108,-18){\makebox(24,-12){$\uparrow$}}
\put (142,0){\makebox(3,0){${\Bigg )}~. $ }}
\end{picture}
\end{figure}
\vspace{1cm}

\noindent The corresponding ground state energy is:
\begin{equation}
\Delta E_m=-\frac{(\mu-\rho)^2}{U_2-J}~.
\label{evb}
\end{equation}
With reference to the picture of the state, this energy lowering (with 
respect to the atomic limit $2(U_2-J)$) is made up of three terms: the 
virtual hopping back and forth of an $e_g$ electron ($-\frac{\mu^2}
{U_2-J}$), the similar process for the $a_{1g}$ electron ($-\frac{\rho^2}
{U_2-J}$) and a sort of correlated hopping in which an $e_g$ electron 
jumps from atom $a$ to atom $b$ while simultaneously an $a_{1g}$ electron 
jumps from atom $b$ to atom $a$ and vice-versa ($\frac{2\rho\mu}{U_2-J}$, 
which is negative due to the opposite sign of $\rho$ and $\mu$, see 
Table III). This latter process is present only due to the ``entangled'' 
orbital nature of the molecular state of Eq. (\ref{psidim}) and is absent 
in its Hartree-Fock approximation, which provides a lowering of only  
$\Delta E_{HF}=-\frac{\mu^2+\rho^2}{U_2-J}$. With the values given in 
Table III, the ratio between the interference and the HF term 
$2\mu\rho/(\mu^2+\rho^2)$ is of the order of 50\%. Therefore the molecular 
correlation energy $\Delta E_m - \Delta E_{HF}=\frac{2\mu\rho}{U_2-J}$
is much bigger than the in-plane exchange energy ($\approx \frac{\alpha^2+ 
\tau^2}{U_2-J}$), so that in this case the best variational wave function 
for the entire crystal should be constructed in terms of molecular states.

Another point worth mentioning here is the quality of the expansion around 
the atomic limit. The exact 
solution of the $2\times 2$ eigenvalue problem for the ferromagnetic 
vertical molecule is given by 
\begin{eqnarray}
\Delta E_m &=& E_m - 2(U_2 - J) \nonumber \\
&=& \frac{(U_2 - J)}{2} 
\left (1 - \sqrt{1 + \frac{4(\mu-\rho)^2}{(U_2-J)^2}}\right)~.
\label{evb1}
\end{eqnarray}
We see from this expression that the expansion parameter is 
$\frac{2(\mu-\rho)}{(U_2-J)}$, which is of the order of one, using 
the standard values of Table III. This value is borderline for a good 
expansion; however, as often happens in perturbation theory, the second 
order term turns out to be a reasonable approximation to the exact result 
($1+0.89$ as compared to $\sqrt{2.78}=1.67$, with a relative error of less than $13\%$ in the worst of the cases). Moreover this problem 
is present only for the vertical pairs, since in the basal plane we are 
well within the values for a rapidly convergent expansion. Notice that, when 
comparing different variational minimal solutions of $H_{\rm eff}$ in section VI, the error in 
the vertical pairs will cancel out and the result will be of the same 
accuracy as the expansion for bonds in the basal plane.

As long as $|\rho|\!\! >\!\! |\mu|$, the ferromagnetic state with $S^M = 2$ in 
Eq. (\ref{psidim}) is the ground state for the vertical pair. However 
it is easy to realize that, in the opposite case $|\rho| < |\mu|$, the orbital part of the
ground state changes to

\begin{equation}
|\psi^o_0 \rangle_{ab}=\frac{1}{\sqrt{2}}(|1 \rangle_a|-1 \rangle_b+|1 \rangle_b|-1 \rangle_a)
\label{phidim}
\end{equation}

\vspace{-0.5cm}

\noindent or, pictorially, including the spin ($S^M_z=2$):

\vspace{0.5cm}

\begin{figure}
\begin{picture}(0,20)
\put (50,0){\makebox(10,0){$|\gamma_0 \rangle_{ab} = \frac{1}{\sqrt{2}}
{\Bigg (} $ }}
\put (100,18){\circle{24}}
\put (100,18){\line (0,1){12}}
\put (100,18){\line(1,0){12}}
\put (100,18){\line(-1,0){12}}
\put (100,6){\line(0,-1){12}}
\put (100,18){\makebox(-12,12){$\uparrow$}}
\put (88,18){\makebox(24,-12){$\uparrow$}}
\put (100,-18){\circle{24}}
\put (100,-18){\line (0,1){12}}
\put (100,-18){\line(1,0){12}}
\put (100,-18){\line(-1,0){12}}
\put (100,-18){\makebox(12,12){$\uparrow$}}
\put (88,-18){\makebox(24,-12){$\uparrow$}}
\put (120,0){\makebox(5,0){$+ $ }}
\put (140,18){\circle{24}}
\put (140,18){\line (0,1){12}}
\put (140,18){\line(1,0){12}}
\put (140,18){\line(-1,0){12}}
\put (140,6){\line(0,-1){12}}
\put (140,18){\makebox(12,12){$\uparrow$}}
\put (128,18){\makebox(24,-12){$\uparrow$}}
\put (140,-18){\circle{24}}
\put (140,-18){\line (0,1){12}}
\put (140,-18){\line(1,0){12}}
\put (140,-18){\line(-1,0){12}}
\put (140,-18){\makebox(-12,12){$\uparrow$}}
\put (128,-18){\makebox(24,-12){$\uparrow$}}
\put (160,0){\makebox(3,0){${\Bigg )} $ }}
\end{picture}
\end{figure}
\vspace{1.0cm}

\noindent with an energy lowering of $\Delta E_m'=-\frac{4\mu^2}{U_2-J}$. 

\vspace{0.1cm}

This state is not orbitally degenerate. However 
it is interesting to note that in this case the percentage of occupation
of the $a_{1g}$ state is 50\%, so that this solution is excluded by the 
findings of Park {\it et al.},\cite{park00} as well as on the basis of 
the theoretical estimates  of  Table III ($\rho\! > \!\mu$).

As long as $\frac{J}{U_2} > 0.22$ the ferromagnetic configuration in Eq. 
(\ref{psidim}) remains the ground state of the pair. By decreasing $J$ 
a transition to an antiferromagnetic ($S^M = 0$) ground state is 
expected, since this spin configuration will maximize the number of virtual hopping processes
 without loosing too much in Hund's energy. This is indeed 
what happens when $\frac{J}{U_2} < 0.22$.

Even in this case we obtain a two-fold orbitally degenerate ground state:

\vspace{-0.2cm}

\begin{equation}
|\psi^o_{|2|} \rangle_{ab}\simeq\frac{1}{\sqrt{2}}(|-1 \rangle_a|-1 \rangle_b - 
|1 \rangle_b|1 \rangle_a)~, 
\label{psiaf1}
\end{equation}

\vspace{-0.2cm}

\begin{figure}
\begin{picture}(0,20)
\put (50,0){\makebox(10,0){$|\psi^o_{|2|} \rangle_{ab}\simeq\frac{1}{\sqrt{2}}
{\Bigg (} $ }}
\put (100,18){\circle{24}}
\put (100,18){\line (0,1){12}}
\put (100,18){\line(1,0){12}}
\put (100,18){\line(-1,0){12}}
\put (100,6){\line(0,-1){12}}
\put (100,18){\makebox(-12,12){$\bullet$}}
\put (88,18){\makebox(24,-12){$\bullet$}}
\put (100,-18){\circle{24}}
\put (100,-18){\line (0,1){12}}
\put (100,-18){\line(1,0){12}}
\put (100,-18){\line(-1,0){12}}
\put (100,-18){\makebox(-12,12){$\bullet$}}
\put (88,-18){\makebox(24,-12){$\bullet$}}
\put (120,0){\makebox(5,0){$ - $ }}
\put (140,18){\circle{24}}
\put (140,18){\line (0,1){12}}
\put (140,18){\line(1,0){12}}
\put (140,18){\line(-1,0){12}}
\put (140,6){\line(0,-1){12}}
\put (140,18){\makebox(12,12){$\bullet$}}
\put (128,18){\makebox(24,-12){$\bullet$}}
\put (140,-18){\circle{24}}
\put (140,-18){\line (0,1){12}}
\put (140,-18){\line(1,0){12}}
\put (140,-18){\line(-1,0){12}}
\put (140,-18){\makebox(12,12){$\bullet$}}
\put (128,-18){\makebox(24,-12){$\bullet$}}
\put (162,0){\makebox(3,0){${\Bigg )} $ }}
\end{picture}
\end{figure}

\vspace{0.8cm}

\noindent and 


\begin{equation}
|\psi^o_0 \rangle_{ab}\simeq\frac{1}{\sqrt{2}}(|-1 \rangle_a|1 \rangle_b+|-1 \rangle_b|1 \rangle_a)~, 
\label{psiaf2}
\end{equation}
\begin{figure}
\begin{picture}(0,20)
\put (50,0){\makebox(10,0){$|\psi^o_0 \rangle_{ab}\simeq\frac{1}{\sqrt{2}}
{\Bigg (} $ }}
\put (100,18){\circle{24}}
\put (100,18){\line (0,1){12}}
\put (100,18){\line(1,0){12}}
\put (100,18){\line(-1,0){12}}
\put (100,6){\line(0,-1){12}}
\put (100,18){\makebox(-12,12){$\bullet$}}
\put (88,18){\makebox(24,-12){$\bullet$}}
\put (100,-18){\circle{24}}
\put (100,-18){\line (0,1){12}}
\put (100,-18){\line(1,0){12}}
\put (100,-18){\line(-1,0){12}}
\put (100,-18){\makebox(12,12){$\bullet$}}
\put (88,-18){\makebox(24,-12){$\bullet$}}
\put (120,0){\makebox(5,0){$+ $ }}
\put (140,18){\circle{24}}
\put (140,18){\line (0,1){12}}
\put (140,18){\line(1,0){12}}
\put (140,18){\line(-1,0){12}}
\put (140,6){\line(0,-1){12}}
\put (140,18){\makebox(12,12){$\bullet$}}
\put (128,18){\makebox(24,-12){$\bullet$}}
\put (140,-18){\circle{24}}
\put (140,-18){\line (0,1){12}}
\put (140,-18){\line(1,0){12}}
\put (140,-18){\line(-1,0){12}}
\put (140,-18){\makebox(-12,12){$\bullet$}}
\put (128,-18){\makebox(24,-12){$\bullet$}}
\put (162,0){\makebox(3,0){${\Bigg )} $ }}
\end{picture}
\end{figure}

\vspace{1cm}

For simplicity of presentation 
we have omitted to show the spin structure of this state, since 
this latter is given in the following section V-B (see states $|\gamma_1\rangle$ and $|\gamma_2\rangle$ of Eq. (\ref{sm0})). The ground state energy is given by:

\begin{equation}
\Delta E^{AF}_m= -\frac{3}{2} \bigg [ \frac{\rho^2}{U_2+4J}+\frac{\rho^2+2\mu^2}{U_2+2J}   \bigg ]  ~.
\label{gseafm}
\end{equation}

Note that the first state (Eq. (\ref{psiaf1})), mixing the values  $\tau_z^M=\pm 2$ does not conserve the value of pseudospin $z$-component, due to the term $H_{0j}^{(4)}$ in Eq. (\ref{h2}). 
\vspace{0.1cm}

The above level scheme is confirmed by the exact treatment of the vertical 
pair on the basis of the original Hubbard Hamiltonian which is reported 
in the following section V-B. There are slight discrepancies, however, due 
to the non optimal conditions for perturbation theory. For example, the 
transition value between the ferromagnetic and antiferromagnetic 
configurations is found at $\frac{J}{U_2} = 0.29$. Moreover the degeneracy of
the two antiferromagnetic states is removed, the second lying always 
lowest (from 1 to 5 meV, in the range of parameters of interest), due to the different mixing with states that have been projected out in the perturbation theory (those with $S_a=S_b=0$ and those with $S_a=S_b=1/2$). In general, though, the exact energy level structure is reasonably 
close to the approximate one.

By switching on the trigonal distortion $\Delta_t$, the ferromagnetic ground state energy 
(\ref{evb}) changes to
\begin{equation}
\Delta E_{mt}=-\frac{(\mu-\rho)^2}{U_2-J}+\Delta_t 
\label{evbdelta}
\end{equation}

and the antiferromagnetic (\ref{gseafm}) becomes

\begin{equation}
\Delta E^{AF}_{mt1}= -\frac{3}{2} \bigg [ \frac{\rho^2}{U_2+4J}+\frac{\rho^2+2\mu^2}{U_2+2J}   \bigg ]+2 \Delta_t   ~.
\label{gseafm2}
\end{equation}

Because of this, the stability 
region of the ferromagnetic state in the parameter $\frac{J}{U_2}$ 
initially increases with $\Delta_t$, since its $a_{1g}$ population is only 25\%, compared
to the 50\% value of the antiferromagnetic states. However for $\Delta_t$ bigger than a
critical value $\overline{\Delta}_t$  there is an 
inversion of tendency and the stability 
region begins to decrease. This is due to the fact that the structure of 
the antiferromagnetic state changes abruptly from a situation in which 
the $a_{1g}$ population is 50\% to one in which is 0\%. Indeed for 
$\Delta_t \geq \overline{\Delta}_t$ the lowest energy configuration for the AF bond is reached when the two electrons on each site occupy both $e_g$ orbitals and are coupled to 
spin $S=1$, with total spin $S^M=0$: in this case the ground state orbital configuration, for $J$ not too close to zero (i.e., $J\ge 0.2$ eV, due to the constraint $2J\ge\Delta_t$ used in our perturbation theory) is given by:

\vspace{0.2cm}

\begin{figure}
\begin{picture}(0,20)
\put (70,0){\makebox(10,0){$|\psi^o_00 \rangle_{ab}\simeq |0 \rangle_a|0 \rangle_b = $ }}
\put (150,18){\circle{24}}
\put (150,18){\line (0,1){12}}
\put (150,18){\line(1,0){12}}
\put (150,18){\line(-1,0){12}}
\put (150,6){\line(0,-1){12}}
\put (150,18){\makebox(12,12){$\bullet$}}
\put (150,18){\makebox(-12,12){$\bullet$}}
\put (150,-18){\circle{24}}
\put (150,-18){\line (0,1){12}}
\put (150,-18){\line(1,0){12}}
\put (150,-18){\line(-1,0){12}}
\put (150,-18){\makebox(-12,12){$\bullet$}}
\put (150,-18){\makebox(12,12){$\bullet$}}
\end{picture}
\end{figure}

\vspace{-1.3cm}

\begin{equation}
\label{psiaf3b}
\end{equation}

\vspace{0.8cm}

The full spin structure of the state (\ref{psiaf3b}) will be given in the next subsection (see state $|\gamma_9\rangle$). The ground state AF energy, in this case, is:

\begin{equation}
\Delta E^{AF}_{mt2}= -3\mu^2 \bigg [ \frac{1}{U_2+4J}+\frac{1}{U_2+2J}   \bigg ]  ~.
\label{gseafm3}
\end{equation}

Finally the estimate of  $\overline{\Delta}_t$, obtained  by comparing Eqs. (\ref{evbdelta}), (\ref{gseafm2}) and (\ref{gseafm3}), and using  Mattheiss parameters of Table III, is $\overline{\Delta}_t \simeq 0.34 ~eV$, in good agreement with the exact value calculated in section V-B ($\overline{\Delta}_t \simeq 0.30 ~eV$).

We shall not dwell anymore on this subject 
since it will be studied more in depth in section V-B.

\subsection{The exact solution using the Hubbard Hamiltonian.}

In constructing the effective Hamiltonian (see section II) we have 
assumed that $J$ is a high energy parameter and therefore have excluded from our zeroth order degenerate manifold singlet spin states, lying at higher energies by $2J$ or more.
In order to assess the range validity of $H_{\rm eff}$ as a function of $J$ 
and the stability of the vertical molecule for $J \rightarrow 0$, 
we examine here the ground state configuration of the vertical molecule  
using Hubbard Hamiltonian ($H_0 + H'$). 
In this case the Hilbert space is made up of 495 atomic states 
with up to four electrons per site and the 
eigen-problem is rather complicated. However, due to the SU(2) 
invariance of the Hubbard Hamiltonian, states with different 
spin $S^M$ and different $S_z^M$  do not mix.
Furthermore, since by assumption only diagonal hopping integrals are 
different from zero, the kinetic and crystal field terms 
$H'$ in Eq. (\ref{h3})
do not change the total molecular pseudospin $\tau_z^M$, while the atomic 
part $H_0$ can only mix states with the same parity of $\tau_z^M$, due
to the $H_{0j}^{(4)}$ term in Eq. (\ref{h2}). 
We can therefore divide the states of our Hilbert space into 6 groups 
which are characterized by the following quantum numbers:

\vspace{0.1cm}

\noindent
1) $S^M=S_z^M=0$ and $\tau_z^M$ even (57 states):\\
\hspace*{3mm} $\tau_z^M=\pm 4$ (2 states); \\
\hspace*{3mm} $\tau_z^M=\pm 2$ (28 states); \\
\hspace*{3mm} $\tau_z^M=0$ (27 states);\\
2) $S^M=S_z^M=0$ and $\tau_z^M$ odd (48 states):\\
\hspace*{3mm} $\tau_z^M=\pm 3$ (8 states); \\
\hspace*{3mm} $\tau_z^M=\pm 1$ (40 states);\\
3) $S^M=1$, $S_z^M$ fixed and $\tau_z^M$ even (49 states):\\
\hspace*{3mm} $\tau_z^M=\pm 2$ (22 states); \\
\hspace*{3mm} $\tau_z^M=0$ (27 states);\\
4) $S^M=1$, $S_z^M$ fixed and $\tau_z^M$ odd (56 states):\\
\hspace*{3mm} $\tau_z^M=\pm 3$ (8 states); \\
\hspace*{3mm} $\tau_z^M=\pm 1$ (48 states);\\
5) $S^M=2$, $S_z^M$ fixed and $\tau_z^M$ even (7 states):\\
\hspace*{3mm} $\tau_z^M=\pm 2$ (2 states);\\
\hspace*{3mm} $\tau_z^M=0$ (5 states);\\
6) $S^M=2$, $S_z^M$ fixed and $\tau_z^M$ odd (8 states):\\
\hspace*{3mm} $\tau_z^M=\pm 1$ (8 states).

\end{multicols}

\vspace{0.7cm}


\parbox[b]{3.3in}
{\begin{figure}
      \epsfysize=55mm
      \centerline{\epsffile{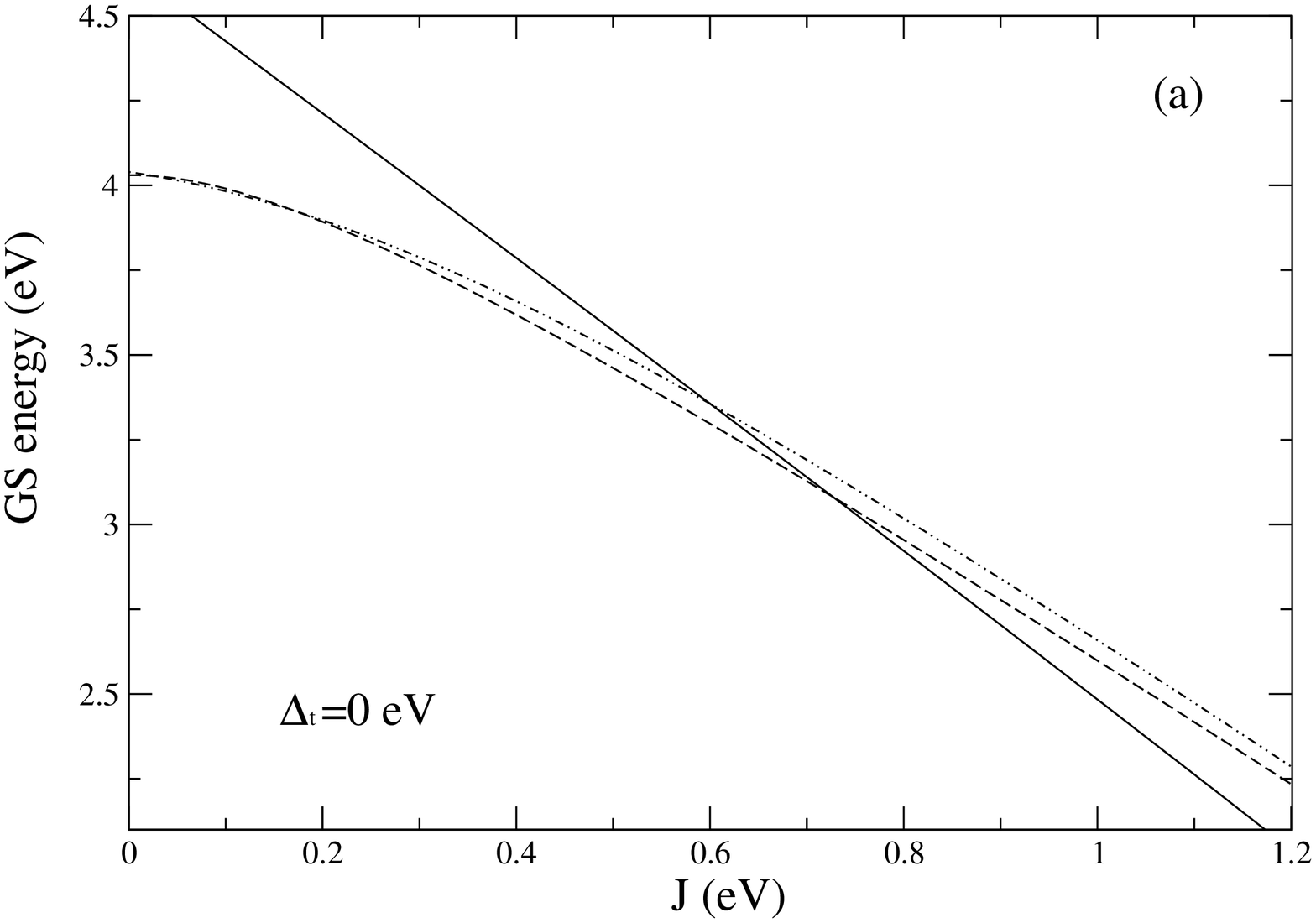}}
\end{figure}}
\parbox[b]{3.3in}
{\begin{figure}
      \epsfysize=55mm
      \centerline{\epsffile{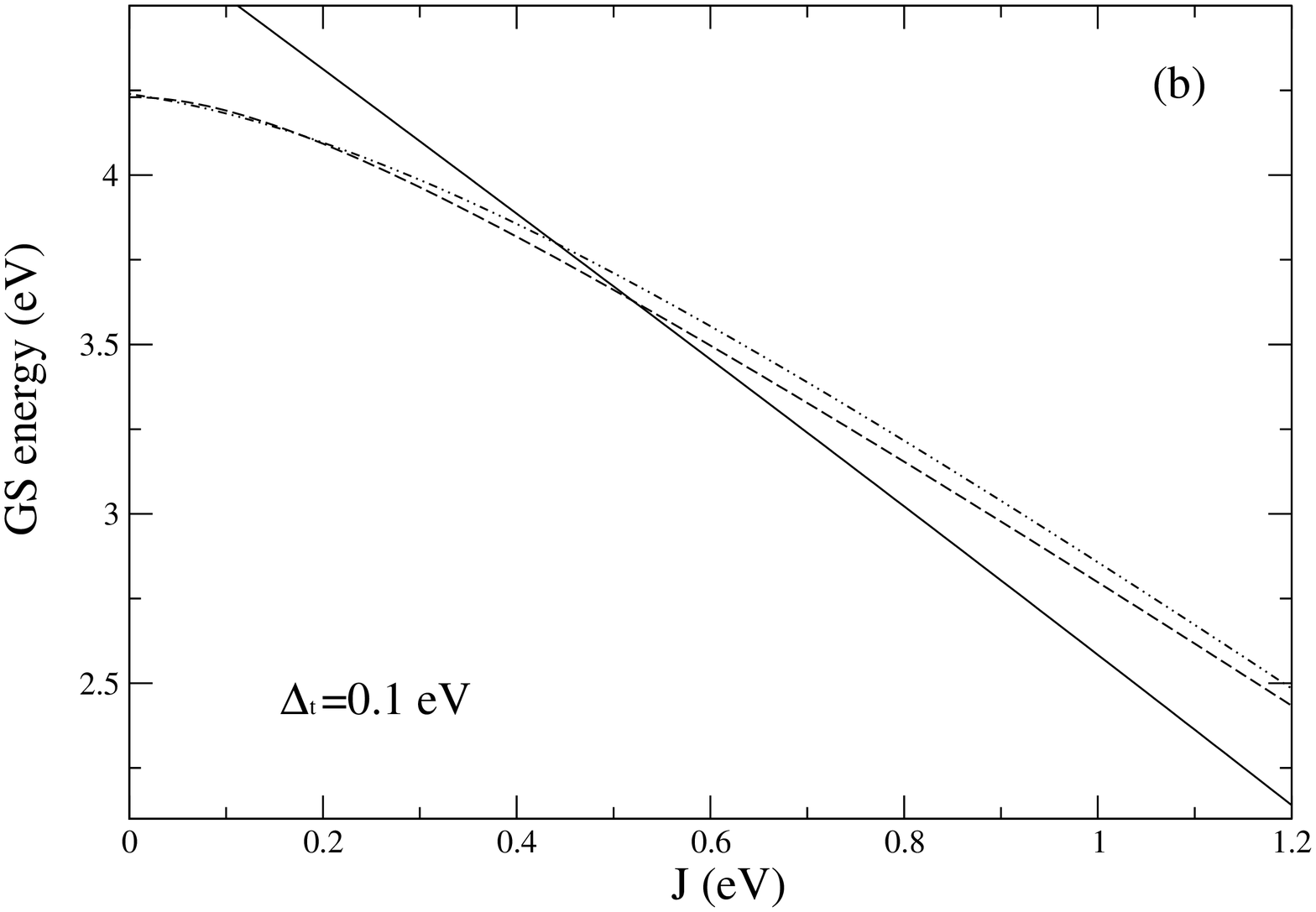}}
\end{figure}}

\parbox[b]{3.3in}
{\begin{figure}
        \epsfysize=55mm
        \centerline{\epsffile{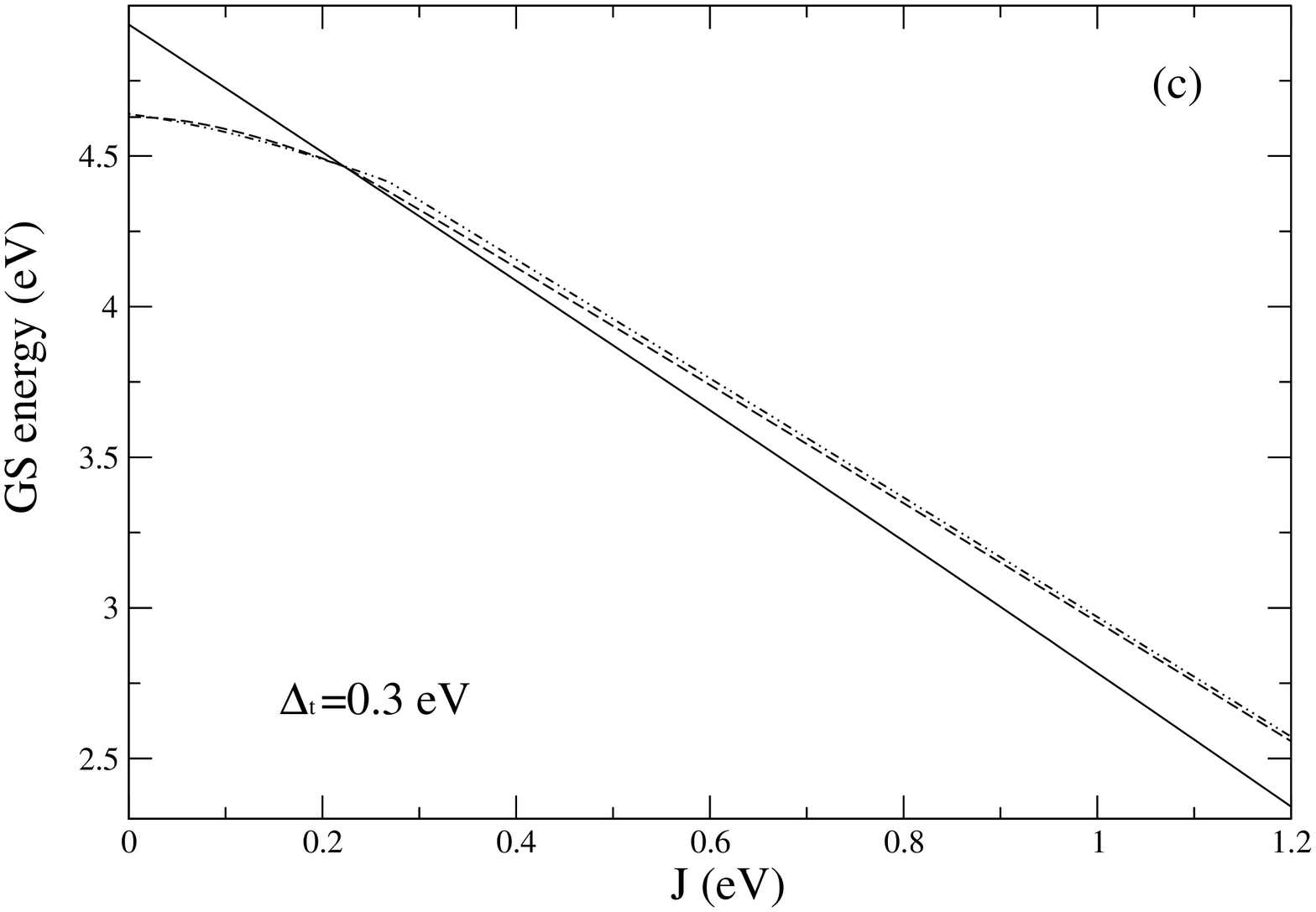}}
\end{figure}}
\parbox[b]{3.3in}
{\begin{figure}
        \epsfysize=55mm
        \centerline{\epsffile{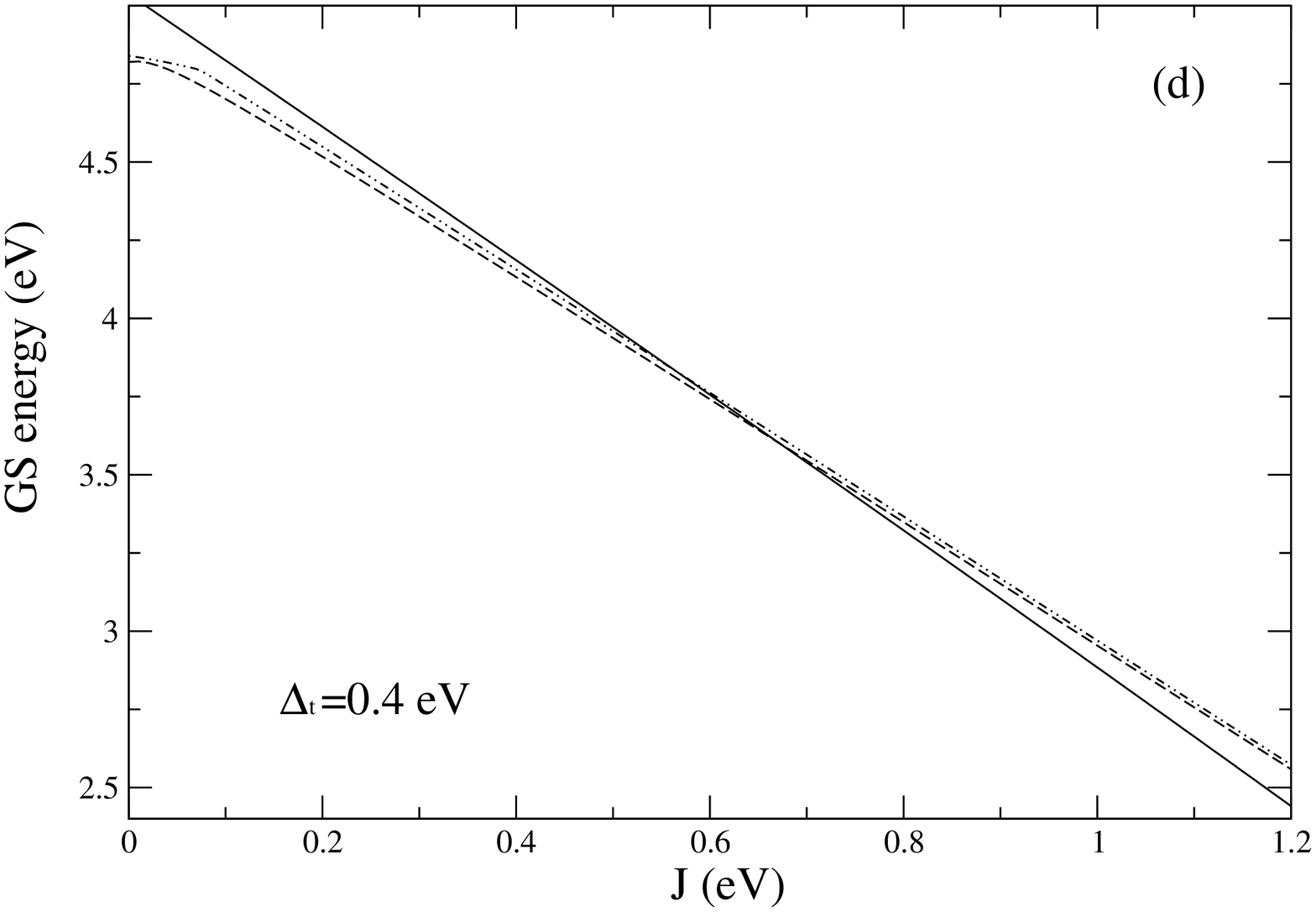}}
\end{figure}}
\vspace{-0.5cm}
\begin{figure}
\caption{Energy eigenvalues of the Hubbard Hamiltonian for the vertical molecule as a 
function of $J$:
solid line -- $S^M=2$ and $\tau_z^M$ odd;
dot-dot-dashed line -- $S^M=1$ and $\tau_z^M$ even;
dashed line -- $S^M=0$ and $\tau_z^M$ even. Panels
$(a), (b), (c)$ and 
$(d)$ correspond to the  trigonal distortion $\Delta_t=0$ eV, $\Delta_t=0.1$ eV,
$\Delta_t=0.3$ eV and $\Delta_t=0.4$ eV, respectively.}
\label{hubd0} 
\end{figure}

\protect\begin{multicols}{2}

\vspace{0.2cm}

\noindent This classifications is valid for each $S_z^M$ component, so that for $S^M=1$ we get $(49+56)\times 3=315$ states and for $S^M=2$ we get $(7+8)\times 5=75$ states. It is worth 
noticing that when the $H_{0j}^{(4)}$ term in Eq. (\ref{h2}) is 
not effective, $\tau_z^M$ is conserved and further reduction of 
the number of mixing states is possible. This is the case for all the states with $S^M=2$, where the 15 states of fixed $S_z^M$ component can be grouped into the orthogonal sets: $\tau_z^M=+2$ (1 state), $\tau_z^M=+1$ (4 states), $\tau_z^M=0$ (5 states), $\tau_z^M=-1$ (4 states), $\tau_z^M=-2$ (1 state).

We take the standard values of the parameters (Mattheiss set) and 
look at the eigenvalue structure as a function of $J$ 
for different subgroups of the Hilbert space (see Fig. \ref{hubd0}).
For this choice of the parameters the three lowest energies are always 
in the groups of states with  
 $S^M=0$ and $\tau_z^M$ even,  $S^M=1$ and
 $\tau_z^M$ even or 
 $S^M=2$ and $\tau_z^M$ odd. They are presented  
in Fig. \ref{hubd0}.

The crystal field degenerate case 
($\Delta_t=0$ eV) is presented in the Fig. \ref{hubd0}$(a)$.
For $ J\rightarrow 0$ eV the ground state 
of the vertical pair belongs to the sector
$S^M=0$ with $\tau_z^M$ even.
By increasing $J$ three transitions occur.
At very low $ J\sim 0.04$ eV  a first transition to a state with  
$S^M=1$ with $\tau_z^M$ even takes place.
This state is reminiscent of the ground state postulated by CNR \cite{cnr1} for 
the vertical molecule (about $55\%$ of its weight is composed by the old CNR state for $J=0.04$) and, because of this, it does not belong to the Hilbert subspace upon which $H_{\rm eff}$ can operate.
In this region the $S^M=0$ with $\tau_z^M$ even state lies only 3 meV 
higher in energy. 
Then at $ J\simeq 0.17$ eV  a second transition takes place again 
toward a state with $S^M=0$ and $\tau_z^M$ even. 
To get an idea of its composition in terms of atomic states, we give  
its expression at $ J= 0.73$ eV, that is the upper boundary of the region of stability of the $S^M=0$ state. Note, that even though
the weight of the particular state depends on the value of $J$, the
tendency of the weight distribution is the same in the whole region
$0.17$ eV$\leq J \leq 0.73$ eV. We get:


\begin{eqnarray}
\begin{array}{l}
\parallel GS \rangle_{J=0.73}\simeq
0.67|\gamma_1\rangle +0.67|\gamma_2\rangle\\[0.3cm]
+0.14|\gamma_3\rangle +0.14|\gamma_{4}\rangle
+0.14|\gamma_{5}\rangle +0.14|\gamma_{6}\rangle ~,
\end{array}
\label{j65}
\end{eqnarray}
where
\end{multicols}

\begin{widetext}
\begin{figure}
\begin{picture}(0,20)
\put (30,0){\makebox(10,0){$|\gamma_{1} \rangle = \frac{1}{2\sqrt{3}}
{\Bigg (}2 $ }}
\put (80,18){\circle{24}}
\put (80,18){\line (0,1){12}}
\put (80,18){\line(1,0){12}}
\put (80,18){\line(-1,0){12}}
\put (80,6){\line(0,-1){12}}
\put (80,18){\makebox(-12,12){$\uparrow$}}
\put (68,18){\makebox(24,-12){$\uparrow$}}
\put (80,-18){\circle{24}}
\put (80,-18){\line (0,1){12}}
\put (80,-18){\line(1,0){12}}
\put (80,-18){\line(-1,0){12}}
\put (80,-18){\makebox(12,12){$\downarrow$}}
\put (68,-18){\makebox(24,-12){$\downarrow$}}
\put (100,0){\makebox(5,0){$- $ }}
\put (120,18){\circle{24}}
\put (120,18){\line (0,1){12}}
\put (120,18){\line(1,0){12}}
\put (120,18){\line(-1,0){12}}
\put (120,6){\line(0,-1){12}}
\put (120,18){\makebox(-12,12){$\uparrow$}}
\put (108,18){\makebox(24,-12){$\downarrow$}}
\put (120,-18){\circle{24}}
\put (120,-18){\line (0,1){12}}
\put (120,-18){\line(1,0){12}}
\put (120,-18){\line(-1,0){12}}
\put (120,-18){\makebox(12,12){$\uparrow$}}
\put (108,-18){\makebox(24,-12){$\downarrow$}}
\put (140,0){\makebox(5,0){$- $ }}
\put (160,18){\circle{24}}
\put (160,18){\line (0,1){12}}
\put (160,18){\line(1,0){12}}
\put (160,18){\line(-1,0){12}}
\put (160,6){\line(0,-1){12}}
\put (160,18){\makebox(-12,12){$\uparrow$}}
\put (148,18){\makebox(24,-12){$\downarrow$}}
\put (160,-18){\circle{24}}
\put (160,-18){\line (0,1){12}}
\put (160,-18){\line(1,0){12}}
\put (160,-18){\line(-1,0){12}}
\put (160,-18){\makebox(12,12){$\downarrow$}}
\put (148,-18){\makebox(24,-12){$\uparrow$}}
\put (180,0){\makebox(5,0){$+ 2$ }}
\put (200,18){\circle{24}}
\put (200,18){\line (0,1){12}}
\put (200,18){\line(1,0){12}}
\put (200,18){\line(-1,0){12}}
\put (200,6){\line(0,-1){12}}
\put (200,18){\makebox(-12,12){$\downarrow$}}
\put (188,18){\makebox(24,-12){$\downarrow$}}
\put (200,-18){\circle{24}}
\put (200,-18){\line (0,1){12}}
\put (200,-18){\line(1,0){12}}
\put (200,-18){\line(-1,0){12}}
\put (200,-18){\makebox(12,12){$\uparrow$}}
\put (188,-18){\makebox(24,-12){$\uparrow$}}
\put (220,0){\makebox(5,0){$- $ }}
\put (240,18){\circle{24}}
\put (240,18){\line (0,1){12}}
\put (240,18){\line(1,0){12}}
\put (240,18){\line(-1,0){12}}
\put (240,6){\line(0,-1){12}}
\put (240,18){\makebox(-12,12){$\downarrow$}}
\put (228,18){\makebox(24,-12){$\uparrow$}}
\put (240,-18){\circle{24}}
\put (240,-18){\line (0,1){12}}
\put (240,-18){\line(1,0){12}}
\put (240,-18){\line(-1,0){12}}
\put (240,-18){\makebox(12,12){$\uparrow$}}
\put (228,-18){\makebox(24,-12){$\downarrow$}}
\put (260,0){\makebox(5,0){$-$ }}
\put (280,18){\circle{24}}
\put (280,18){\line (0,1){12}}
\put (280,18){\line(1,0){12}}
\put (280,18){\line(-1,0){12}}
\put (280,6){\line(0,-1){12}}
\put (280,18){\makebox(-12,12){$\downarrow$}}
\put (268,18){\makebox(24,-12){$\uparrow$}}
\put (280,-18){\circle{24}}
\put (280,-18){\line (0,1){12}}
\put (280,-18){\line(1,0){12}}
\put (280,-18){\line(-1,0){12}}
\put (280,-18){\makebox(12,12){$\downarrow$}}
\put (268,-18){\makebox(24,-12){$\uparrow$}}
\put (300,0){\makebox(3,0){${\Bigg )} $ }}
\end{picture}
\end{figure}

\vspace{1.0cm}
\begin{figure}
\begin{picture}(0,20)
\put (30,0){\makebox(10,0){$|\gamma_{2} \rangle = \frac{1}{2\sqrt{3}}
{\Bigg (}2 $ }}
\put (80,18){\circle{24}}
\put (80,18){\line (0,1){12}}
\put (80,18){\line(1,0){12}}
\put (80,18){\line(-1,0){12}}
\put (80,6){\line(0,-1){12}}
\put (80,18){\makebox(12,12){$\uparrow$}}
\put (68,18){\makebox(24,-12){$\uparrow$}}
\put (80,-18){\circle{24}}
\put (80,-18){\line (0,1){12}}
\put (80,-18){\line(1,0){12}}
\put (80,-18){\line(-1,0){12}}
\put (80,-18){\makebox(-12,12){$\downarrow$}}
\put (68,-18){\makebox(24,-12){$\downarrow$}}
\put (100,0){\makebox(5,0){$- $ }}
\put (120,18){\circle{24}}
\put (120,18){\line (0,1){12}}
\put (120,18){\line(1,0){12}}
\put (120,18){\line(-1,0){12}}
\put (120,6){\line(0,-1){12}}
\put (120,18){\makebox(12,12){$\uparrow$}}
\put (108,18){\makebox(24,-12){$\downarrow$}}
\put (120,-18){\circle{24}}
\put (120,-18){\line (0,1){12}}
\put (120,-18){\line(1,0){12}}
\put (120,-18){\line(-1,0){12}}
\put (120,-18){\makebox(-12,12){$\uparrow$}}
\put (108,-18){\makebox(24,-12){$\downarrow$}}
\put (140,0){\makebox(5,0){$- $ }}
\put (160,18){\circle{24}}
\put (160,18){\line (0,1){12}}
\put (160,18){\line(1,0){12}}
\put (160,18){\line(-1,0){12}}
\put (160,6){\line(0,-1){12}}
\put (160,18){\makebox(12,12){$\uparrow$}}
\put (148,18){\makebox(24,-12){$\downarrow$}}
\put (160,-18){\circle{24}}
\put (160,-18){\line (0,1){12}}
\put (160,-18){\line(1,0){12}}
\put (160,-18){\line(-1,0){12}}
\put (160,-18){\makebox(-12,12){$\downarrow$}}
\put (148,-18){\makebox(24,-12){$\uparrow$}}
\put (180,0){\makebox(5,0){$+ 2$ }}
\put (200,18){\circle{24}}
\put (200,18){\line (0,1){12}}
\put (200,18){\line(1,0){12}}
\put (200,18){\line(-1,0){12}}
\put (200,6){\line(0,-1){12}}
\put (200,18){\makebox(12,12){$\downarrow$}}
\put (188,18){\makebox(24,-12){$\downarrow$}}
\put (200,-18){\circle{24}}
\put (200,-18){\line (0,1){12}}
\put (200,-18){\line(1,0){12}}
\put (200,-18){\line(-1,0){12}}
\put (200,-18){\makebox(-12,12){$\uparrow$}}
\put (188,-18){\makebox(24,-12){$\uparrow$}}
\put (220,0){\makebox(5,0){$- $ }}
\put (240,18){\circle{24}}
\put (240,18){\line (0,1){12}}
\put (240,18){\line(1,0){12}}
\put (240,18){\line(-1,0){12}}
\put (240,6){\line(0,-1){12}}
\put (240,18){\makebox(12,12){$\downarrow$}}
\put (228,18){\makebox(24,-12){$\uparrow$}}
\put (240,-18){\circle{24}}
\put (240,-18){\line (0,1){12}}
\put (240,-18){\line(1,0){12}}
\put (240,-18){\line(-1,0){12}}
\put (240,-18){\makebox(-12,12){$\uparrow$}}
\put (228,-18){\makebox(24,-12){$\downarrow$}}
\put (260,0){\makebox(5,0){$-$ }}
\put (280,18){\circle{24}}
\put (280,18){\line (0,1){12}}
\put (280,18){\line(1,0){12}}
\put (280,18){\line(-1,0){12}}
\put (280,6){\line(0,-1){12}}
\put (280,18){\makebox(12,12){$\downarrow$}}
\put (268,18){\makebox(24,-12){$\uparrow$}}
\put (280,-18){\circle{24}}
\put (280,-18){\line (0,1){12}}
\put (280,-18){\line(1,0){12}}
\put (280,-18){\line(-1,0){12}}
\put (280,-18){\makebox(-12,12){$\downarrow$}}
\put (268,-18){\makebox(24,-12){$\uparrow$}}
\put (300,0){\makebox(3,0){${\Bigg )} $ }}
\end{picture}
\end{figure}

\end{widetext}
\vspace{1.0cm}
\protect\begin{multicols}{2}


\begin{figure}
\begin{picture}(0,20)
\put (30,0){\makebox(10,0){$|\gamma_{3} \rangle =\frac{1}{\sqrt{2}}{\Bigg (} $ }}
\put (80,18){\circle{24}}
\put (80,18){\line (0,1){12}}
\put (80,18){\line(1,0){12}}
\put (80,18){\line(-1,0){12}}
\put (80,6){\line(0,-1){12}}
\put (80,18){\makebox(-12,12){$\uparrow$}}
\put (68,18){\makebox(24,-12){$\downarrow\uparrow$}}
\put (80,-18){\circle{24}}
\put (80,-18){\line (0,1){12}}
\put (80,-18){\line(1,0){12}}
\put (80,-18){\line(-1,0){12}}
\put (80,-18){\makebox(12,12){$\downarrow$}}
\put (100,0){\makebox(5,0){$- $ }}
\put (120,18){\circle{24}}
\put (120,18){\line (0,1){12}}
\put (120,18){\line(1,0){12}}
\put (120,18){\line(-1,0){12}}
\put (120,6){\line(0,-1){12}}
\put (120,18){\makebox(-12,12){$\downarrow$}}
\put (108,18){\makebox(24,-12){$\downarrow\uparrow$}}
\put (120,-18){\circle{24}}
\put (120,-18){\line (0,1){12}}
\put (120,-18){\line(1,0){12}}
\put (120,-18){\line(-1,0){12}}
\put (120,-18){\makebox(12,12){$\uparrow$}}
\put (142,0){\makebox(3,0){${\Bigg )} $ }}
\end{picture}
\end{figure}
\vspace{1cm}
\begin{figure}
\begin{picture}(0,20)
\put (30,0){\makebox(10,0){$|\gamma_{4} \rangle =\frac{1}{\sqrt{2}}{\Bigg (} $ }}
\put (80,18){\circle{24}}
\put (80,18){\line (0,1){12}}
\put (80,18){\line(1,0){12}}
\put (80,18){\line(-1,0){12}}
\put (80,6){\line(0,-1){12}}
\put (80,18){\makebox(12,12){$\uparrow$}}
\put (80,-18){\circle{24}}
\put (80,-18){\line (0,1){12}}
\put (80,-18){\line(1,0){12}}
\put (80,-18){\line(-1,0){12}}
\put (80,-18){\makebox(-12,12){$\downarrow$}}
\put (68,-18){\makebox(24,-12){$\downarrow\uparrow$}}
\put (100,0){\makebox(5,0){$- $ }}
\put (120,18){\circle{24}}
\put (120,18){\line (0,1){12}}
\put (120,18){\line(1,0){12}}
\put (120,18){\line(-1,0){12}}
\put (120,6){\line(0,-1){12}}
\put (120,18){\makebox(12,12){$\downarrow$}}
\put (120,-18){\circle{24}}
\put (120,-18){\line (0,1){12}}
\put (120,-18){\line(1,0){12}}
\put (120,-18){\line(-1,0){12}}
\put (120,-18){\makebox(-12,12){$\uparrow$}}
\put (108,-18){\makebox(24,-12){$\downarrow\uparrow$}}
\put (142,0){\makebox(3,0){${\Bigg )} $ }}
\end{picture}
\end{figure}

\vspace{1.0cm}

\begin{figure}
\begin{picture}(0,20)
\put (30,0){\makebox(10,0){$|\gamma_{5} \rangle =\frac{1}{\sqrt{2}}{\Bigg (} $ }}
\put (80,18){\circle{24}}
\put (80,18){\line (0,1){12}}
\put (80,18){\line(1,0){12}}
\put (80,18){\line(-1,0){12}}
\put (80,6){\line(0,-1){12}}
\put (80,18){\makebox(12,12){$\uparrow$}}
\put (68,18){\makebox(24,-12){$\downarrow\uparrow$}}
\put (80,-18){\circle{24}}
\put (80,-18){\line (0,1){12}}
\put (80,-18){\line(1,0){12}}
\put (80,-18){\line(-1,0){12}}
\put (80,-18){\makebox(-12,12){$\downarrow$}}
\put (100,0){\makebox(5,0){$- $ }}
\put (120,18){\circle{24}}
\put (120,18){\line (0,1){12}}
\put (120,18){\line(1,0){12}}
\put (120,18){\line(-1,0){12}}
\put (120,6){\line(0,-1){12}}
\put (120,18){\makebox(12,12){$\downarrow$}}
\put (108,18){\makebox(24,-12){$\downarrow\uparrow$}}
\put (120,-18){\circle{24}}
\put (120,-18){\line (0,1){12}}
\put (120,-18){\line(1,0){12}}
\put (120,-18){\line(-1,0){12}}
\put (120,-18){\makebox(-12,12){$\uparrow$}}
\put (142,0){\makebox(3,0){${\Bigg )} $ }}
\end{picture}
\end{figure}

\vspace{1.0cm}

\begin{figure}
\begin{picture}(0,20)
\put (30,0){\makebox(10,0){$|\gamma_{6} \rangle =\frac{1}{\sqrt{2}}{\Bigg (} $ }}
\put (80,18){\circle{24}}
\put (80,18){\line (0,1){12}}
\put (80,18){\line(1,0){12}}
\put (80,18){\line(-1,0){12}}
\put (80,6){\line(0,-1){12}}
\put (80,18){\makebox(-12,12){$\uparrow$}}
\put (80,-18){\circle{24}}
\put (80,-18){\line (0,1){12}}
\put (80,-18){\line(1,0){12}}
\put (80,-18){\line(-1,0){12}}
\put (80,-18){\makebox(12,12){$\downarrow$}}
\put (68,-18){\makebox(24,-12){$\downarrow\uparrow$}}
\put (100,0){\makebox(5,0){$- $ }}
\put (120,18){\circle{24}}
\put (120,18){\line (0,1){12}}
\put (120,18){\line(1,0){12}}
\put (120,18){\line(-1,0){12}}
\put (120,6){\line(0,-1){12}}
\put (120,18){\makebox(-12,12){$\downarrow$}}
\put (120,-18){\circle{24}}
\put (120,-18){\line (0,1){12}}
\put (120,-18){\line(1,0){12}}
\put (120,-18){\line(-1,0){12}}
\put (120,-18){\makebox(12,12){$\uparrow$}}
\put (108,-18){\makebox(24,-12){$\downarrow\uparrow$}}
\put (142,0){\makebox(3,0){${\Bigg )} $ }}
\end{picture}
\end{figure}

\vspace{1.0cm}

This state is essentially composed by
\begin{equation} 
\parallel GS \rangle _{J=0.73}\simeq \frac {1}{\sqrt{2}}(|\gamma_{1} \rangle + |\gamma_{2} \rangle) 
\label{sm0}
\end{equation}
i.e., the $\tau^M_z=0$ combination of nonpolar atomic spin-1 states 
coupled to $S^M=0$. Its orbital part is the same as the state $|\psi^o_0 \rangle_{ab}$ mentioned in Eq. (\ref{psiaf2}), of which it is the complete spin orbital representation. Note that the same spin structure belongs also to the state $|\psi^o_{|2|} \rangle_{ab}$ given by Eq. (\ref{psiaf1}), even if the orbital part is different.

At still greater $J$, the value $ J\sim 0.73$ marks the final transition 
to the doubly degenerate ferromagnetic state with 
$S^M=2$ and $\tau_z^M=\pm 1$, which is therefore stable for  
$J/U_2 \geq 0.29$. We have two $2\times 2$ eigenvalue equations for both $\tau_z^M=\pm 1$.  
Choosing $\tau_z^M=-1$ and solving for the ground state, we get:

\begin{eqnarray}
\parallel GS \rangle  _{J\ge 0.73} = N \left (~ |\gamma_7\rangle + C~ |\gamma_8\rangle 
~\right) 
\label{j7}
\end{eqnarray}
where $C = \frac{U_2-J}{2(\mu - \rho)} 
\left(1 - \sqrt{1 + \frac{4(\mu-\rho)^2}{(U_2-J)^2}} \right)$, $N$ 
is an appropriate normalization factor and, for $S^M_z=2$,

\vspace{3mm}
\begin{figure}
\begin{picture}(0,20)
\put (30,0){\makebox(10,0){$|\gamma_{7} \rangle =\frac{1}{\sqrt{2}}{\Bigg (} $ }}
\put (80,18){\circle{24}}
\put (80,18){\line (0,1){12}}
\put (80,18){\line(1,0){12}}
\put (80,18){\line(-1,0){12}}
\put (80,6){\line(0,-1){12}}
\put (80,18){\makebox(-12,12){$\uparrow$}}
\put (68,18){\makebox(24,-12){$\uparrow$}}
\put (80,-18){\circle{24}}
\put (80,-18){\line (0,1){12}}
\put (80,-18){\line(1,0){12}}
\put (80,-18){\line(-1,0){12}}
\put (80,-18){\makebox(-12,12){$\uparrow$}}
\put (80,-18){\makebox(12,12){$\uparrow$}}
\put (100,0){\makebox(5,0){$+ $ }}
\put (120,18){\circle{24}}
\put (120,18){\line (0,1){12}}
\put (120,18){\line(1,0){12}}
\put (120,18){\line(-1,0){12}}
\put (120,6){\line(0,-1){12}}
\put (120,18){\makebox(-12,12){$\uparrow$}}
\put (120,18){\makebox(12,12){$\uparrow$}}
\put (120,-18){\circle{24}}
\put (120,-18){\line (0,1){12}}
\put (120,-18){\line(1,0){12}}
\put (120,-18){\line(-1,0){12}}
\put (120,-18){\makebox(-12,12){$\uparrow$}}
\put (108,-18){\makebox(24,-12){$\uparrow$}}
\put (142,0){\makebox(3,0){${\Bigg )} $ }}
\end{picture}
\end{figure}

\vspace{1cm}
 
\begin{figure}
\begin{picture}(0,20)
\put (30,0){\makebox(10,0){$|\gamma_{8} \rangle =\frac{1}{\sqrt{2}}{\Bigg (} $ }}
\put (80,18){\circle{24}}
\put (80,18){\line (0,1){12}}
\put (80,18){\line(1,0){12}}
\put (80,18){\line(-1,0){12}}
\put (80,6){\line(0,-1){12}}
\put (80,18){\makebox(-12,12){$\uparrow$}}
\put (80,18){\makebox(12,12){$\uparrow$}}
\put (68,18){\makebox(24,-12){$\uparrow$}}
\put (80,-18){\circle{24}}
\put (80,-18){\line (0,1){12}}
\put (80,-18){\line(1,0){12}}
\put (80,-18){\line(-1,0){12}}
\put (80,-18){\makebox(-12,12){$\uparrow$}}
\put (100,0){\makebox(5,0){$+ $ }}
\put (120,18){\circle{24}}
\put (120,18){\line (0,1){12}}
\put (120,18){\line(1,0){12}}
\put (120,18){\line(-1,0){12}}
\put (120,6){\line(0,-1){12}}
\put (120,18){\makebox(-12,12){$\uparrow$}}
\put (120,-18){\circle{24}}
\put (120,-18){\line (0,1){12}}
\put (120,-18){\line(1,0){12}}
\put (120,-18){\line(-1,0){12}}
\put (120,-18){\makebox(-12,12){$\uparrow$}}
\put (120,-18){\makebox(12,12){$\uparrow$}}
\put (108,-18){\makebox(24,-12){$\uparrow$}}
\put (142,0){\makebox(3,0){${\Bigg )}~. $ }}
\end{picture}
\end{figure}

\vspace{1cm}

Notice that the state in Eq. (\ref{psidim}) is nothing else that the 
atomic limit of $|GS\rangle _{J \ge 0.73}$ (actually, $|\gamma_{7} \rangle\equiv |\gamma_{-} \rangle_{ab}$). For the chosen values of the 
parameters and $J=0.73$ eV the component $|\gamma_7\rangle$ represents 83\% of the total weight and increases with increasing $J$.

To demonstrate the role of the trigonal distortion $\Delta_t$ on the 
stability region of the various ground states, we consider different 
values up to 0.4 eV, which is the value suggested by Ezhov {\it et al.}
\cite{ezhov99} 
As seen from Fig. \ref{hubd0}$(a)$ to Fig. \ref{hubd0}$(c)$, for values of 
$\Delta_t$ up to 0.3 eV the role of the trigonal splitting  is essentially 
to decrease the value of $J$ at which the $S^M=0\rightarrow S^M=2$ 
transition takes place, i.e., to increase the stability region of the 
ferromagnetic state. As already anticipated in the previous subsection, this fact can be easily 
explained by looking at the structure of the $S^M=2$ 
state, essentially composed by $|\gamma_{7} \rangle$ with 25\% of $a_{1g}$ 
occupancy, and the $S^M=0$ state, given in Eq. (\ref{sm0})
and composed by $|\gamma_{1} \rangle$ and $|\gamma_{2} \rangle$, with 50\% of $a_{1g}$ 
occupancy. 
At $\Delta_t \simeq 0.3$ eV an abrupt transition in the composition of the 
$S^M=0$ ground state takes place such that the preferred orbital occupation change to the $e_g$ states (no $a_{1g}$ orbitals, see state $|\gamma_9 \rangle$), while the $S^M=2$ remains 
the same. As a consequence the stability region of this latter starts
decreasing, as shown in Figs. \ref{hubd0}$(c)$, $(d)$. This transition in the composition of the $S^M=0$ state had to be expected, since for $\frac{\Delta_t}{J}, \frac{\Delta_t}{\rho} \rightarrow \infty$ the $a_{1g}$ occupation must go to zero.

\end{multicols}

\parbox[b]{3.3in}
{\begin{figure}
      \epsfysize=50mm
     \centerline{\epsffile{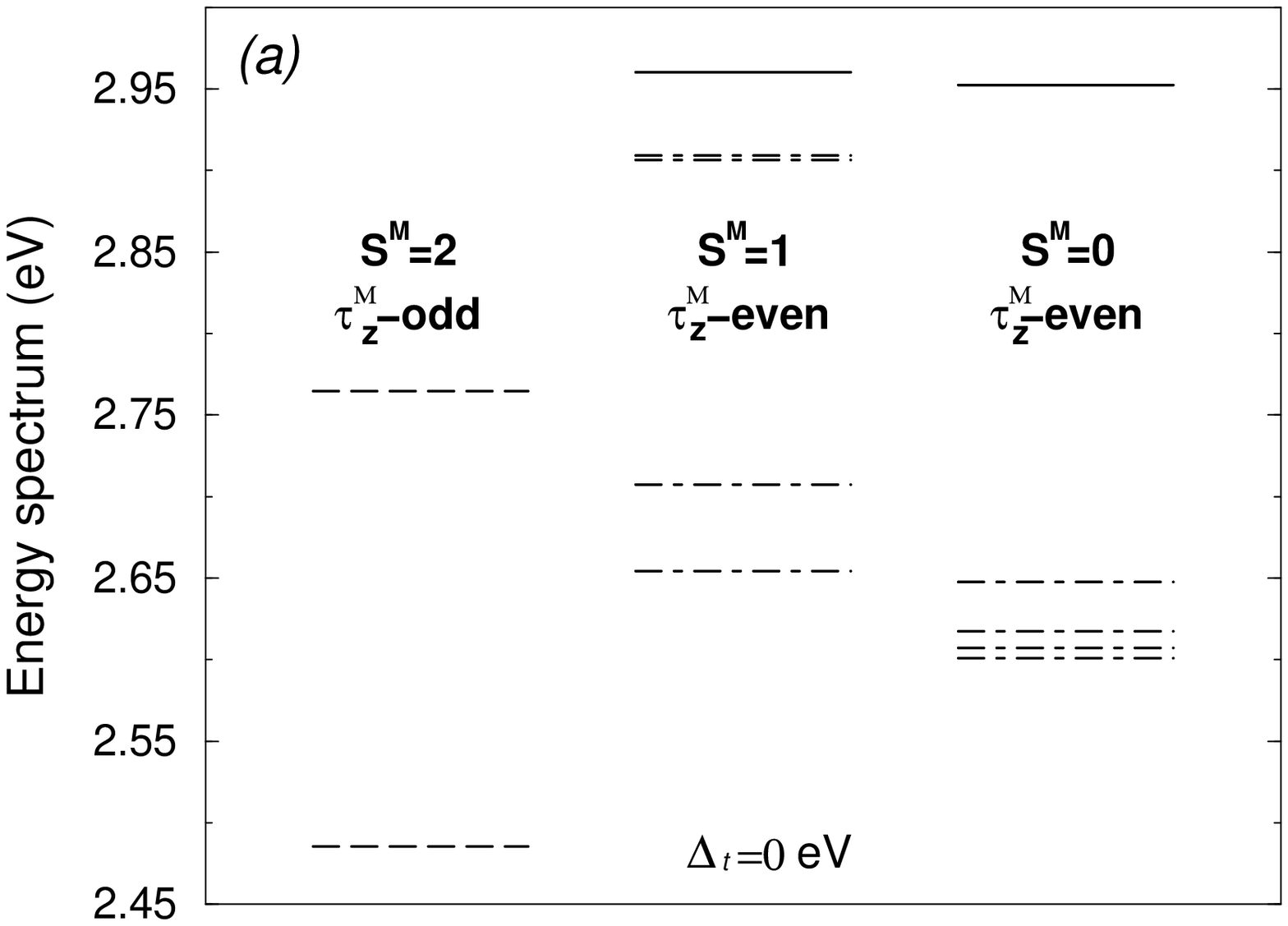}}
\end{figure}}
\parbox[b]{3.3in}
{\begin{figure}
 \epsfysize=50mm
      \centerline{\epsffile{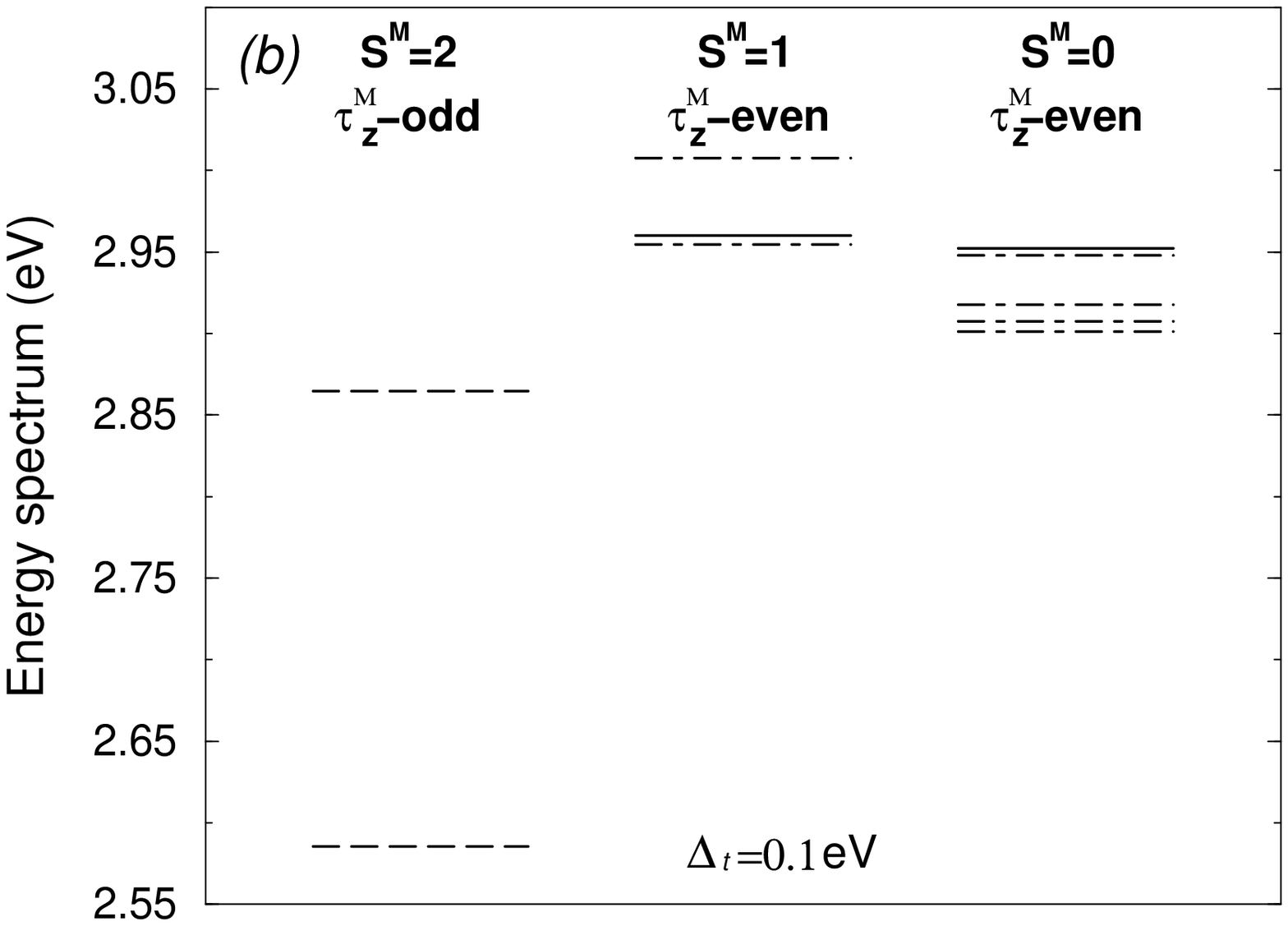}}
\end{figure}}

\parbox[b]{3.3in}
{\begin{figure}
        \epsfysize=50mm
        \centerline{\epsffile{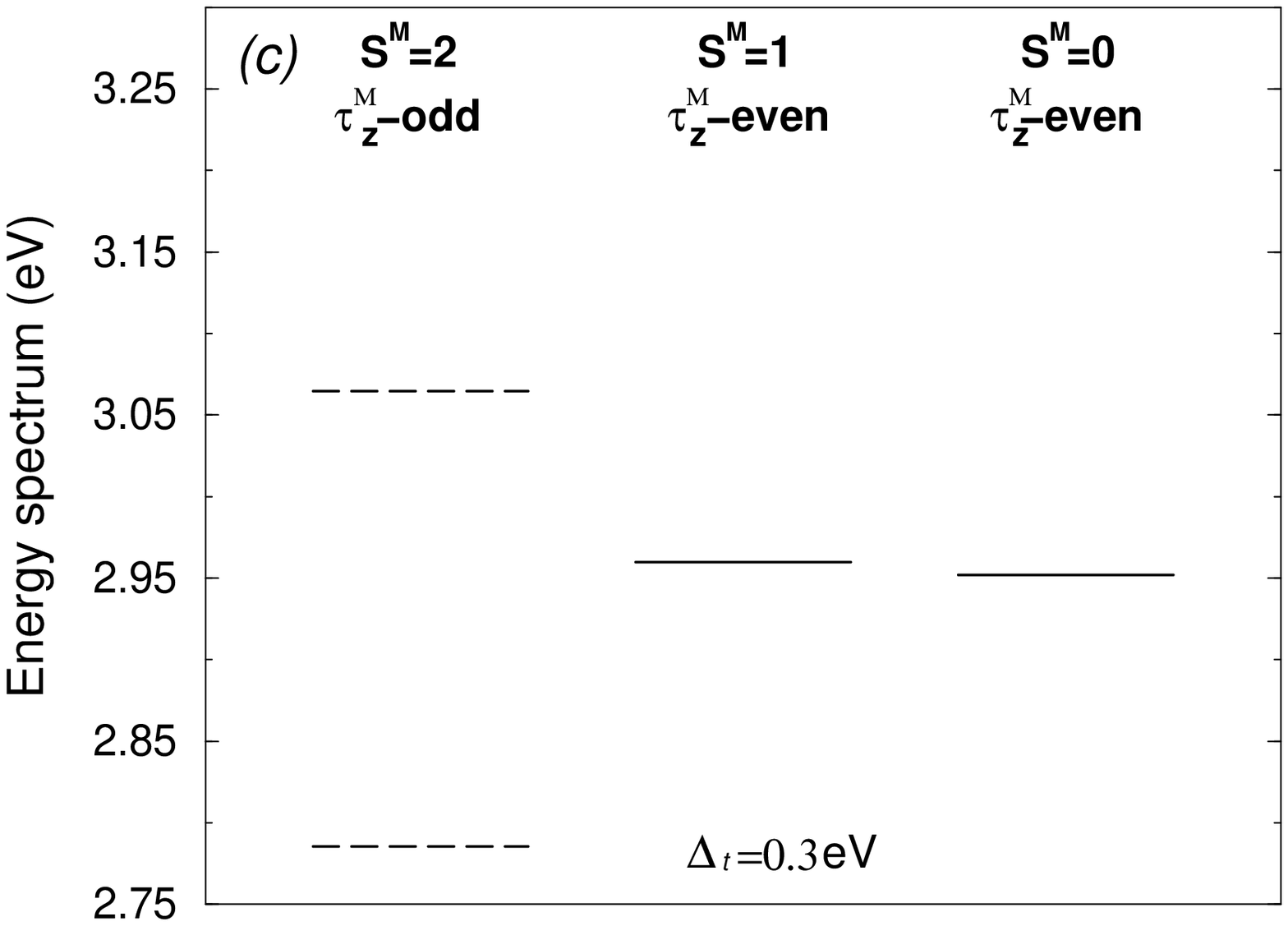}}
\end{figure}}
\parbox[b]{3.3in}
{\begin{figure}
        \epsfysize=50mm
        \centerline{\epsffile{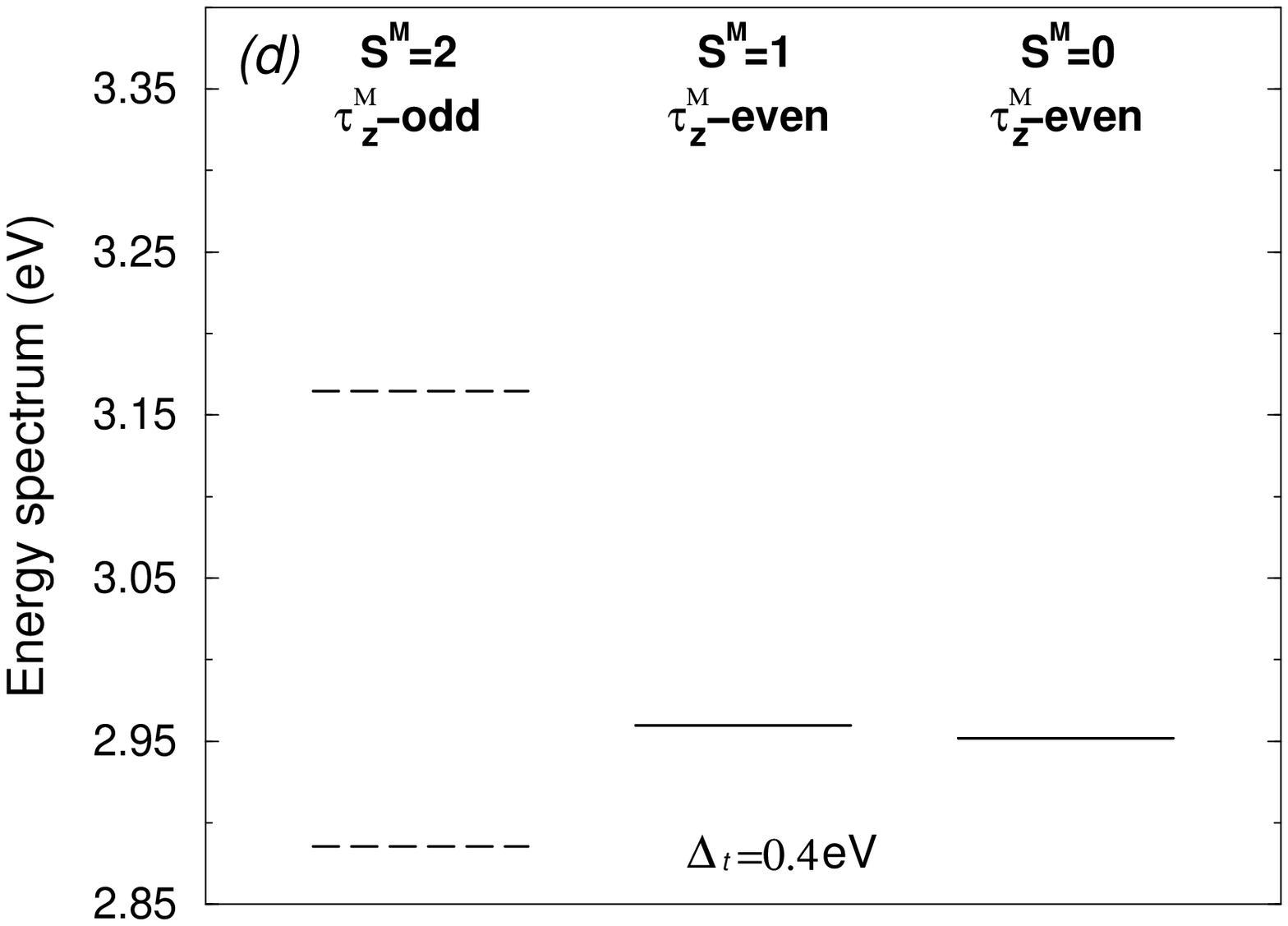}}
\end{figure}}
\vspace{-0.5cm}
\begin{figure}
\caption{
Energy level structure  at $J=1.0$ eV 
for molecular states with $S^M=2$, $\tau_z^M$ odd; 
$S^M=1$, $\tau_z^M$ even and
$S^M=0$, $\tau_z^M$ even. Panels $(a), (b), (c)$ and 
$(d)$ correspond to the  trigonal distortion $\Delta_t=0$ eV, $\Delta_t=0.1$ eV,
$\Delta_t=0.3$ eV and $\Delta_t=0.4$ eV, respectively. Dot-dashed, dashed and solid lines indicate two, one or zero atomic $a_{1g}$ orbital occupancy in the ground state.}
\label{eigen}
\end{figure}

\protect\begin{multicols}{2}

Figure \ref{eigen}$(a)$-$(d)$ illustrate how the low lying level structure 
and the composition of the ground state of the vertical pair changes as a 
function of $\Delta_t$. We fix $J=1.0~eV$, for reasons that will be apparent in the next section, although the same results are qualitatively valid for all the physical values of $J$. The main difference is that below the chosen value of $1.0$ eV, the energy gap between the $S^M=2$ ground state and the excited $S^M=0$ and $S^M=1$ states reduces, while above $J=1.0$ eV, it increases.
At low values of $\Delta_t$ there are many low-lying excited levels
 for $S^M=0$, $\tau_z^M$ even states (dot-dashed lines of Fig.\ref{eigen}$(a)$ and $(b)$). The orbital part of the two lowest states in this situation is given by Eqs. (\ref{psiaf1}) and (\ref{psiaf2}) and their energy difference is about $6$ meV. Notice that the lowest state is exactly given by Eq. (\ref{sm0}). For the chosen value of $J=1.0$ eV, at $\Delta_t \sim 0.18$ eV there is a
drastic redistribution of the weight in the atomic configuration of the
 $ S^M=0$ molecule state, i.e., the non polar state is now given by
\end{multicols}

\begin{figure}
~~~~~~~~~~~~~~~~~~~~~~
\begin{picture}(0,20)
\put (30,0){\makebox(10,0){$|\gamma_{9} \rangle = \frac{1}{2\sqrt{3}}
{\Bigg (}2 $ }}
\put (80,18){\circle{24}}
\put (80,18){\line (0,1){12}}
\put (80,18){\line(1,0){12}}
\put (80,18){\line(-1,0){12}}
\put (80,6){\line(0,-1){12}}
\put (80,18){\makebox(-12,12){$\uparrow$}}
\put (80,18){\makebox(12,12){$\uparrow$}}
\put (80,-18){\circle{24}}
\put (80,-18){\line (0,1){12}}
\put (80,-18){\line(1,0){12}}
\put (80,-18){\line(-1,0){12}}
\put (80,-18){\makebox(-12,12){$\downarrow$}}
\put (80,-18){\makebox(12,12){$\downarrow$}}
\put (100,0){\makebox(5,0){$- $ }}
\put (120,18){\circle{24}}
\put (120,18){\line (0,1){12}}
\put (120,18){\line(1,0){12}}
\put (120,18){\line(-1,0){12}}
\put (120,6){\line(0,-1){12}}
\put (120,18){\makebox(-12,12){$\uparrow$}}
\put (120,18){\makebox(12,12){$\downarrow$}}
\put (120,-18){\circle{24}}
\put (120,-18){\line (0,1){12}}
\put (120,-18){\line(1,0){12}}
\put (120,-18){\line(-1,0){12}}
\put (120,-18){\makebox(-12,12){$\uparrow$}}
\put (120,-18){\makebox(12,12){$\downarrow$}}
\put (140,0){\makebox(5,0){$- $ }}
\put (160,18){\circle{24}}
\put (160,18){\line (0,1){12}}
\put (160,18){\line(1,0){12}}
\put (160,18){\line(-1,0){12}}
\put (160,6){\line(0,-1){12}}
\put (160,18){\makebox(-12,12){$\uparrow$}}
\put (160,18){\makebox(12,12){$\downarrow$}}
\put (160,-18){\circle{24}}
\put (160,-18){\line (0,1){12}}
\put (160,-18){\line(1,0){12}}
\put (160,-18){\line(-1,0){12}}
\put (160,-18){\makebox(-12,12){$\downarrow$}}
\put (160,-18){\makebox(12,12){$\uparrow$}}
\put (180,0){\makebox(5,0){$+ 2$ }}
\put (200,18){\circle{24}}
\put (200,18){\line (0,1){12}}
\put (200,18){\line(1,0){12}}
\put (200,18){\line(-1,0){12}}
\put (200,6){\line(0,-1){12}}
\put (200,18){\makebox(-12,12){$\downarrow$}}
\put (200,18){\makebox(12,12){$\downarrow$}}
\put (200,-18){\circle{24}}
\put (200,-18){\line (0,1){12}}
\put (200,-18){\line(1,0){12}}
\put (200,-18){\line(-1,0){12}}
\put (200,-18){\makebox(-12,12){$\uparrow$}}
\put (200,-18){\makebox(12,12){$\uparrow$}}
\put (220,0){\makebox(5,0){$- $ }}
\put (240,18){\circle{24}}
\put (240,18){\line (0,1){12}}
\put (240,18){\line(1,0){12}}
\put (240,18){\line(-1,0){12}}
\put (240,6){\line(0,-1){12}}
\put (240,18){\makebox(-12,12){$\downarrow$}}
\put (240,18){\makebox(12,12){$\uparrow$}}
\put (240,-18){\circle{24}}
\put (240,-18){\line (0,1){12}}
\put (240,-18){\line(1,0){12}}
\put (240,-18){\line(-1,0){12}}
\put (240,-18){\makebox(-12,12){$\uparrow$}}
\put (240,-18){\makebox(12,12){$\downarrow$}}
\put (260,0){\makebox(5,0){$-$ }}
\put (280,18){\circle{24}}
\put (280,18){\line (0,1){12}}
\put (280,18){\line(1,0){12}}
\put (280,18){\line(-1,0){12}}
\put (280,6){\line(0,-1){12}}
\put (280,18){\makebox(-12,12){$\downarrow$}}
\put (280,18){\makebox(12,12){$\uparrow$}}
\put (280,-18){\circle{24}}
\put (280,-18){\line (0,1){12}}
\put (280,-18){\line(1,0){12}}
\put (280,-18){\line(-1,0){12}}
\put (280,-18){\makebox(-12,12){$\downarrow$}}
\put (280,-18){\makebox(12,12){$\uparrow$}}
\put (300,0){\makebox(3,0){${\Bigg )}~, $ }}
\end{picture}
\end{figure}
\vspace{1cm}

\protect\begin{multicols}{2}
\noindent which is nothing else that the complete spin-orbit representation of the state (\ref{psiaf3b}).
By further increasing $\Delta_t$ 
(see Fig. \ref{hubd0}$(c)$ and $(d)$) this state becomes more and more favorable and 
finally  at $\Delta_t \simeq 0.47$ becomes the overall ground state.
At this value of $\Delta_t$ the weight of the non polar state 
$|\gamma_9\rangle$ is more than 99\%. 
Note that the value of the crystal field splitting where we have the transition from a $S^M=2$ to a $S^M=0$ overall ground state does depend on $J$ as is clear from Figs. \ref{hubd0}.

For  $J=1.0$ eV the $S^M=1$, $\tau_z^M$ even state is never 
the ground state of the vertical molecule.  Nevertheless at $\Delta_t=0.4$ eV 
(see Fig. \ref{eigen}$(d)$) it lies only about 80 meV above the ground
state with $S^M=2$ and its composition is made up for more than 99\% of the following state (e.g., with $ S^M_z=1$):

\vspace{0.8cm}

\begin{figure}
\begin{picture}(0,20)
\put (30,0){\makebox(10,0){$|\gamma_{10} \rangle =\frac{1}{2}{\Bigg (} $ }}
\put (80,18){\circle{24}}
\put (80,18){\line (0,1){12}}
\put (80,18){\line(1,0){12}}
\put (80,18){\line(-1,0){12}}
\put (80,6){\line(0,-1){12}}
\put (80,18){\makebox(-12,12){$\uparrow$}}
\put (80,18){\makebox(12,12){$\downarrow$}}
\put (80,-18){\circle{24}}
\put (80,-18){\line (0,1){12}}
\put (80,-18){\line(1,0){12}}
\put (80,-18){\line(-1,0){12}}
\put (80,-18){\makebox(-12,12){$\uparrow$}}
\put (80,-18){\makebox(12,12){$\uparrow$}}
\put (100,0){\makebox(5,0){$+ $ }}
\put (120,18){\circle{24}}
\put (120,18){\line (0,1){12}}
\put (120,18){\line(1,0){12}}
\put (120,18){\line(-1,0){12}}
\put (120,6){\line(0,-1){12}}
\put (120,18){\makebox(-12,12){$\downarrow$}}
\put (120,18){\makebox(12,12){$\uparrow$}}
\put (120,-18){\circle{24}}
\put (120,-18){\line (0,1){12}}
\put (120,-18){\line(1,0){12}}
\put (120,-18){\line(-1,0){12}}
\put (120,-18){\makebox(-12,12){$\uparrow$}}
\put (120,-18){\makebox(12,12){$\uparrow$}}
\put (140,0){\makebox(5,0){$- $ }}
\put (160,18){\circle{24}}
\put (160,18){\line (0,1){12}}
\put (160,18){\line(1,0){12}}
\put (160,18){\line(-1,0){12}}
\put (160,6){\line(0,-1){12}}
\put (160,18){\makebox(-12,12){$\uparrow$}}
\put (160,18){\makebox(12,12){$\uparrow$}}
\put (160,-18){\circle{24}}
\put (160,-18){\line (0,1){12}}
\put (160,-18){\line(1,0){12}}
\put (160,-18){\line(-1,0){12}}
\put (160,-18){\makebox(-12,12){$\uparrow$}}
\put (160,-18){\makebox(12,12){$\downarrow$}}
\put (180,0){\makebox(5,0){$- $ }}
\put (200,18){\circle{24}}
\put (200,18){\line (0,1){12}}
\put (200,18){\line(1,0){12}}
\put (200,18){\line(-1,0){12}}
\put (200,6){\line(0,-1){12}}
\put (200,18){\makebox(-12,12){$\uparrow$}}
\put (200,18){\makebox(12,12){$\uparrow$}}
\put (200,-18){\circle{24}}
\put (200,-18){\line (0,1){12}}
\put (200,-18){\line(1,0){12}}
\put (200,-18){\line(-1,0){12}}
\put (200,-18){\makebox(-12,12){$\downarrow$}}
\put (200,-18){\makebox(12,12){$\uparrow$}}
\put (220,0){\makebox(3,0){${\Bigg )} $ }}
\end{picture}
\end{figure}
\vspace{1cm}

As is clear from Fig. \ref{hubd0}$(d)$, its excitation energy decreases with $J$. Note that $|\gamma_{10}\rangle$ is made up of two atomic $S=1$ states, so that it belongs to the subspace of $H_{\rm eff}$.

We shall make use of these findings later on in order to 
determine the various parameters in the AFI phase of V$_2$O$_3$.

\section{The minimization procedure.}

In this section we look for all the possible orbital and magnetic 
ground-state configurations of the effective Hamiltonian by using a variational procedure. 
The trial wave function can be written in general as follows:
\begin{equation}
|\Psi\rangle =\Pi_{n}~|\Psi_n\rangle =\Pi_{n}~|\psi^o_n\rangle |\phi^s_n\rangle
\label{variational}
\end{equation} 
where the state $|\psi^o_n\rangle$ refers to orbital occupancy and 
$|\phi^s_n\rangle$ refers to spin occupancy on site $n$. In the following, 
we will use as a variational wave function $|\Psi_n\rangle$ either an 
atomic state or a molecular one, with $n$ labeling an 
atomic or a molecular site, respectively.

Discarding for the moment the single site crystal field part in Eq. 
(\ref{hcf}) which will be easily dealt with, we observe that 
$H_{\rm eff}$ acts only onto two atomic sites at a time and 
factors  into an orbital $H^o_{\rm eff}$ and a spin $H^s_{\rm eff}$ 
part. Therefore its average value over the above state takes the form:
\begin{eqnarray}
\begin{array}{c}
\langle\Psi_n|\langle\Psi_m|H_{\rm eff}|\Psi_m\rangle |\Psi_n\rangle =
\\[0.2cm]
\langle\psi^o_n|\langle\psi^o_m|H^o_{\rm eff}|\psi^o_m\rangle |\psi^o_n\rangle
\times
\langle\phi^s_n|\langle\phi^s_m|H^s_{\rm eff}|\phi^s_m\rangle |\phi^s_n\rangle
\end{array}
\label{av}
\end{eqnarray}

Whereas orbital averaging in the first term will require some algebra, 
the second average in this equation, referring to spin variables, 
is straightforward in a mean field treatment. For a ferromagnetic bond
$\langle \vec{S}_n\cdot\vec{S}_m+2\rangle_{HF}=3$ and 
$\langle \vec{S}_n\cdot\vec{S}_m-1\rangle_{HF}=0$, while for an 
antiferromagnetic coupling, $\langle \vec{S}_n\cdot\vec{S}_m+2\rangle_{HF}=1$
and $\langle \vec{S}_n\cdot\vec{S}_m-1\rangle_{HF}=-2$.
 
As discussed in section V-A, the correlation energy of the 
ferromagnetic state of the vertical pair, defined as the difference 
between the exact ground state energy and its Hartree-Fock approximation, 
is given by 
$\Delta E_m-\Delta E_{HF}=2\rho\mu/(U_2-J)$.
Therefore we can have two qualitatively different regimes of  solutions:

\noindent {\it i)} If this difference is much higher than the 
interaction in the basal plane, then the most appropriate 
variational wave function for the whole $H_{\rm eff}$ must be constructed in 
terms of molecular units, taking into account exactly  the molecular binding 
energy, with orbital wave functions $|\psi^o_n\rangle$ given by (\ref{psidim}). This means that the whole crystal consists of some ordered sequence 
of molecular units, whose internal energy is so high that it is 
energetically more favorable for the system  not to break this structure.
This seems to be the case for the values of parameters given in Table II. 
This state will be called the crystal ''molecular'' variational state.

\noindent {\it ii)} If instead it is the values of the exchange energy in the basal plane to be bigger 
than the correlation energy $2\rho\mu/(U_2-J)$,
then the most natural variational wave function can be written in terms 
of single site atomic states, as will be shown in section VI-B.

\subsection{The case of the molecular variational state.}

As mentioned above, when the molecular correlation energy is much bigger
than the in-plane exchange energy, then 
the best candidate for $| \psi^o_n\rangle$ 
in Eq. (\ref{variational}) is given by a linear combination of the wave 
functions of the type shown in Eq. (\ref{psidim}). 

Therefore for any molecular site $n$ ($ab$, $cc'$, $dd'$ or $ee'$ with 
reference to Fig. \ref{bonds}) the orbital part of the trial molecular 
electronic wave function can be written as:  

\begin{eqnarray}
| \psi^o_{n}\rangle=\cos\psi_{n} |  \psi^o_-\rangle_{n}+
\sin\psi_{n} | \psi^o_+\rangle_{n} ~.
\label{mixhor}
\end{eqnarray}

In order to construct the expectation value of $H_{\rm eff}$ over the 
crystal molecular wave function, we need to know the result of its  
application over states of the form (\ref{av}), for $n$ and $m$ molecular labels. Considering, for example, the two vertical pairs $ab$ and $cc'$ of Fig.  \ref{bonds}, this state can be written as:

\begin{eqnarray}
\begin{array}{c}
|  \psi^o_{cc'}\rangle | \psi^o_{ab}\rangle = \\[0.2cm]
{\big (}\cos \psi_{ab} | \psi_-\rangle_{ab}
 +\sin \psi_{ab} | \psi_+\rangle_{ab}
{\big )} \\[0.2cm]
\times {\big (}\cos \psi_{cc'} | \psi_-\rangle_{cc'}
 +\sin \psi_{cc'} | \psi_+\rangle_{cc'}
{\big )} ~.
\end{array}
\label{st1}
\end{eqnarray}

The details of the calculations can be found in Appendix D. Then the contribution coming from
a ferromagnetic bond along $\delta_1$ is found to be:  

\begin{eqnarray}
\begin{array}{c}
\langle\Psi_{ab} |  \langle\Psi_{cc'}|  H_{\rm eff}(\delta_1) |
\Psi_{cc'}\rangle | \Psi_{ab}\rangle=\\[0.2cm]
\cos^2\psi_{ab}\cos^2\psi_{cc'}G_1
+(\sin^2\psi_{ab}\cos^2\psi_{cc'}\\[0.2cm]
+\cos^2\psi_{ab}\sin^2\psi_{cc'})G_2
+\sin^2\psi_{ab}\sin^2\psi_{cc'}G_3\\[0.2cm]
+\sin 2\psi_{ab}\sin 2\psi_{cc'} G_4 ~,
\end{array}
\label{san1}
\end{eqnarray}
where 
\begin{eqnarray}
\begin{array}{l}
G_1=\frac{u}{4}(2\beta^2+2\sigma^2+4\tau^2+4\chi^2+4\theta^2)\\[0.2cm]
G_2=\frac{u}{4}(2\alpha^2+2\beta^2+2\sigma^2+4\tau^2+2\chi^2+4\theta^2)\\
     [0.2cm]
G_3= \frac{u}{4}(2\alpha^2+2\sigma^2+4\tau^2+4\chi^2+4\theta^2)\\[0.2cm]
G_4=\frac{u}{4}\alpha\beta\\[0.2cm]
u=-\frac{1}{U_2-J} ~.
\end{array}
\label{G}
\end{eqnarray}
 Note that the spin contribution has been already taken into account. We have also retained 
for future use the values of the hopping integrals $\theta=t^{13}$ and 
$\chi=t^{12}$, which are zero in the corundum phase\cite{nebenzahl71}
and can be different from zero in the monoclinic one.

For the AF bond along the same direction we obtain: 
\begin{eqnarray}
\begin{array}{l}
\langle\Psi_{ab}|  \langle\Psi_{cc'}|  H_{\rm eff}(\delta_1) |  
\Psi_{cc'}\rangle|  \Psi_{ab}\rangle=\\[0.2cm]
\cos^2\psi_{ab}\cos^2\psi_{cc'}F_1+
(\sin^2\psi_{ab}\cos^2\psi_{cc'}\\[0.2cm]
+\cos^2\psi_{ab}\sin^2\psi_{cc'})F_2+
\sin^2\psi_{ab}\sin^2\psi_{cc'}F_3\\[0.2cm]
+\sin 2\psi_{ab}\sin 2\psi_{cc'}F_4 ~,
\end{array}
\label{san2}
\end{eqnarray}
with the notations:
\begin{eqnarray}
\begin{array}{ll}
F_1=& \frac{G_1}{3}+\frac{v}{4}
(4\alpha^2+\beta^2+\sigma^2+2\tau^2+4\chi^2+4\theta^2) \nonumber \\[0.2cm]
&+\frac{w}{12}
(12\alpha^2+7\beta^2+7\sigma^2+14\tau^2+20\chi^2+20\theta^2) 
\nonumber\\[0.2cm]
F_2=&\frac{G_2}{3}+\frac{v}{4}
(2\alpha^2+2\beta^2+\sigma^2+3\tau^2+5\chi^2+3\theta^2) \nonumber \\[0.2cm]
&+\frac{w}{12}
(10\alpha^2+10\beta^2+7\sigma^2+17\tau^2+19\chi^2+17\theta^2) 
 \nonumber \\[0.2cm]
F_3=&\frac{G_3}{3}+\frac{v}{4}
(\alpha^2+4\beta^2+\sigma^2+4\tau^2+4\chi^2+2\theta^2) \nonumber \\[0.2cm]
&+\frac{w}{12}
(7\alpha^2+12\beta^2+7\sigma^2+20\tau^2+20\chi^2+14\theta^2) 
\nonumber \\[0.2cm]
F_4=&  \frac{G_4}{3}-\frac{1}{8}\alpha\beta\left ( v + \frac{w}{3} \right )
\\[0.2cm]
v=&-\frac{1}{U_2+4J} \nonumber \\[0.2cm]
w=&-\frac{1}{U_2+2J} ~. \nonumber 
\end{array}
\end{eqnarray}

\vspace{-2.0cm}

\begin{eqnarray} 
\label{F}
\end{eqnarray} 

\vspace{1.0cm}

Again, we already included the spin contribution.
To  evaluate the averages of $H_{\rm eff}$ along $\delta_{2}$ and 
$\delta_{3}$, it is convenient to  use
the invariance properties of the Hamiltonian (see CNR\cite{cnr1}) under the 
trigonal $D_{3d}^6$ symmetry. 
Performing a $C_3$ rotation around the vertical axis, the state 
along $\delta_{1}$ is transformed 
in the corresponding one along $\delta_{3}$: 
\begin{eqnarray}
\begin{array}{c}
C_3 |  \Psi_{dd'}\rangle | \Psi_{ab}\rangle=\\[0.2cm]
[\cos (\psi_{dd'}-2\pi /3) |  \Psi_-\rangle_{cc'}+\sin (\psi_{dd'}-
2\pi /3) |  \Psi_+\rangle_{cc'}]\\[0.2cm]
\times [\cos (\psi_{ab}-2\pi /3) |  \Psi_-
\rangle_{ab}+\sin (\psi_{ab}-2\pi /3) |  \Psi_+\rangle_{ab}] ~.
\nonumber
\end{array}
\end{eqnarray}

In the same way, applying $C_3^{-1}$ to the same state, we get the 
corresponding one along $\delta_{2}$. Then the expectation value of 
$H_{\rm eff}$ can be easily obtained directly from Eq. (\ref{san1}) 
and Eq. (\ref{san2}). 

By summing  over the three  nearest neighbors of the molecular site $n$
in the horizontal plane  ($\delta_1$, $\delta_2$, and $\delta_3$ of Fig. \ref{bonds}), we can write the energy of the cluster in the  compact form:

\begin{eqnarray}
\lefteqn{\langle \Psi|\langle \Psi_n| H_{\rm eff}|\Psi_n\rangle |\Psi\rangle =} \nonumber \\[0.2cm]
 & & \begin{array}{l}
\sum_{m=1,2,3}
[\cos^2(\psi_{n}+\gamma _m)\cos^2(\psi_{m}+\gamma _m)\\[0.2cm]
\times (s G_1+\bar s F_1)+(\sin^2(\psi_{n}+\gamma _m)\cos^2(\psi_{m}+\gamma _m)\\[0.2cm]
+\cos^2(\psi_{n}+\gamma _m)\sin^2(\psi_{m}+\gamma _m))(s G_2+\bar s F_2)\\[0.2cm]
+\sin^2(\psi_{n}+\gamma _m)\sin^2(\psi_{m}+\gamma _m)(s G_3+\bar s F_3)\\[0.2cm]
+\sin 2(\psi_{n}+\gamma _m)\sin 2(\psi_{m}+\gamma _m)(s G_4+\bar s F_4) ] ~,
\end{array}
\label{averh}
\end{eqnarray}

\noindent adopting the following notations:
 $s=1$ ($\bar s=0$) if the horizontal bond is ferromagnetic (F) and $s=0$ ($\bar s=1 $) 
for the AF bond,
 $\gamma_m=0$ when $m=1$ and $\gamma_m=\pm 2\pi/3$ when $m=2,3$. 
 
In the AFI phase the unit cell contains four of these clusters and eight V atoms. Therefore the energy per V atom, $E_V$, is given by the sum of  these four energy contributions plus the four molecular energies $\Delta E_{mt}$ of Eq (\ref{evbdelta}), divided by eight. Referring to the atom numbering of Fig. \ref{boh} we have 
the following expression for $E_V$:                                                                       

\begin{eqnarray}
\label{ev}
\begin{array}{ll}
E_V=&\frac{1}{8}[(E_{12}+E_{13}+E_{13'})\\[0.2cm]
&+(E_{37}+E_{27}+E_{27'})\\[0.2cm]
&+(E_{45}+E_{46}+E_{46'})\\[0.2cm]
&+(E_{68}+E_{58}+E_{58'})] + \frac{1}{2} \Delta E_{mt} ~.
\end{array}
\end{eqnarray}

\noindent where $E_{nm}$ is the appropriate energy of the horizontal bond $nm$, whether F or AF.
Note that in Eq. (\ref{ev}) each of the four terms $(E_{12}+E_{13}+E_{13'})$, $(E_{37}+E_{27}+E_{27'})$, $(E_{45}+E_{46}+E_{46'})$, or $(E_{68}+E_{58}+E_{58'})$ is given by Eq. (\ref{averh}).

The term $\frac{1}{2} \Delta E_{mt}=-\frac{1}{2}\frac{(\rho-\mu)^2}{U_2-J} + \frac{1}{2}\Delta_t$ is constant with respect to the 
minimization angles because it is half the binding
energy of the vertical molecule.
For this reason in the following we shall consider $E_V' \equiv E_V -\frac{1}{2}\Delta E_{mt} $, that represents the energy gain per V atom due to the 
intermolecular basal plane interactions with respect to this reference energy. 

We have then performed the numerical minimization of this expression 
with respect to all the four independent angular variational angles  
of Eq. (\ref{mixhor}) in the unit cell, using the standard parameters as
given in section IV or reasonable variations around them. As in CNR\cite{cnr1} 
the minimizing angular values will provide the 
molecular orbital occupancy throughout the crystal.

We have examined the following four magnetic phases:
 
\begin{itemize}
\item
AF phase --  all three bonds $\delta_1$, 
$\delta_2$ and $\delta_3$ are antiferromagnetic;
\item
RS phase --
$\delta_1$ is ferromagnetic and $\delta_2$ and $\delta_3$ are 
antiferromagnetic;
 (this is the spin structure actually observed in V$_2$O$_3$); 
\item
ARS phase --
$\delta_1$ is antiferromagnetic and $\delta_2$, $\delta_3$ are ferromagnetic;
\item
F phase --  all three bonds $\delta_1$, 
$\delta_2$ and $\delta_3$ are ferromagnetic.
\end{itemize}

\noindent which correspond, respectively, to phases G, C, A, F in the work 
of Mila {\it et al.}\cite{mila00}

\begin{figure}
     \epsfysize=60mm
     \centerline{\epsffile{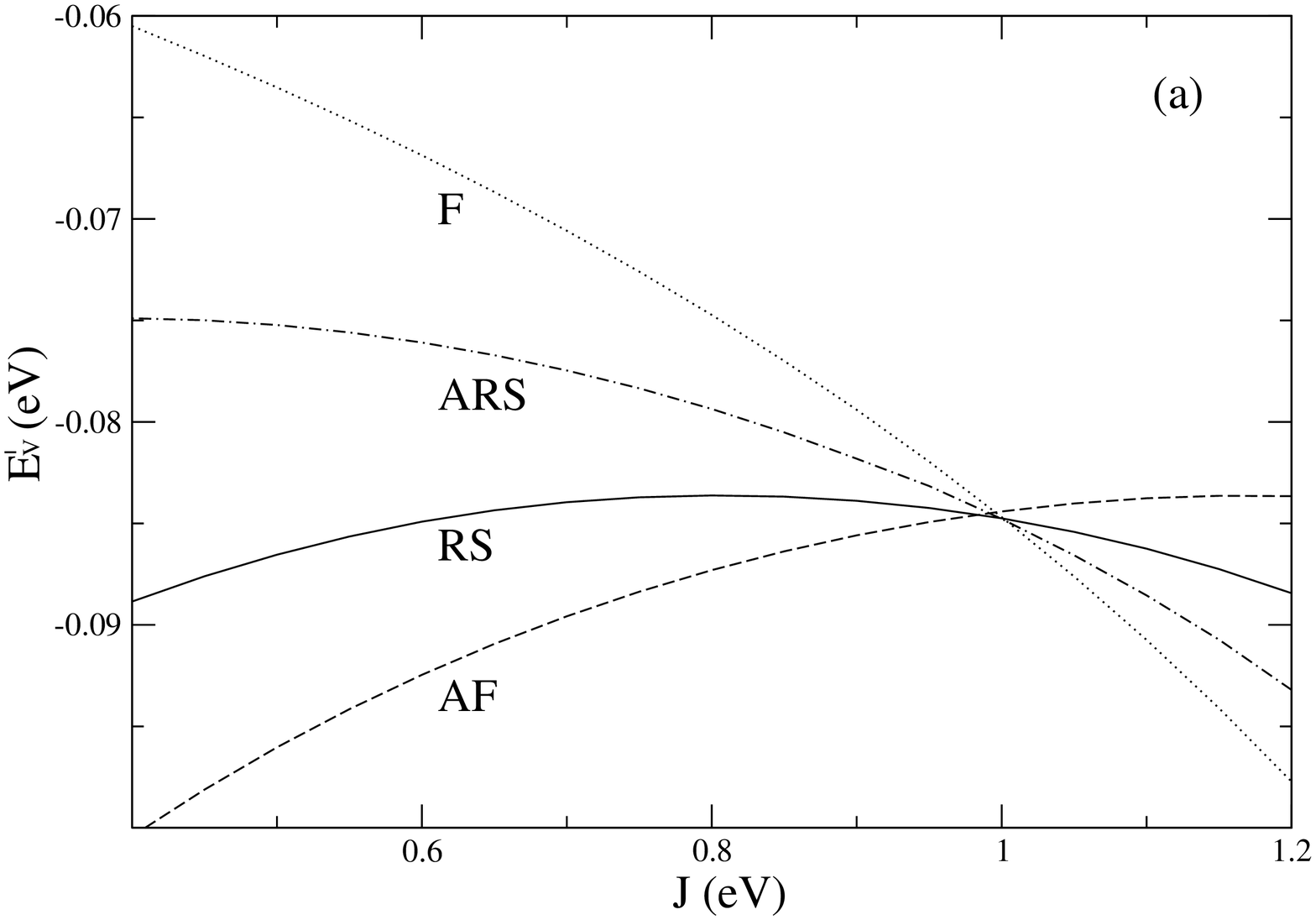}}
\vspace{0.5cm}
 \epsfysize=60mm
      \centerline{\epsffile{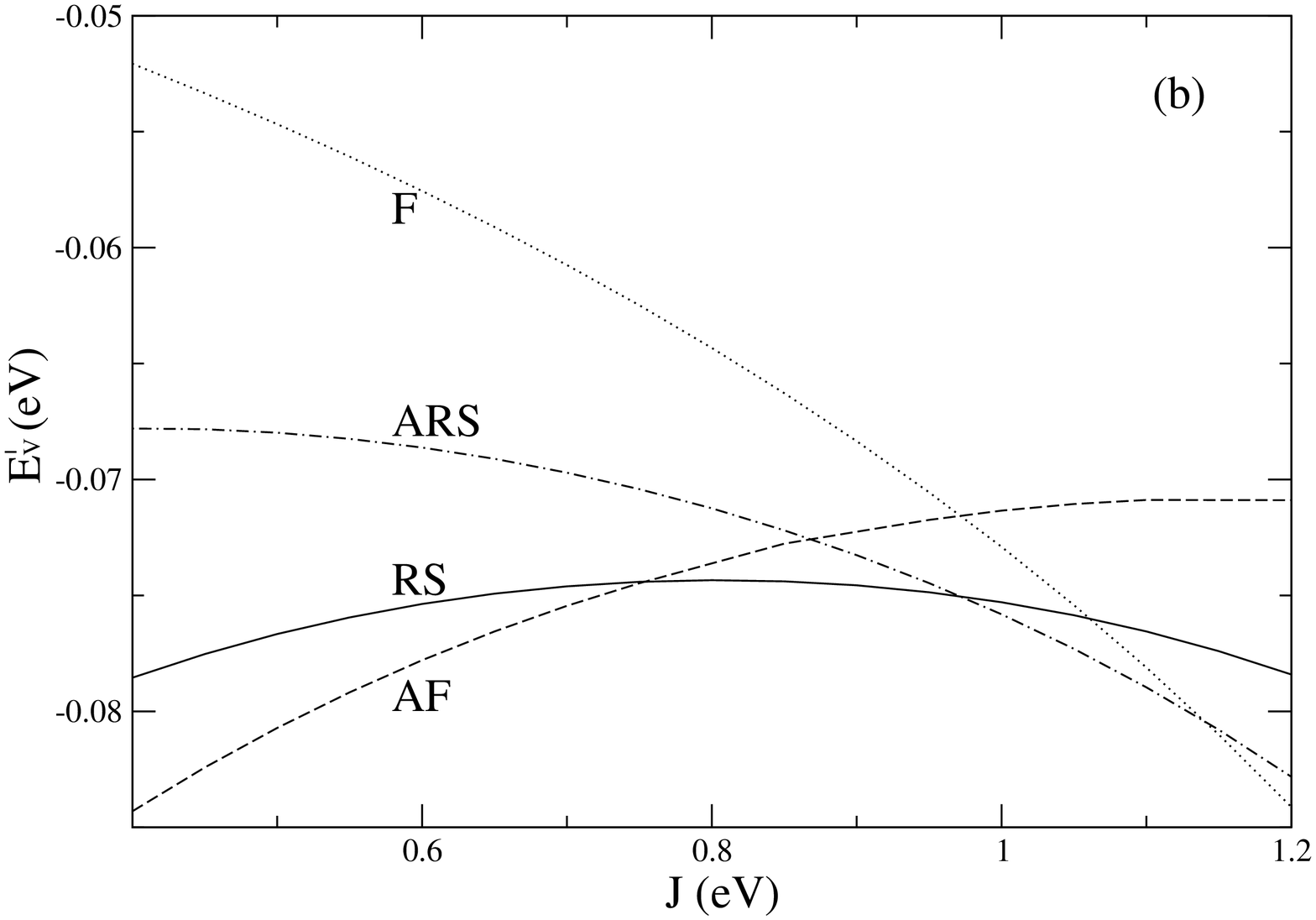}}
\caption{The energy gain $E'_V$ per V atom as a function of $J$ for different
spin configurations (AF, RS, ARS and F type), represented by
dashed, solid, dot-dashed and dotted lines, respectively.
In panel (a) the hopping parameters are the standard ones (Mattheiss set),
while in panel (b) the set considered by Mila {\it et al.} is used 
($\tau = 0.27$ and all others hopping integrals equal to zero.)}
\label{mol}
\end{figure}

Figures \ref{mol}$(a)$ and $(b)$ show a plot of $E_V'$ as a function 
of $J$ for all these magnetic phases. Notice that for fixed $\Delta_t$, $E_V'$ 
depends only on the ratio $J/U_2$ and scales like $\tau^2/U_2$, if $\tau$ 
is the largest hopping integral in the basal plane.

\vspace{1cm}

One general feature that is apparent from these figures and the 
next Fig. \ref{phase} is that 
the stability region for the RS phase is very much reduced in the 
two parameter space of the hopping integrals and the ratio $J/U_2$, in 
contrast with the spin $S=1/2$ case. \cite{cnr1} The AF and F phases 
occupy nearly all phase space in such a way that  
$J/U_2 \approx 0.4$ almost marks the transition from a stable 
antiferromagnetic in-plane spin structure to a ferromagnetic one 
(see, for example, Fig. \ref{mol}$(a)$). This behavior depends 
on the fact that for low $J$ values, the system wants to maximize the 
number of electron jumps (which occur more easily in an AF structure) to the 
detriment of on-site Hund's energy gain, whereas for high $J$ values this 
last mechanism is prevalent. In between, in a small range of $J/U_2$ 
slightly depending on the values of the hopping integrals, typically 
$0.30 < J/U_2 < 0.45$, find their place the RS and the ARS phases, 
each occupying about half of the interval. Even in the most favorable case 
(see Fig. \ref{mol}$(b)$), their 
stabilization energy, due to the competing presence of the AF and F 
phases, is very small, of the order of 2 meV, to be compared 
with a transition temperature corresponding to 15 meV. Even though the 
stabilization energy scales like $\tau^2/U_2$, there is not enough room 
to improve substantially the situation by a reasonable variation of the 
parameters. 

More specifically, we see from Fig. \ref{mol}($a$) that assuming 
Mattheiss'  parameters for the hopping integrals 
we achieve a stable solution for the RS structure only for a very small 
window of the ratio $J/U_2$ around $0.4$. The most 
favorable situation for this latter is obtained with the choice made by 
Mila {\it et al.},\cite{mila00} by putting $\alpha=\beta=\sigma=0$ and 
$\tau \neq 0$ (Fig. \ref{mol}$(b)$), but even in this case, as already stated, the 
stabilization energy is of the order of $\approx 2$ meV. We have also 
tried to investigate the role of the monoclinic distortion in stabilizing 
the RS structure. To introduce it, we have assumed that after the setting 
in of the broken symmetry phase, the hopping parameters  
$\chi=t^{12}$ and $\theta=t^{13}$ take a value different from zero along
the bond $\delta_1$, whose length increases by 0.1 \AA, and remain 
substantially zero along the other two directions, where the bond distance 
is unchanged after the transition. The result is essentially negative as 
there is a little but not significant improvement.  

\begin{figure}
      \epsfysize=30mm
      \centerline{\epsffile{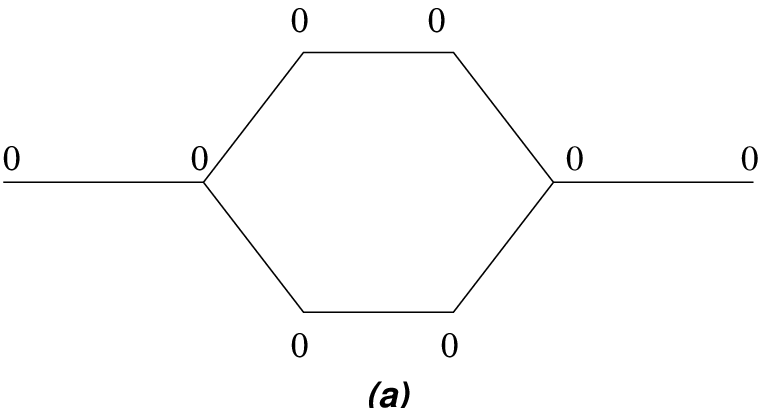}}
\vspace{0.5cm}
      \epsfysize=30mm
      \centerline{\epsffile{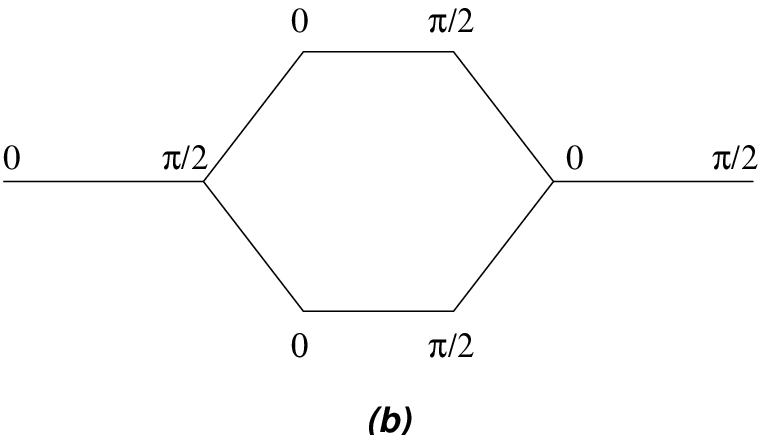}}
\vspace{0.5cm}
      \epsfysize=30mm
      \centerline{\epsffile{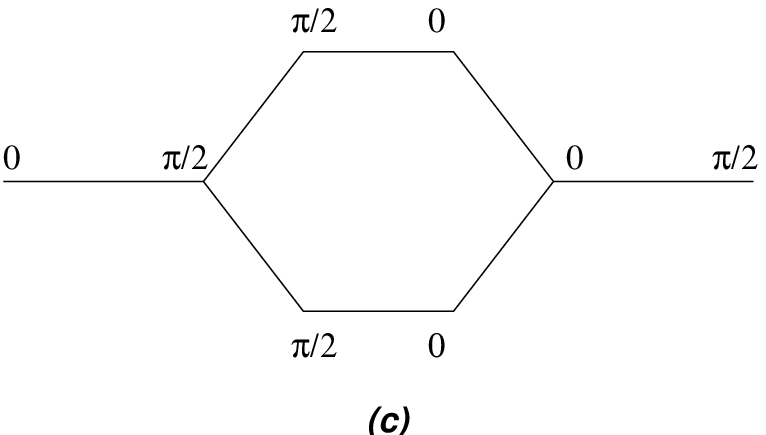}}
\vspace{0.5cm}
\caption{The angles indicate the orbital occupancy of the 
vertical molecule at a given site. $(a)$ Orbital ordering 
in the RS(FO) phase. $(b)$ Orbital ordering in the RS(AO) phase. 
$(c)$ Orbital ordering in the RS(ME) excited phase.}
\label{oordering}
\end{figure}

Turning now to the orbital structure, Fig. \ref{oordering}$(a)$ gives 
the minimizing values of the orbital mixing angles $\psi_{n}$ of
Eq. (\ref{mixhor}) in the basal plane at all molecular sites for the 
RS configurations found in Fig. \ref{mol}. The value $\psi_{n} = 0$ for 
all sites means a uniform occupation throughout the crystal of the 
molecular state $\|\psi_-  \rangle$, i.e., the ferro-orbital solution 
found by Mila {\it et al.}\cite{mila00} In agreement with them, also for 
the ARS configuration (their A phase) we find a uniform solution with a 
minimizing angle $\psi_{n} = \pi/2$, i.e., a uniform occupation 
throughout the crystal of the molecular state $\|\psi_+  \rangle$.

Moreover for the AF and F phases we find a continuum of orbital degeneracies of antiferro-orbital type, 
in the sense that all the orbital configurations with any mixing angle 
$\psi_a$ on the central molecule and a mixing angle of $\psi_a + \pi/2$ 
on the three in-plane neighboring molecules, have the same energy (see Fig. \ref{oordering}$(b)$ for the particular case $\psi_a = 0$). This feature can also be deduced analytically from 
Eq. (\ref{averh}).

As anticipated in Section IV, the magnetic group of the ferro-orbital 
solution is not in keeping with the experimental findings of Goulon 
{\it et al.}\cite{goulon00} We have therefore analyzed the orbital 
order of the excited configurations within a range of $\simeq$ 4 meV 
from the ground state. Referring to Fig. \ref{mol}$(b)$, for $J=0.85$ meV, 
the ferro-orbital RS(FO) phase is at $E'_V = -0.0744$ meV, all the 
degenerate AF(AO) 
phases with antiferro-orbital ordering at $E'_V = -0.0729$ meV and a phase 
that will be called RS(ME), for reasons that will shortly become apparent  
(ME stands for magneto-electric), with the orbital ordering depicted in 
Fig. \ref{mol}$(c)$ lies at $E'_V = -0.0707$.

We have also explored the consequences of varying the ratio $\alpha/\tau$ 
from the zero value assumed in Fig. \ref{mol}$(b)$ to about one. From the
picture following Eq. (\ref{psidim})
it is in fact evident that an $\alpha = t_{11}$ value of the same order of 
$\tau = t_{23}$ would favor an antiferro-orbital coupling along the 
$\delta_1$ bond in the RS phase.

\begin{figure}
      \epsfysize=55mm
      \centerline{\epsffile{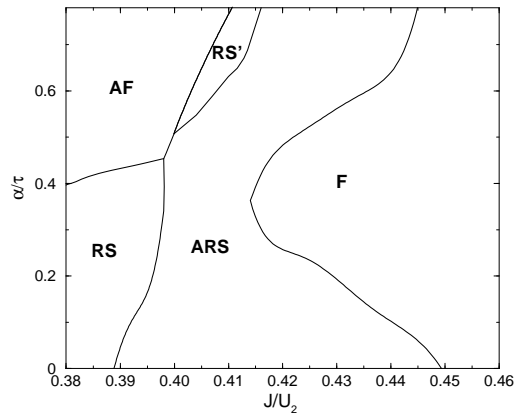}}
\caption{Phase diagram in the ($\alpha/\tau$, $J/U_2$) parameter plane. 
Here AF, F, ARS denote 
the corresponding type of magnetic order; 
 RS and RS'  denote the RS(FO) and RS(AO) phase with orbital structures shown in Fig. \ref{oordering}.
The solid lines indicate the  phase boundaries.
We put for simplicity $\beta=\sigma=0$.}
\label{phase}
\end{figure}

\vspace{-0.3cm}

To this purpose we have drawn the phase diagram for the various magnetic
configurations in the plane $\alpha/\tau$ versus $J/U_2$. The relevant 
result is shown in Fig. \ref{phase}.
As expected another RS' phase 
with the same spin configuration as the RS phase appears, 
centered around the values $\alpha/\tau = 0.7$ and $J/U_2 = 0.40$, 
with the molecular orbital ordering depicted in Fig. \ref{oordering}$(b)$
(in-plane antiferro-orbital (AO) ordering). Again an energy analysis of the 
excited phases for $J = 1.0$ eV and $\alpha = -0.19$ eV leads to the 
following sequency: 
\begin{eqnarray*}
E'_V [RS(AO)] & = & -0.0909 ~{\rm eV} \\
E'_V [AF(AO)] & = & -0.0908 ~{\rm eV} \\
E'_V [RS(ME)] & = & -0.0905 ~{\rm eV}  
\end{eqnarray*}

where in brackets we have indicated the spin and orbital configurations. 
As seen, the energy spreading is now much more reduced, only 0.4  
meV separating the orbital ME phase from the ground state.

In order to establish the magnetic group for the RS phases, 
of interest here, we observe 
that, because of the entangled nature of the molecular state, the 
average orbital type of the two atoms constituting the 
vertical pair is the same. With respect to the 
$D_{3d}$ corundum symmetry point group, the state $|0 \rangle$ 
transforms according to the totally symmetric representation ($A_{1g}$ 
in Schoenflies notation) while $|-1 \rangle$ and $|1 \rangle$ are partner 
functions of the bidimensional $E_g$ representation and transform, 
respectively, 
like the basis $e^{(1)}_{g}$ and $e^{(2)}_{g}$ in CNR.\cite{cnr1,cnr2}

We can then use Table I to see which 
symmetry operations conserve the colored magnetic structure, obtained 
by adding to the lattice sites not only a spin label but also a color 
label given by the type of orbital occupation at that site. 
We find that the magnetic group for the RS(AO) phase is 
$C_2\otimes\hat T$ (group n. 4 in Section IV), whereas that for the RS(ME) 
phase is $C_{2h}(C_s)$ (group n. 7). Therefore the only phase compatible 
with the findings of Goulon {\it et al.} \cite{goulon00} is this latter. 
As it will be argued in Section VII, it might be possible 
that the combined effect of the symmetry breaking of the spin and 
orbital degrees of freedom lead to a favorable
coupling to the lattice in such a way as to stabilize the 
RS(ME) phase with respect to the competing configurations. In this case the 
role of the monoclinic distortion would be essential to achieve the ground 
state with the correct symmetry.

\subsection{The case of the atomic variational state.}

Even though the values of the hopping parameters as shown in Table III seem to 
favor what we called the molecular regime, nonetheless we think it could 
be useful to analyze, for the sake of completeness, also the other regime. 
In this case the orbital on-site part of the variational wave function is atomic-like and
should be written as 
\begin{eqnarray}
\parallel\psi^o_i\rangle
=\cos\theta_i|0\rangle_{i}+\sin\theta_i
(\cos\psi_i|1\rangle_{i}+\sin\psi_i|-1\rangle_{i}) ~, \nonumber 
\end{eqnarray}


\noindent allowing all the three states $|0\rangle_{i}$, $|1\rangle_{i}$ and 
$|-1\rangle_{i}$ to be present without any {\it a priori} restriction 
on their relative weight, contrary to the molecular case. 
The relative weight of the three states is then determined through the 
minimization procedure with respect to the variational parameters 
$\theta_i$ and $\psi_i$.
The only restriction we impose is that the solutions must fill the whole 
crystal, with a periodicity not less than that of the monoclinic cell. 
This is indeed a quite reasonable request, as the solutions with periodicity 
of more than the unit cell should describe excited states. 
As the unit cell of V$_2$O$_3$ is formed by 8 V atoms, there are in 
principle 16 minimization angles. In order to simplify the problem, 
we use the symmetry relations between the variational angles dictated 
by all the possible magnetic space groups for V$_2$O$_3$ described 
in section IV. Indeed for each group, the states of the V atoms inside 
the cell are not independent, but are related by the symmetry 
operations, thus providing a reduction of the number of parameters. By 
taking the absolute minimum we shall determine the orbital and spin nature
of the ground state, together with the corresponding magnetic group.
In this way we exclude solutions not invariant with respect to the chosen 
groups, but we note that all the interesting subgroups for V$_2$O$_3$ 
have been taken into account.

\vspace{0.3cm}

\noindent Table IV. Number of independent variational angles in the unit cell 
according to the various possible magnetic groups.

\begin{eqnarray}
\begin{array}{llll}

1.&  \hat E, \hat I, \hat C_2, \hat \sigma_b & 
\Rightarrow & ({\rm V}_1, {\rm V}_2, {\rm V}_6, {\rm V}_8) \equiv 
(\theta_1,\psi_1)\\&
 & & ({\rm V}_3, {\rm V}_4, {\rm V}_5, {\rm V}_7 ) \equiv (\theta_2,\psi_2) \\
2. &  \hat E, \hat I, \hat T, \hat T \hat I   &  \Rightarrow   &   ({\rm V}_1,
 {\rm V}_2, {\rm V}_3, {\rm V}_7)\equiv(\theta_1,\psi_1) \\&
 & &  ({\rm V}_4, {\rm V}_5, {\rm V}_6, {\rm V}_8)\equiv(\theta_2,\psi_2) \\
3. &  \hat E, \hat I, \hat T\hat C_2, \hat T\hat \sigma_b  &  \Rightarrow  &  
({\rm V}_1, {\rm V}_2, {\rm V}_4, {\rm V}_5)\equiv(\theta_1,\psi_1) \\&
 & &  ({\rm V}_3, {\rm V}_6, {\rm V}_7, {\rm V}_8)\equiv(\theta_2,\psi_2) \\
4. &  \hat E, \hat T, \hat C_2, \hat T \hat  C_2  &  \Rightarrow  &  
({\rm V}_1, {\rm V}_4, {\rm V}_7, {\rm V}_8)\equiv(\theta_1,\psi_1) \\&
 & &  ({\rm V}_2, {\rm V}_3, {\rm V}_5, {\rm V}_6)\equiv(\theta_2,\psi_2) \\
5. &  \hat E, \hat C_2, \hat T \hat I,\hat T \hat \sigma_b  &  \Rightarrow  & 
 ({\rm V}_1, {\rm V}_3, {\rm V}_5, {\rm V}_8)\equiv(\theta_1,\psi_1) \\&
 & &  ({\rm V}_2, {\rm V}_4, {\rm V}_6, {\rm V}_7)\equiv(\theta_2,\psi_2) \\
6. &  \hat E, \hat T, \hat \sigma_b, \hat T\hat \sigma_b   &  \Rightarrow  & 
 ({\rm V}_1, {\rm V}_5, {\rm V}_6, {\rm V}_7)\equiv(\theta_1,\psi_1) \\&
 & &  ({\rm V}_2, {\rm V}_3, {\rm V}_4, {\rm V}_8)\equiv(\theta_2,\psi_2) \\
7. &  \hat E,  \hat \sigma_b, \hat T\hat I, \hat T  \hat C_2   &
 \Rightarrow  &   ({\rm V}_1, {\rm V}_3, {\rm V}_4, {\rm V}_6)\equiv
(\theta_1,\psi_1) \\&
 & &  ({\rm V}_2, {\rm V}_5, {\rm V}_7, {\rm V}_8)\equiv(\theta_2,\psi_2) \\
8.& C_{2h} \otimes \hat T &   \Rightarrow   & 
{\rm ~same ~for ~all~ V_i} \equiv ( \theta ,\psi  ) 
\end{array}
\nonumber
\end{eqnarray}


In Table IV we list, for each magnetic group, the number of 
independent angles associated with the corresponding atomic sites. 
This number is obviously given by 16 divided the order of the group.

Given the full expression for $H_{\rm eff}$ reported in Appendix C, we can
then evaluate the matrix elements 
for $H^F_{\rm eff}$ and $H^{A}_{\rm eff}$ (as defined by Eqs. (\ref{hf}) and (\ref{haf})) along vertical ($\delta_4$) and horizontal 
($\delta_1, \delta_2, \delta_3$) bonds. The spin averages are again calculated and included in the formulas as in the previous subsection.

In the case of $H^F_{\rm eff}$ the matrix element for the $\delta_4$ bond is:

\begin{eqnarray}
\begin{array}{l}
\langle\Psi_{b}|  \langle\Psi_{a}|  H_{\rm eff}^{F}(\delta_4) |  
\Psi_{a}\rangle|  \Psi_{b}\rangle=\\
(\cos ^2\theta_a\sin ^2\theta_b+\sin ^2\theta_a\cos ^2\theta_b) \tilde G_1\\
+\sin ^2\theta_a\sin ^2\theta_b (\cos ^2\psi_a\sin ^2\psi_b +
\sin ^2\psi_a\cos ^2\psi_b ) \tilde G_2\\
+\sin ^2\theta_a\sin ^2\theta_b\sin 2\psi_a\sin 2\psi_b\tilde G_3\\
+\sin 2\theta_a\sin 2\theta_b(\cos \psi_a\cos  \psi_b+
\sin \psi_a\sin \psi_b) \tilde G_4 ~,
\end{array}
\nonumber
\end{eqnarray}

\vspace{-2.5cm}

\begin{eqnarray}
\label{fmvert}
\end{eqnarray}

\vspace{1.5cm}

\noindent where
\begin{eqnarray}
\begin{array}{l}
\tilde G_1=u(\mu ^2+\rho ^2)\\
\tilde G_2=2u\mu ^2\\
\tilde G_3=-u\mu ^2\\
\tilde G_4=-u\mu\rho\\
u=-\frac{1}{U_2-J} ~.
\end{array}
\label{fmvertbis}
\end{eqnarray}

For $H^{A}_{\rm eff}$ it can be written as
\begin{eqnarray}
\begin{array}{l}
\langle\Psi_{b}|  \langle\Psi_{a}|  H_{\rm eff}^{A}(\delta_4) |  
\Psi_{a}\rangle|  \Psi_{b}\rangle=\\
\cos ^2\theta_a\cos ^2\theta_b \tilde F_1\\
+(\cos ^2\theta_a\sin ^2\theta_b+\sin ^2\theta_a\cos ^2\theta_b) \tilde F_2\\
+\sin ^2\theta_a\sin ^2\theta_b \tilde F_3\\
+\sin ^2\theta_a\sin ^2\theta_b (\cos ^2\psi_a\cos ^2\psi_b +
\sin ^2\psi_a\sin ^2\psi_b ) \tilde F_4\\
+\sin ^2\theta_a\sin ^2\theta_b (\cos ^2\psi_a\sin ^2\psi_b +
\sin ^2\psi_a\cos ^2\psi_b ) \tilde F_5\\
+\sin ^2\theta_a\sin ^2\theta_b\sin 2\psi_a\sin 2\psi_b\tilde F_6\\
+\sin 2\theta_a\sin 2\theta_b(\cos \psi_a\cos  \psi_b+
\sin \psi_a\sin \psi_b) \tilde F_7 ~,
\end{array}
\nonumber
\end{eqnarray}

\vspace{-3.0cm}

\begin{eqnarray}
\label{afmvert}
\end{eqnarray}

\vspace{2.0cm}

\noindent where
\begin{eqnarray}
\begin{array}{l}
\tilde F_1=2(v+w)\mu ^2\\
\tilde F_2=(v+\frac{5w}{3})\mu ^2+\frac{2w}{3}\rho ^2\\
\tilde F_3=(v+w)\rho ^2\\
\tilde F_4=(v+\frac{7w}{3})\mu ^2\\
\tilde F_5=\frac{4w}{3}\mu ^2\\
\tilde F_6=(\frac{v}{2}-\frac{w}{6})\mu ^2\\
\tilde F_7=(\frac{v}{2}-\frac{w}{6})\mu\rho\\
v=-\frac{1}{U_2+4J}\\
w=-\frac{1}{U_2+2J} ~.
\end{array}
\label{tildeF}
\end{eqnarray}

For the horizontal bond $\delta_1$, the $H^{F}_{\rm eff}$ average value takes the form:

\begin{eqnarray}
\begin{array}{l}
\langle\Psi_{c}|  \langle\Psi_{a}|  H_{\rm eff}^{F}(\delta_1) |  
\Psi_{a}\rangle|  \Psi_{c}\rangle \\
=[\sin ^2\theta_a\cos ^2\psi_a(\cos ^2\theta_c+\sin ^2\theta_c\sin ^2\psi_c) \\
+\sin ^2\theta_c\cos ^2\psi_c(\cos ^2\theta_a+\sin ^2\theta_a\sin ^2\psi_a)]\bar G_1 \\
+[\sin ^2\theta_a\sin ^2\psi_a(\cos ^2\theta_c+\sin ^2\theta_c\cos ^2\psi_c)\\
+\sin ^2\theta_c\sin ^2\psi_c(\cos ^2\theta_a+\sin ^2\theta_a\cos ^2\psi_a)] 
\bar G_2\\
+[2\cos ^2\theta_a\cos ^2\theta_c+\sin ^2\theta_a\cos ^2\psi_a\cos ^2\theta_c \\
+\cos ^2\theta_a\sin ^2\theta_c\cos ^2\psi_c+
\sin ^2\theta_a\sin ^2\theta_c(\sin ^2\psi_a\\
+\sin ^2\psi_c)-
\sin 2\theta_a\sin 2\theta_c\sin \psi_a\sin \psi_c]\bar G_3\\
+\sin ^2\theta_a\sin ^2\theta_c\sin 2\psi_a\sin 2\psi_c\bar G_4\\
+\sin 2\theta_a\sin 2\theta_c\cos \psi_a\cos \psi_c\bar G_5\\
+(\sin 2\theta_a\sin ^2\theta_c\cos \psi_a\sin 2\psi_c\\
+\sin 2\theta_c\sin ^2\theta_a\cos \psi_c\sin 2\psi_a)\bar G_6\\
+\sin 2\theta_a\sin 2\theta_c\sin \psi_a\sin \psi_c\bar G_7\\
+(\cos ^2\theta_a\sin ^2\theta_c+\cos ^2\theta_c\sin ^2\theta_a)\bar G_8\\
+(\cos ^2\theta_a\sin 2\theta_c\sin \psi_c\\
+\cos ^2\theta_c\sin 2\theta_a\sin \psi_a)\bar G_9\\
+[\sin 2\theta_a\sin \psi_a(\cos ^2\theta_c+\sin ^2\theta_c\cos 2\psi_c)\\
+\sin 2\theta_c\sin \psi_c(\cos ^2\theta_a+\sin ^2\theta_a\cos 2\psi_a)]\bar G_{10} ~,
\end{array}
\label{fmhor}
\end{eqnarray}

\noindent where
\begin{eqnarray}
\begin{array}{l}
\bar G_1=u\alpha^2\\
\bar G_2=u\beta^2\\
\bar G_3=u\tau^2\\
\bar G_4=u\alpha\beta\\
\bar G_5=u\alpha\sigma\\
\bar G_6=-u\alpha\tau\\
\bar G_7=-u\beta\sigma\\
\bar G_8=u\sigma ^2\\
\bar G_9=-u\sigma \tau\\
\bar G_{10}=u\beta \tau ~.
\end{array}
\label{barG}
\end{eqnarray}

For the same bond in the case of $H^{A}_{\rm eff}$ we obtain:

\begin{eqnarray}
\begin{array}{l}
\langle\Psi_{c}|  \langle\Psi_{a}|  H_{\rm eff}^{A}(\delta_1) |  
\Psi_{a}\rangle|  \Psi_{c}\rangle=\\
\cos ^2\theta_a\cos ^2\theta_c \bar F_1\\
+(\cos ^2\theta_a\sin ^2\theta_c\sin ^2\psi_c
+\sin ^2\theta_a\cos ^2\theta_c \sin ^2\psi_a) \bar F_2\\
+(\cos ^2\theta_a\sin ^2\theta_c
+\sin ^2\theta_a\cos ^2\theta_c ) \bar F_3\\
+\sin ^2\theta_a\sin ^2\theta_c (\cos ^2\psi_a\sin ^2\psi_c +
\sin ^2\psi_a\cos ^2\psi_c ) \bar F_4\\
+\sin ^2\theta_a\sin ^2\theta_c\bar F_5\\
+\sin^2\theta_a\sin ^2\theta_c(\cos ^2\psi_a+\cos ^2\psi_c)\bar F_6\\
+\sin ^2\theta_a\sin ^2\theta_c\sin 2\psi_a\sin 2\psi_c\bar F_7\\
+\sin 2\theta_a\sin 2\theta_c\sin \psi_a\sin  \psi_c \bar F_8\\
+(\sin^ 2\theta_a\sin 2\theta_c \sin 2\psi_a\cos \psi_c\\
+\sin^ 2\theta_c\sin 2\theta_a \sin 2\psi_c\cos \psi_a)\bar F_{9}\\
+\sin 2\theta_a\sin 2\theta_c \cos \psi_a\cos \psi_c\bar F_{10}\\
+[\sin 2\theta_a\sin \psi_a(1+\sin^ 2\theta_c\sin ^2 \psi_c)\\
+\sin 2\theta_c\sin \psi_c(1+\sin^ 2\theta_a\sin ^2 \psi_a)]\bar F_{11}\\
+[\sin 2\theta_a\sin \psi_a(2-\sin ^2\theta_c)\\
+\sin 2\theta_c\sin \psi_c(2-\sin ^2\theta_a)]\bar F_{12} ~,
\end{array}
\nonumber
\end{eqnarray}

\vspace{-4.0cm}

\begin{eqnarray}
\label{afmhor}
\end{eqnarray}

\vspace{3.0cm}

\noindent with the definitions
\begin{eqnarray}
\begin{array}{l}
\bar F_1=v(\alpha ^2+\beta^2)+\frac{w}{3}(3(\alpha ^2+\beta^2)+5\tau^2)\\

\bar F_2=v(-\alpha ^2+\beta^2)+\frac{w}{3}(\alpha ^2-\beta^2-2\tau^2)\\

\bar F_3=v(\alpha^2 +\tau^2)+\frac{w}{3}(2\alpha ^2+3\beta^2+2\sigma ^2+5\tau^2)\\

\bar F_4=v(\alpha ^2+\beta^2)+\frac{1}{6}(\alpha ^2+\beta^2)\\

\bar F_5=v\sigma ^2+\frac{w}{3}(3\alpha ^2+\sigma ^2+5 \tau^2)\\

\bar F_6=v\tau^2+\frac{w}{6}(-3\alpha ^2+3\beta^2+2 \tau^2)\\

\bar F_7=(-\frac{v}{2}+\frac{w}{6})\alpha\beta\\

\bar F_8=\frac{v}{2}(\beta\sigma +\tau^2)-\frac{w}{6}(\beta\sigma+\tau^2) \\

\bar F_{9}=-(\frac{v}{2}-\frac{w}{6})\alpha \tau\\

\bar F_{10}=-(\frac{v}{2}-\frac{w}{6})\alpha \sigma\\

\bar F_{11}=-\frac{w}{3}\beta \tau \\

\bar F_{12}=-\frac{w}{3} \sigma \tau ~.

\end{array}
\label{barF}
\end{eqnarray}

The matrix elements along $\delta_2$ and $\delta_{3}$ are easily 
derived from the expressions (\ref{fmhor})-(\ref{barF}) using the 
symmetry properties of $H_{\rm eff}$ under the trigonal $C_{3}$ 
symmetry, as done previously in the molecular case.

In this way the average of the Hamiltonian on a ferromagnetic bond is given by the sole $H_{\rm eff}^F$ contribution, while for an antiferromagnetic bond we have to add $\frac{1}{3}H_{\rm eff}^F$ to the contribution of $H_{\rm eff}^A$.

The term due to the trigonal field splitting (\ref{hcf}) can be also taken into account. Its energy contribution per V-atom is given by: $\frac{\Delta_t}{8}\sum_{i=1}^8 \sin^2{\theta_i}$.

\begin{center}
\begin{figure}
      \epsfysize=55mm
      \centerline{\epsffile{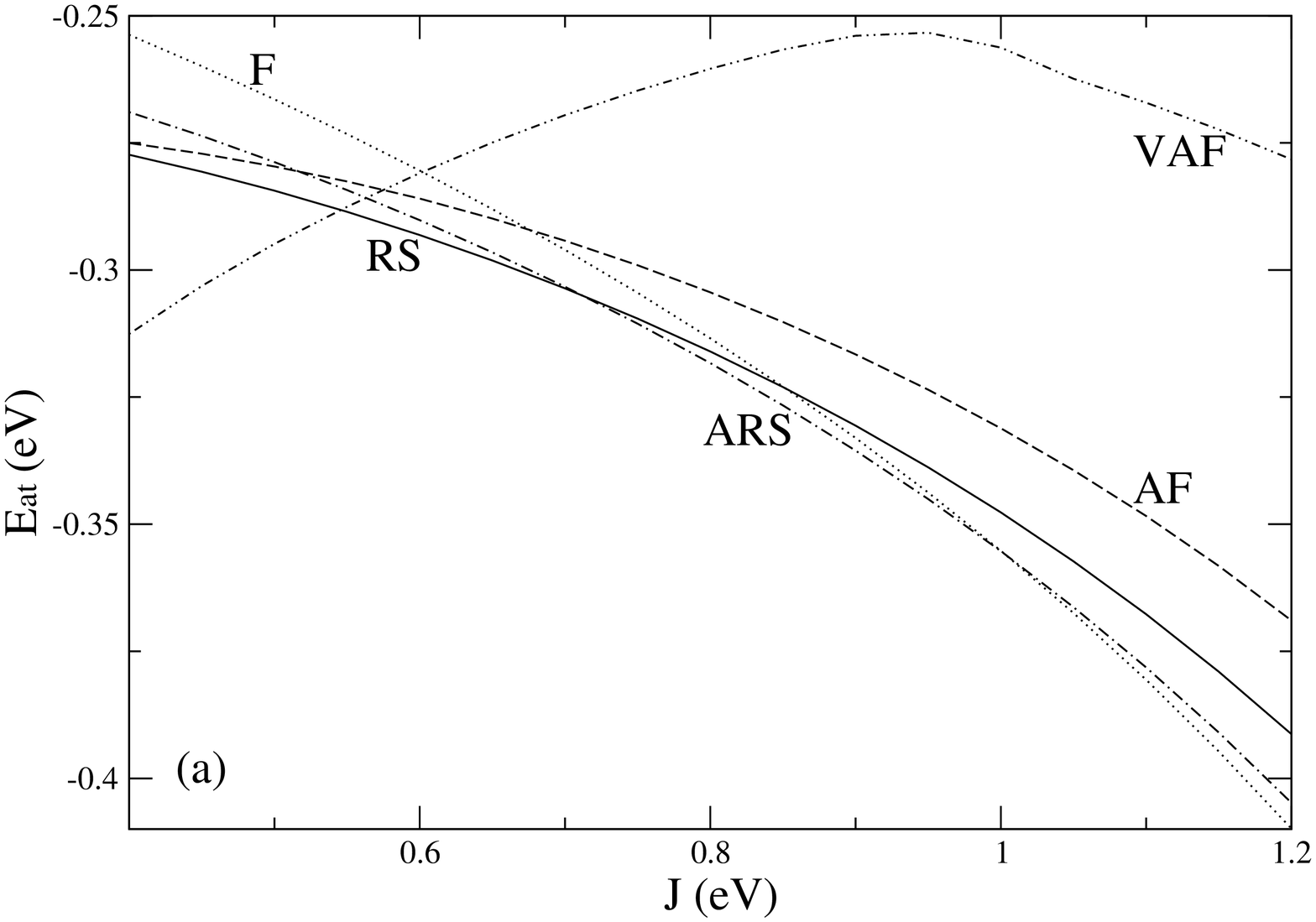}}
\vspace{0.8cm}
      \epsfysize=55mm
      \centerline{\epsffile{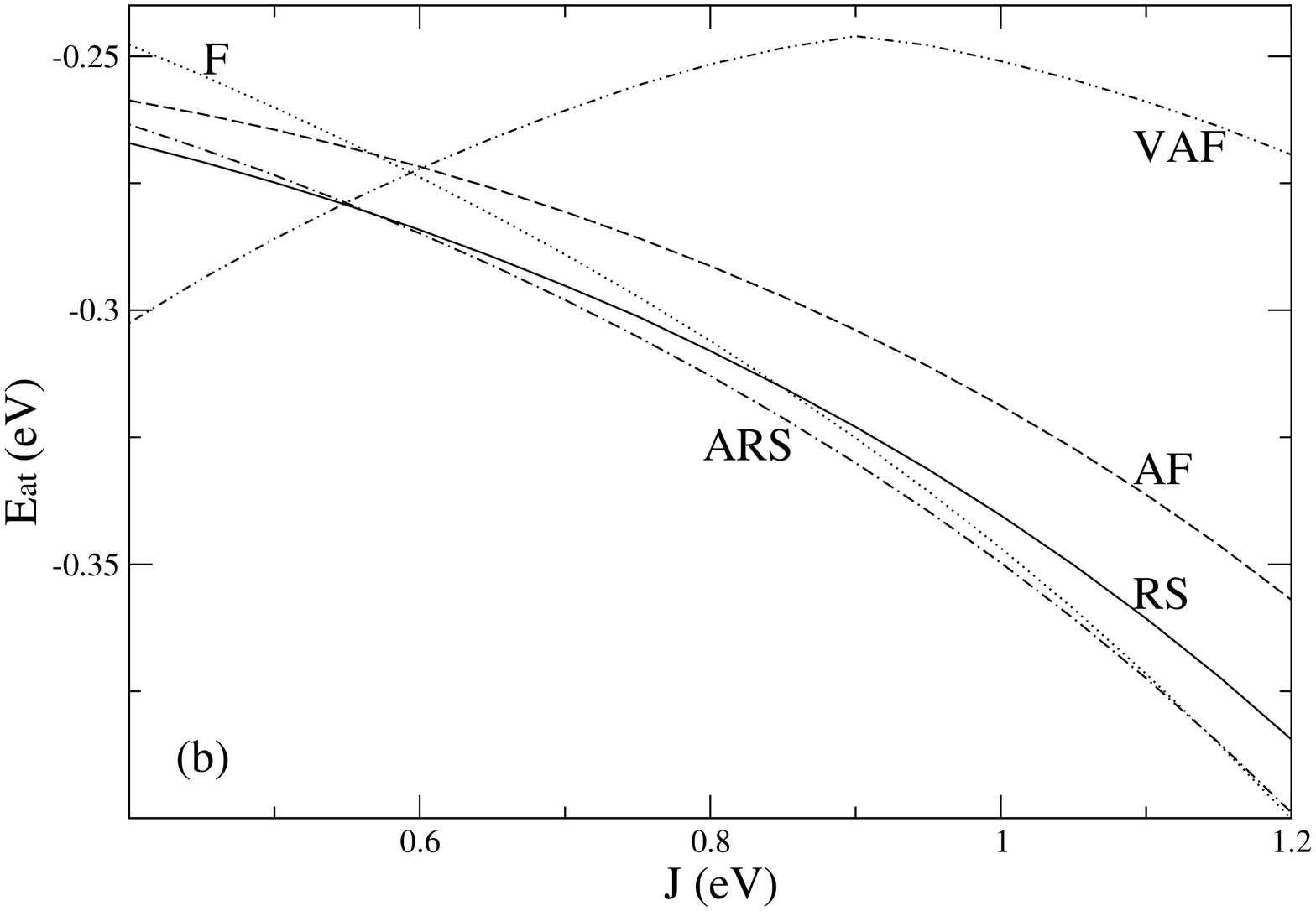}}
\vspace{0.8cm}
      \epsfysize=55mm
      \centerline{\epsffile{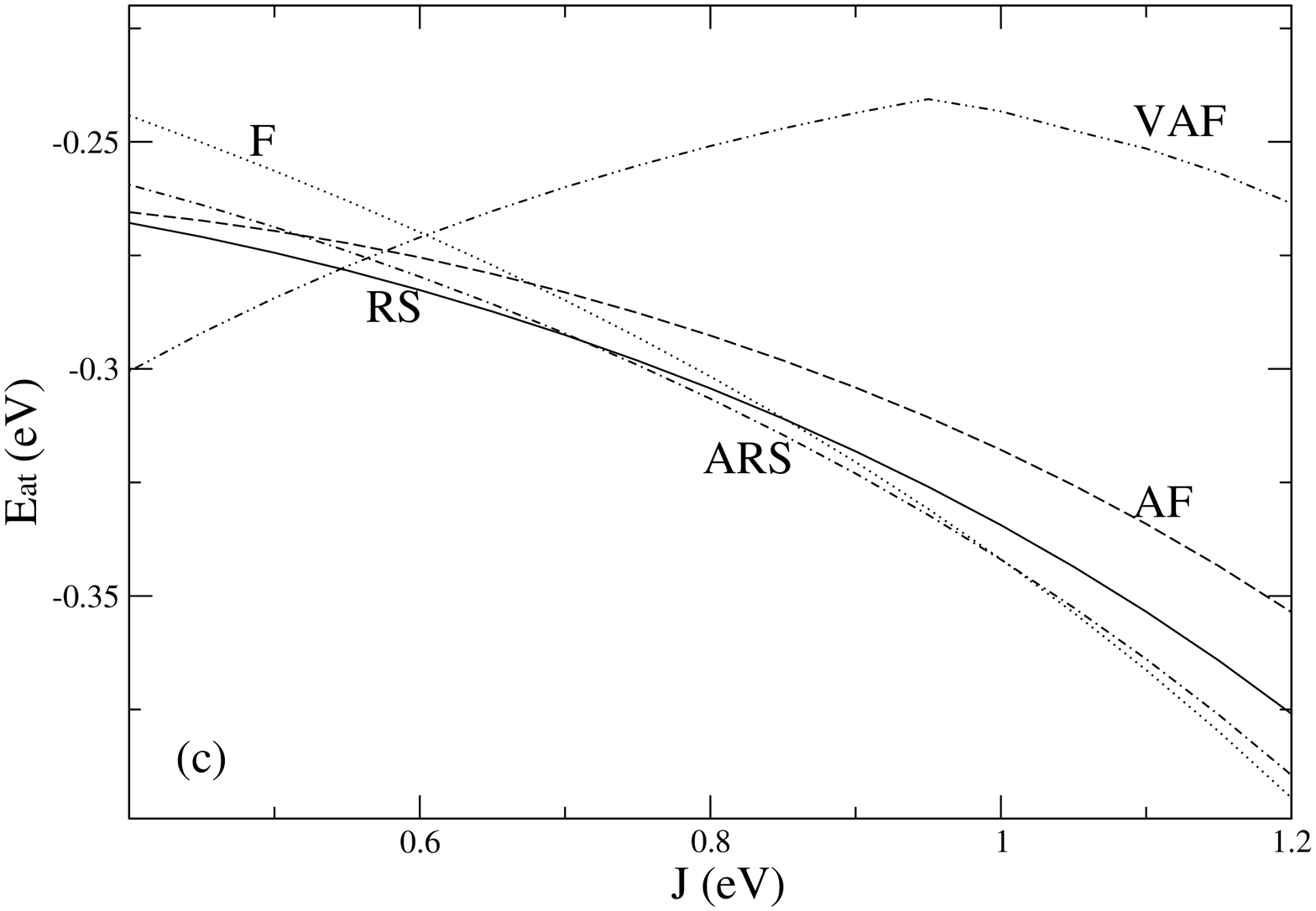}}
\caption{The ground state energy $E_{at}$ as a function of $J$ 
for different magnetic configurations.
The hopping parameters are taken as follows:
(a)  $\rho=-0.82$, $\mu=0.2$, $\alpha=0.14$, $\beta=-0.05$, $\sigma=0.05$, 
     $\tau=0.27$;
(b)  $\rho=-0.82$, $\mu=0.2$, $\alpha=0$, $\beta=0$, $\sigma=0$, $\tau=0.27$;
(c)  $\rho=-0.82$, $\mu=0$, $\alpha=0.14$, $\beta=-0.05$, $\sigma=0.05$, 
     $\tau=0.27$.}
\label{atom}
\end{figure}
\end{center}

Given all these ingredients, we can now evaluate the ground state 
energy for all 8 groups: the ``true'' ground state energy per V atom, $E_{at}$, is then 
obtained as the absolute minimum among the 8 minima, determining 
in this way also the magnetic group for V$_2$O$_3$. 
The results are presented for two choices of hopping parameters 
corresponding to those of Fig. \ref{mol}$(a)$, $(b)$ in the molecular regime and $\rho = -0.82$ 
and $\mu = 0.2$ eV (Mattheiss set, see Table III).
We consider for the moment only the case $\Delta_t = 0$.
As apparent from Fig. \ref{atom}$(a)$, $(b)$, we have in the 
atomic regime one more curve (labeled VAF) giving the ground state energy
of the system in a magnetic configuration in which the two atoms along
$\delta _4$ are coupled antiferromagnetically, independently of all the 
other spin couplings along $\delta _1$, $\delta _2$ and $\delta _3$.
Of course, such ground state configuration was absent in the 
molecular regime, where  we started from a $S^M=2$ state. It relates 
to the $S^M=0$ molecular solution in Fig. \ref{hubd0}.

Notice that a direct comparison between Fig. \ref{mol} and Fig. \ref{atom} 
may be misleading, since in the atomic regime the vertical bond 
$\delta _4$ is included in the ground state energy $E_{at}$, while in the molecular 
regime we have subtracted from $E_V$ the binding energy of the molecule: 
$\frac{\Delta E_m}{2}$. This means that the energy 
$\frac{\Delta E_m}{2}$ ($\simeq -0.33~eV$, for $J \simeq 1.0$)) must be added to $E'_V$ 
of Fig. \ref{mol}, thus restoring the correct numerical
correspondence between the two cases. 

In particular, for values of $J$ such that the VAF is not the stable phase, we can write the atomic ground state energy per V atom as
\begin{equation}
E_{at}=E'_{at}-\frac{1}{2}\frac{\rho^2+\mu^2}{U_2-J}~,
\label{eat}
\end{equation}

\noindent where we separated the in-plane contribution $E'_{at}$ from the energy gain along $\delta_4$, i.e., $-\frac{1}{2}\frac{\rho^2+\mu^2}{U_2-J}$.

\vspace{0.4cm}

In this way, dropping all the common Hartree-Fock terms along the vertical bond $\delta_4$, we can write the two inequalities:
\begin{eqnarray}
\left \{ 
\begin{array}{l}
E'_{at} < E'_V \\[0.3cm]
E'_{at} > E'_V - \frac{\rho \mu}{U_2-J} ~.
\end{array}
\right. 
\label{orden}
\end{eqnarray}

The reason for these inequalities is the following. The first of Eqs. (\ref{orden}) says that the in-plane energy gain with the atomic variational wave functions is always lower than the corresponding molecular one.
This is to be  expected, since the variational space in the molecular case is reduced with respect to the atomic one, where the states are not constraint to satisfy the form of Eq. (\ref{psidim}).

The second of Eqs. (\ref{orden}) states, instead, that  the molecular ground state lies lower than the atomic one, because of the correlation energy. Note that, while the first Equation is always valid, the validity of the second is limited to sufficiently high values of $\rho$ and $\mu$ (for example, the standard set) and its breakdown marks the transition point between the molecular and the atomic regime.


\vspace{-0.6cm}

The analysis of the two Figs. \ref{atom}$(a)$, $(b)$ shows that the RS phase is realized in both cases (though in Fig. \ref{atom}$(b)$ it is very small). Indeed in the case of the standard set of parameters, the atomic solution is even more stable than the corresponding molecular one, the first excited state lying $\simeq 3$ meV above. Nonetheless it cannot be considered a good ground state solution for the reasons stated above (see Eqs. (\ref{orden})). Moreover, the magnetic space group of the solution is not ME, being group 2. of Table IV. The orbital pattern is made of planes of V-atoms, orthogonal to the $c_H$-axis, in $|0\rangle$ state, alternating with planes of V atoms in $|-1\rangle$ state.

Figure \ref{atom}$(c)$ presents the case $\rho = -0.82$ eV and $\mu = 0$ eV, in
order to satisfy the criterion for using an atomic variational function, i.e., the correlation energy less than the in-plane exchange energy. In fact, in this case the condition $\mu = 0$ implies that the second of Eqs. (\ref{orden}) is not satisfied, and then $E_{at}<E_{V}$. Thus the RS solution obtained in this case is the overall ground state of the system for this value of the parameters. Unfortunately, this solution has the usual drawbacks already analyzed (stability energy of only $3$ meV, space magnetic group 2. of Table IV, not ME) and moreover the ME solution lies very far from the ground state ($\simeq 100$ meV), thus confirming 
the fact that the choice $\mu=0$ is not suitable to describe V$_2$O$_3$.

The effect of the crystal field in this situation is to favor the occupancy of the $|0\rangle$ states on all the atoms, as expected. In the case of high trigonal distortion $\Delta_t=0.4$ eV, we obtain (not shown) a direct transition from the VAF to the F phase, with no stable RS solution. In VAF phase, all the V atoms are in the $|0\rangle$ configuration, while in the F phase, the percentage of $|0\rangle$ states lowers to $85\%$, the remaining $15\%$ being essentially $|1\rangle$, to allow the hopping process.

Note that even in the case of such a high value of the trigonal distortion ($\Delta_t =0.4$ eV), we checked that the molecular phase is by far the more stable state (for example, for the standard set of parameters and $J=1.0$ eV, the molecular ground state lies 60 meV lower than the atomic one).


\section{Discussion and conclusions.}

In this concluding section we shall review the implications of the spin 
$S=1$ model for V$_2$O$_3$ described above in relation to the present 
experimental evidence. The comparison will allow us to focus on merits 
and drawbacks of the various solutions obtained in the previous section.

\subsection{Occupancy of $a_{1g}$ orbital and spin.}

In the molecular regime, the variational wave function was assumed to be 
of the form given in Eq. (\ref{st1}), in which the occupancy of the $a_{1g}$ 
orbital is 25\%. 
However Park {\it et al.}\cite{park00} report an occupancy of 17\% for this
orbital in the AFI phase. This is an indication that other states with only 
$e_g$ occupancy are mixed in the ground state of this phase. In reality, 
even though the state in Eq. (\ref{st1}) is the main component of the 
wave function, in doing the variational procedure we have neglected 
excited states of the vertical molecule lying within the range of the 
exchange energy in the basal plane. Now, since for all  
the RS phases the ratio $J/U_2$ is
around $0.35 \div 0.40$, our study of the vertical molecule in section V-B 
(see Fig. (\ref{eigen})) shows that we should take $\Delta_t$ around $0.4$ eV 
if we want to stabilize the ferromagnetic 
coupling and find excited states with only $e_g$ occupancy close enough to the 
ground state. Fig. \ref{eigen}$(d)$ illustrates the level structure in the 
case $J=1.0$ eV, from which one can infer 
that two more states with $S^M = 1$ and $S^M = 0$ can be mixed with the 
$S^M = 2$ wave function. Notice from Fig. \ref{hubd0}$(d)$ that for $J=0.85$ eV
these states are even closer to the $S^M = 2$ ground state.
A more general calculation will be done in the future. 
We can however estimate in an 
approximate way the amount of mixing by writing the variational state 
$|\Psi \rangle = \alpha |S^M = 2 \rangle + \beta |S^M = 1 \rangle + 
\gamma |S^M = 0 \rangle $ and imposing the condition that the average 
value of the molecular spin be 1.7, as derived from neutrons and non 
magnetic resonant scattering data (see Introduction). Therefore
$$ 2\alpha^2 + 1\beta^2 + 0\gamma^2 = 1.7 ~,$$
which, together with the normalization condition 
$\alpha^2~+~\beta^2~+~\gamma^2~=~1$, gives 
$$ \alpha^2 = 0.85 - \beta^2 /2 ~~;~~\gamma^2 = 0.15 - \beta^2 /2 ~.$$

In the situation shown in Fig. \ref{eigen}$(c)$, even the small spin-orbit 
interaction of the AFI phase of V$_2$O$_3$ ($\simeq 25 ~meV$\cite{}), 
which  couples only to $S^M=1,2$ states and not with $S^M=0$ state, can 
make the ground state energies for $S^M=1$ and $S^M=0$ comparable.
This means that we can reasonably take $\beta^2=\gamma^2$, thus obtaining 
$\alpha^2 = 0.8$ and $\beta^2 = \gamma^2 = 0.1$. 
Therefore the $a_{1g}$ orbital occupancy reduces to 
$0.8 \times 25\% = 20\%$, in good agreement with 
Park {\it et al.}\cite{park00} findings.
This means that also Park's findings point to a value of the trigonal distortion around $0.4$ eV.

\subsection{Trigonal distortion.}

In CNR\cite{cnr1} the trigonal distortion $\Delta_t$ in the 
corundum phase (and therefore also in the monoclinic phase, since the 
cage of the oxygens remains almost unaltered at the metal to insulator 
transition \cite{dernier70}) was assumed to be negligible, in keeping 
with the suggestion by Rubistein.\cite{rubin70}
This latter was based on the following experimental evidence in the paramagnetic phases of V$_2$O$_3$:
\begin{itemize}
\item
There is no detectable anisotropy in the Knight shift of the NMR spectrum;
\item
There is no observable quadrupole splitting in the NMR spectrum.
\item
Zero anisotropy was observed in single-crystal susceptibility measurements for various crystalline orientations with respect to the 
magnetic field. A large value for $\Delta_t$ would result in highly 
anisotropic Van Vleck susceptibility.
\end{itemize}

However the theoretical estimate by Ezhov {\it et al.}\cite{ezhov99} and 
the previous discussion on the $a_{1g}$ occupancy
seem to point out to a value of $\Delta_t$ around $0.4$ eV.
The problem is how to reconcile this high value with the above experimental 
findings. The concept of an entangled state for the vertical pair described 
by the wave function in Eq. (\ref{psidim}) solves this apparent inconsistency 
in an elegant way. Indeed,  the short range magnetic fluctuations observed 
by neutron scattering experiments \cite{bao98} in the paramagnetic phases 
imply a statistical occupancy of the two orbitally degenerate molecular states of Eq. (\ref{psidim})
and x-ray absorption spectra \cite{park00} show evidence of an average 25\% 
occupancy of the $a_{1g}$ and 75\% of $e_g$ orbitals. 
This is very near to the totally symmetric occupation of the three orbitals 
(1/3 each), restoring in such a way the cubic symmetry. Notice that in a 
static picture the conclusion of a negligible trigonal distortion would 
be unescapable.

\subsection{Monoclinic distortion.}

In their work Mila {\it et al.}\cite{mila00} point out that the 
ferro-orbital molecular ordering found for the C phase (our RS(FO)) causes 
an effective uniaxial stress on the lattice degrees of freedom, leading 
to a uniform rotation of the vertical V-V pairs, in agreement with the 
monoclinic distortion proposed by Dernier and Marezio.\cite{dernier70}
At first sight this explanation seems convincing. However on second 
thought it is not clear why atoms of two adjacent vertical pairs 
belonging to the same basal plane (e.g., V$_1$ and V$_2$, or 
V$_4$ and V$_5$ with reference to Fig. \ref{boh}) should go on opposite 
sides of the line joining them (x-axis in the figure). Since the
two V atoms with their local oxygen environment are on average in the 
same electronic state, the initial 
displacive force should have the same sign along the x-axis, by an 
application of the Feynman theorem to an Hamiltonian depending on 
parameters (in this case the position coordinates of the V atoms). While
{\it a posteriori} we see
that the lattice energy is lowered if the two atoms move toward 
the oxygen voids, the initial drive should go on the correct direction.
This is indeed what happens if one assume the orbital ordering found in 
the RS(ME) phase. The two atoms in the basal plane are in different electronic 
states, one odd with respect to the operation x $\rightarrow$ $-$x 
(state $|\psi^o_- \rangle$) and the other even (state $|\psi^o_+ \rangle$), 
and this situation is reproduced for each pair of atoms in the basal planes 
throughout the crystal.

We observe that in the RS(ME) phase the stress arising from the
magnetostrictive forces due to the broken symmetry of the spin degrees
of freedom and the one originating from the orbital coupling to the
lattice, presenting the same pattern of broken symmetry, act along
the same axis, in contrast to the other two phases.
This fact might give an advantage to the RS(ME) spin-orbital
configuration
in its coupling to the lattice so as to become stabilized with respect
to them. In such an instance the monoclinic distortion would be essential
to achieve the ground state with the correct symmetry indicated by the
Goulon experiment. Clearly more investigation is needed on this point.

\subsection{Anomalous diffraction.}

In their paper Mila {\it et al.}\cite{mila00} claim that the ferro-orbital 
ordering  they propose, while different from that invoked by Paolasini
{\it et al.}\cite{paolasini99}, is also consistent with the experimental 
findings. However we feel that their argument was rather scant and we 
try to present here a more in depth analysis of the implications of X-ray 
resonant scattering data, deferring a more complete treatment to a future 
work. We shall see that the space
magnetic symmetry group of the AFI ground state plays an essential role 
in the discussion of the structure factor, since it can discriminate among 
different theoretical models. 

In the anomalous diffraction at the V K-edge, the structure factor $F$ is 
given by
\begin{equation} 
\label{csf}
F = \sum_n f_n e^{i\vec{Q}\cdot\vec{R_n}}
\end{equation}

\noindent where the sum is performed over the eight V-atoms of the unit cell, 
$f_n$ is the resonant atomic scattering factor (RASF), $\vec{Q}$ the 
momentum transfer at the chosen given reflection and $\vec{R}_n$ the 
atomic position inside the cell.

We follow the numbering of V atoms as given in Fig.  \ref{boh}, and divide 
the eight atoms in the monoclinic unit cell into two groups with opposite 
orientation of the magnetic moment, $4 \; (u, v, w), \; 5 \; (-u, -v, -w), 
\; 2 \; (1/2+u, -v, w), \; 1 \; (1/2-u, v, -w)$, 
and 8, 6, 3, 7, with coordinates obtained by adding the 
vector (1/2, 1/2, 1/2) to the first group.
Here $u = 0.3438, \; v = 0.0008, \; w = 0.2991$ are the fractional 
coordinates of the atoms in unit of the monoclinic axis\cite{dernier70} and the origin is taken between the atoms V$_4$ and V$_5$ in the same basal plane.
It is important to note for the following discussion that the bc monoclinic 
unit cell (space group $I2/a$) is not primitive, so
that, neglecting the magnetic moments, the two groups of atoms with their 
oxygen environment are translationally equivalent.

At the $(h,k,l)$ monoclinic reflection, putting for brevity $b = e^{i\pi (h+k+l)}$, and 
$\phi^{\pm} = 2\pi (hu \pm hv + lw)$,   
the structure factor F in Eq. (\ref{csf}) can be written as

\begin{eqnarray} 
\label{csff}
F & = & {\bigg [} (f_8 + bf_4)e^{i\phi^+} + (f_6 + bf_5)e^{-i\phi^+} {\bigg ]} \\
  & + & b (-1)^h {\bigg [} (f_2 + bf_3)e^{i\phi^-} + (f_1 + bf_7)e^{-i\phi^-} {\bigg ]} \nonumber
\end{eqnarray}

The RASF $f_n$'s in this equation are tensorial quantities given by the 
following expression\cite{blume94}:

\begin{equation} 
\label{fi}
f_n(\omega) = \sum_m \frac{<1s|O^{\dagger}|\Psi_m> 
<\Psi_m|O|1s>}
{\hbar \omega + E_{1s} - E_m - i \Gamma_n/2}  
\end{equation}

The transition operator $O$ is the electronic part of  
$\vec{\epsilon} \cdot \vec{r} (1 + i \vec{k} \cdot \vec{r}) $ with the
usual notation for the photon energy $\omega$, the polarization 
$\vec{\epsilon}$ and
the propagation vector $\vec{k}$, so that $f_n(\omega)$ may bear two, 
three or four Cartesian 
indices according to dipole-dipole (DD), dipole-quadrupole (DQ) and 
quadrupole-quadrupole (QQ) transitions. Then the atomic scattering amplitude is 
obtained by saturating with the appropriate $\vec{\epsilon}$ and 
$\vec{k}$ components.

We can now use the operations of the magnetic symmetry group of the 
crystal to transform the $f_n$'s into one another. In doing so we 
remember that a symmetry operation changes the site index according to
Table I of Section IV and simultaneously acts on the tensor components. 
This action reduces to a multiplicative factor $\pm 1$ 
since $\hat I ^2 = \hat C_2^2 = \hat \sigma_b^2 = \hat T^2 = E$. Thus  
$\hat I f_n = f_n$ for
DD, QQ  and $\hat I f_n = - f_n$ for DQ transitions. Moreover, we find 
it expedient to split the RASF tensor into a 
symmetric (charge) part and an antisymmetric (magnetic) part\cite{blume94}:
$f_n = f^{+}_n + f^{-}_n$. In this way, $\hat T f_n^{\pm} = 
\pm f_n^{\pm}$. Notice that in establishing the site correspondence and 
the transformation properties of the tensor components it is essential 
that the Wannier functions of adjacent hexagonal planes be defined as 
in CNR\cite{cnr2} (Eqs. (2.4) and (2.7)), i.e., they should 
be related by the same symmetry operation connecting the sites. For example,
since the sites 2 and 4 are related by the glide plane $\hat \sigma_b$, 
the definition of the Wannier function at site 2 should be obtained from 
that at site 4 by changing $y$ into $-y$, if we take the reference frame 
as depicted in Fig. \ref{boh}.

Consider now the magnetic group $C_{2h} \otimes \hat T$, which is 
appropriate to the ferro-orbital AFI phase of V$_2$O$_3$ as proposed 
by  Mila {\it et al.}\cite{mila00} Defining 
$A(\phi^{\pm}) = (e^{i\phi^{\pm}} + \hat I e^{-i\phi^{\pm}})$ and using
the symmetry operation $\hat I$ to connect sites 5 and 6 to 4 and 8, and sites 1 
and 7 to 2 and 3, and the operation $\hat T \hat \sigma_b$ to connect 
sites 2 and 3 to 4 and 8, we can express the structure factor as:

\begin{eqnarray}
 F_{hkl} & = & [A(\phi^{+})+ A(\phi^{-}) (-1)^h \hat T \hat \sigma_b] 
               (f_8^+ + bf_4^+)  
\nonumber \\
         & + & [A(\phi^{+})+ A(\phi^{-}) (-1)^h \hat T \hat \sigma_b] 
(f_8^- + bf_4^-)
\label{fc1}
\end{eqnarray}

Using now the peculiar fact that in V$_2$O$_3$ $v \approx 0$, 
defining $\phi= 2\pi (hu + lw)$ and $A(\phi^{\pm}) \equiv A(\phi)$ 
and remembering that $\hat T f_n^{\pm} = \pm f_n^{\pm}$, we finally 
obtain:

\begin{eqnarray}
 F_{hkl} & = &(e^{i\phi} + \hat I e^{-i\phi}) [1 + (-1)^h \hat \sigma_b] 
(f_8^+ + bf_4^+)  \nonumber \\
         & + &(e^{i\phi} + \hat I e^{-i\phi}) [1 - (-1)^h \hat \sigma_b] 
(f_8^- + bf_4^-)
\label{fc2}
\end{eqnarray}

In this expression we assume that the operators have already acted on the
site indices, so that they should act only on the tensor components.
We notice that, considering the $f_n$ as scalars, we recover 
the crystallographic extinction rules (the reflections 
with $h+k+l$ odd ($b=-1$) are lattice forbidden) and the selection rules 
for non resonant magnetic scattering (for $b=-1$, those with $h$ even are 
magnetically allowed, those with $h$ odd are forbidden).
The classification by Paolasini {\it et al.}\cite{paolasini99} of the 
lattice forbidden reflections into purely orbital and purely magnetic 
was actually based on the non resonant magnetic scattering and might 
therefore be only approximate.

Returning to the resonant regime at lattice forbidden reflections with 
$h$ odd, one can easily see from Eq. (\ref{fc2}) 
that there cannot be any charge (and therefore orbital) signal if 
sites 8 and 4 are in the same orbital state, since they are completely 
equivalent from the point of view of the lattice. Therefore what Mila 
{\it et al.}\cite{mila00} are really saying is that the 
(1,1,1) reflection observed by Paolasini {\it et al.} can only be of 
purely magnetic origin, which however has nothing to do with orbital
ordering.

A similar argument is also valid for the $C_2 \otimes {\hat T}$ symmetry 
of the RS(AO) phase. In this case, using  ${\hat T}$ and ${\hat C_2}$ to 
relate the various sites,  as shown in Table I, we obtain:

\begin{eqnarray}
 F_{hkl} & = &(e^{i\phi} + b(-1)^h {\hat C}_2 e^{-i\phi}) (b + 1) 
(f_3^+ + b(-1)^h f_4^+)  \nonumber \\
         & + &(e^{i\phi} - b(-1)^h {\hat C}_2 e^{-i\phi}) (b - 1) 
(f_3^- + b(-1)^h f_4^-) \nonumber 
\end{eqnarray}

\vspace{-0.5cm}

\begin{equation}
\label{ftc2}
\end{equation}

\noindent and again it is impossible to observe a charge scattering when $b=-1$.

We believe however that the claim made by Paolasini {\it et al.}  for 
evidence of orbital ordering has some substance, despite their approximate 
classification. Indeed it is well known that the magnetic part of 
the tensor for DD transition is only active in the $\sigma-\pi$ channel 
and its imaginary part can be shown to be proportional to the magnetic 
circular dichroism in absorption. At the K-edge this latter is proportional
to $\langle L\rangle$, the average of the orbital momentum operator in the 
final state.
\cite{thole,natoli} Due to the fact that in V$_2$O$_3$ the moment lies
\cite{moon70} on the glide plane, a reflection with respect to this 
plane change the sign of $\langle L\rangle$, so that 
$\hat \sigma f_n^-(DD)=-f_n^-(DD)$.
This implies, from Eq. (\ref{fc2}), that the DD tensor is magnetically 
active only at 
reflections with $h$ even, but cannot contribute at $h$ odd. 
The experimental evidence from Paolasini {\it et al.}
\cite{paolasini99,paolasini00} is in keeping with this conclusion 
except for the observed intensity at 5464 eV, corresponding to 
$1s \rightarrow 3d$ transitions, where at both types of reflections 
($h$ even and $h$ odd) signals of comparable magnitude were detected. 

Notice that we expect a strong contribution from the DD transitions at 
this energy for $h$ even, since the V-atoms are off-center 
with respect to the oxygen octahedra, 
which are furthermore distorted. At the same time, 
the ratio of the transition matrix elements between DD and DQ transitions 
in the case of V$_2$O$_3$ is estimated by Fabrizio {\it et al.}
\cite{fabrizio98} to be of the order of 7, in good agreement with the 
experimental data by Goulon {\it et al.}\cite{goulon00}. Therefore the 
observed comparable intensities at 5464 eV for reflections with $h$ even 
and odd cannot have the same magnetic origin. This argument also invalidates 
the explanation proposed by Lovesey {\it et al.}\cite{lovesey00} that the 
(1,1,1) signal is due to the octupolar component of the QQ transition, 
a tensor of rank 3, apart from problems related to its use of an approximate 
symmetry. At this reflection the signal should indeed be much smaller than 
at an allowed DD magnetic reflection ($h$ even). Therefore much more work is 
needed to establish the exact implications of the resonant diffraction 
experiment on the nature of the AFI ground state of V$_2$O$_3$, since this 
complex question cannot be dismissed by too simple arguments.

If we now consider the reduced group $C_{2h}(C_s)$, 
using $\hat T \hat I$ and $\hat \sigma_b$ to relate the various sites, 
again according to Table I, we find for the scattering factor : 
 
\begin{eqnarray}
F_{hkl} & = &(e^{i\phi} + b \hat T \hat I e^{-i\phi}) 
             [1 + b (-1)^h \hat \sigma_b] (f_8 + bf_4)  \nonumber \\
        & = &(e^{i\phi} + b \hat I e^{-i\phi}) [1 + b (-1)^h \hat \sigma_b] 
             (f_8^+ + bf_4^+)  \nonumber \\
        & + &(e^{i\phi} - b \hat I e^{-i\phi}) [1 + b (-1)^h \hat \sigma_b] 
             (f_8^- + bf_4^-)
\label{fc3}
\end{eqnarray}

At reflections forbidden by the lattice ($b = -1$), the charge scattering 
behavior of expressions (\ref{fc2}), (\ref{ftc2})  and (\ref{fc3}) is substantially 
different. As already emphasized in the case of the two groups $C_{2h} \otimes \hat T$ 
and $C_{2} \otimes \hat T$ there cannot be charge scattering,
while for $C_{2h}(C_s)$, $f_8^+$ is different from  $f_4^+$, 
so that a charge signal might appear. 
In particular this is the case for 
the orbitally ordered state RS(ME), due to the different 
molecular states of the molecules straddling atoms 8-7 and 4-1. This fact 
would readily explain the comparable intensity observed at reflections 
with $h$ even and odd, since the transition is always dipolar in nature, 
and stronger than that due to DQ or QQ processes. Again more work is 
needed to establish whether the predictions of the ground state proposed 
here for V$_2$O$_3$ are in keeping with the experimental findings.

Finally we want to point out that the RS(AO) configuration, due to the orbital ordering, can give rise to 
magnetic reflections of dipolar character, with $b = -1$ 
and $h$ even or odd, as apparent from Eq. (\ref{ftc2}), since atoms 3 and 4
belong to vertical pairs in different molecular states. However this phase 
cannot be an eligible candidate for being the ground state of V$_2$O$_3$, 
due to the presence of $\hat T$ in the associated space magnetic group, that forbids magnetoelectricity.

\subsection{Conclusions.}

In this paper we have derived 
the strong coupling limit of an Hubbard Hamiltonian with three degenerate 
$t_{2g}$ states containing two electrons coupled to spin $S = 1$, and have
re-examined the low-temperature ground-state properties of V$_2$O$_3$. 
Given the present experimental evidence in favor of a spin-1 state of 
the V$^{3+}$-ions, there is no doubt that the original suggestion  by 
J. Allen, \cite{allen76} later taken up and developed by 
Mila {\it et al.}\cite{mila00}, 
that the vertical molecules constitute the building blocks of this compound 
in all three phases of the phase diagram, is the key concept to understand 
the  physics of the phase transitions observed in V$_2$O$_3$. In fact the 
molecular unit reconciles the existence of orbital degeneracy, required to 
explain experimental evidence ranging from inelastic neutron 
scattering data to the relaxation time of the nuclear spins in NMR studies, 
with the spin-1 state of the V$^{3+}$-ions.
Focusing on the AFI phase and analyzing the parameter space of the problem, 
we have indeed found that one has to consider two regimes, depending on the 
relative size of the intramolecular correlation energy and the in-plane 
exchange energy. These two regimes dictate the form of the variational wave 
function. In both cases we found minimizing solutions in a reduced region 
of the Hamiltonian parameters (hopping integrals versus $\frac{J}{U_2}$) 
with the real spin configuration. The molecular solutions are more suited 
to explain certain experimental facts, like, for example, the lack of 
detectable anisotropy in various spectroscopies in the paramagnetic phases 
of V$_2$O$_3$. However there remain two orders of problems which will 
require future investigation. 

The first one is the rather small stability 
region in the phase space of the parameters 
$\frac{\alpha}{\tau}$ versus $\frac{J}{U_2}$ and the small stabilization 
energy of the RS phases compared to neighboring competing configurations 
(in the range $1.5 \div 2$ meV, using the standard set of parameters). 
The AF(AO) phase, with its continuum degeneracy and the in-plane spin-orbital
antiferromagnetic coupling with trigonal symmetry, is very reminiscent of 
the paramagnetic insulating (PI) phase at higher temperature, setting aside
the short range of the correlations.\cite{bao98} 
Therefore one would expect an energy gap at least of the order of the 
transition temperature (15 meV), and we know\cite{bao98}  that the estimated Neel 
temperature of the AFI phase is about 20 meV.
The problem shows up again if one tries to calculate 
the exchange integrals for spin wave excitations. Even though, looking 
at Fig. \ref{mol}(b), one can find a value of Hund parameter $J$ such 
that the exchange integrals along $\delta_1$ is ferromagnetic and about 
half the one antiferromagnetic along $\delta_2$ or $\delta_3$, as found by inelastic neutron scattering measurements,\cite{word81}
their actual values are off by a factor of five.
Moreover it does not seem that longer range hopping parameters, neglected
in our calculations, might cure this drawback.
Since the energy scale is set by $\frac{\tau^2}{U_2}$, in order to get 
the energetics correct
one should increase $\tau$ by more than a factor of two. However this 
would be an easy way to escape a problem that has more profound roots.
We tend to believe in fact that the difficulty 
stems from the mean field treatment of this highly correlated electron 
system on one side and on the other side from the neglect of the lattice 
degrees of freedom. 

The second order of problems regards the consistency of the RS phases 
with the experimental findings of x-ray synchrotron radiation spectroscopies,
namely the anomalous diffraction and non-reciprocal x-ray gyrotropy.
We have argued that the comparable intensity of the observed signal
at the Vanadium K-edge in resonant diffraction at both reflections 
with $h$ even and odd cannot be interpreted as being purely magnetic 
in origin, but finds a natural explanation if one assume that the
ground state magnetic symmetry group is such as to allow charge scattering,
i.e., if there is the appropriate orbital ordering. This group is the same
as the one indicated by the non-reciprocal spectroscopy. However none of 
the stable RS phases found admit this group, so that they only allow resonant magnetic 
scattering. This is the reason that prompted us to indicate the excited 
configurations RS(ME) as possible candidates for the correct ground state 
of the AFI phase, invoking an optimal coupling with the lattice degrees of
freedom. This assignment makes also a definite prediction for a transverse 
magneto-electric effect in this phase which could be subjected to 
experimental test.

\acknowledgements

We thank P. Carra for stimulating criticism, M. Cuoco for a precious suggestion, V.Yushankhai for useful comments and especially G. Jackeli for many valuable discussions and a careful final reading of the manuscript. \\
S.D.M. acknowledges the support by the grant from University of Salerno, D.R. n. 578/2000 .


\end{multicols}
\begin{widetext}

\appendix

\section{The effective Hamiltonian in terms of fermion operators.}

Here we report the results of the calculations performed to derive the effective Hamiltonian 
in terms of the fermion operators. 

The explicit form of the projection operators $X^{(i)}_j$, ($i=1,2,3$) defined in Eqs. (\ref{h6}), (\ref{h7}) is given by:

\begin{eqnarray}
X^{(1)} = \sum_{\sigma} \bigg[n_{1\sigma}n_{2\sigma}n_{3\sigma}+
\frac{1}{3}(n_{1\sigma}n_{2\sigma}n_{3{\bar \sigma}}+n_{1\sigma}n_{2{\bar \sigma}}
n_{3 \sigma}+n_{1{\bar \sigma}}n_{2\sigma}n_{3\sigma})+
\frac{1}{3}\sum_{{m,m',m''}\atop{m\neq m'\neq m''}}
n_{m\sigma}c_{m'\sigma}^{+}c_{m'{\bar \sigma}}
c^{+}_{m''{\bar \sigma}}c_{m''\sigma}\bigg]  
\label{h8}
\end{eqnarray}
\begin{eqnarray}
X^{(2)} = \sum_{\sigma}\bigg[ \frac{1}{2}  \sum_{m,m'\neq m}
n_{m\sigma}n_{m'\sigma}n_{m'{\bar\sigma}}
- \sum_{{m,m',m''}\atop{m\neq m'\neq m''}}
n_{m\sigma}c_{m'\sigma}^{+}c_{m'{\bar\sigma}}^{+}
c_{m''{\bar\sigma}}c_{m''\sigma} \nonumber 
\end{eqnarray}
\begin{eqnarray}
 +\frac{2}{3}(n_{1\sigma}n_{2\sigma}n_{3{\bar \sigma}}+n_{1\sigma}n_{2{\bar \sigma}}n_{3 \sigma}+
n_{1{\bar \sigma}}n_{2\sigma}n_{3\sigma})-\frac{1}{3} \left (n_{1\sigma}(c_{2\sigma}^{+}c_{2{\bar\sigma}}
c_{3{\bar\sigma}}^{+}c_{3\sigma}+c_{3\sigma}^{+}c_{3{\bar\sigma}}
c_{2{\bar\sigma}}^{+}c_{2\sigma}) \right.  
\label{h81}
\end{eqnarray}
\begin{eqnarray}
+n_{2\sigma}(c_{1\sigma}^{+}c_{1{\bar\sigma}}
\left. c_{3{\bar\sigma}}^{+}c_{3\sigma}+c_{3\sigma}^{+}c_{3{\bar\sigma}}
c_{1{\bar\sigma}}^{+}c_{1\sigma})+n_{3\sigma}(c_{1\sigma}^{+}c_{1{\bar\sigma}}
c_{2{\bar\sigma}}^{+}c_{2\sigma}+c_{2\sigma}^{+}c_{2{\bar\sigma}}
c_{1{\bar\sigma}}^{+}c_{1\sigma})\right )\bigg] \nonumber 
\end{eqnarray}
\begin{eqnarray}
X^{(3)} = \frac{1}{2}\sum_{\sigma}\sum_{{m,m'\neq m}\atop{m''\neq m}}
n_{m\sigma}c_{m'\sigma}^{+}c_{m'{\bar\sigma}}^{+}c_{m''{\bar\sigma}}c_{m''\sigma} 
\label{h82}
\end{eqnarray}

For simplicity we omitted the site index  as all
operators act on site $j$.

Dividing $H_{\rm eff}$ into three parts, according to the $X^{(i)}$ 
definitions, we obtain the following expressions:

\begin{eqnarray}
H_{\rm eff}^{(1)}=-\frac{1}{U_2-J}\sum_{ij\sigma}\sum_{nn'}\sum_{mm'm''}
(1-\delta_{mm'})(1-\delta_{mm''})(1-\delta_{m'm''})c^{+}_{in\sigma}
c_{in'\sigma} \nonumber 
\end{eqnarray}
\begin{eqnarray}
\times {\bigg [}t_{ij}^{nm} t_{ji}^{mn'}n_{jm'\sigma}n_{jm''\sigma}-
t_{ij}^{nm}t_{ji}^{m'n'}c_{jm'\sigma}^{+}c_{jm\sigma}n_{jm''\sigma}{\bigg ]} \nonumber 
\end{eqnarray}
\begin{eqnarray}
-\frac{1}{3(U_2-J)}\sum_{ij\sigma}\sum_{nn'}\sum_{mm'm''}    
(1-\delta_{mm'})(1-\delta_{mm''})(1-\delta_{m'm''}) \nonumber
\end{eqnarray}
\begin{eqnarray}
\times {\bigg [}t_{ij}^{nm} t_{ji}^{mn'}
(c_{in\sigma}^{+}c_{in'\sigma}n_{jm'\sigma}n_{jm''{\bar \sigma}}+
\frac{1}{2}c_{in{\bar \sigma}}^{+}c_{in'{\bar \sigma}}
n_{jm'\sigma}n_{jm'' \sigma}) \nonumber 
\end{eqnarray}
\begin{eqnarray}
-t_{ij}^{nm} t_{ji}^{m'n'}
(c_{in\sigma}^{+}c_{in'\sigma}c_{jm'\sigma}^{+}c_{jm\sigma}
n_{jm''{\bar \sigma}}+
c_{in\sigma}^{+}c_{in'{\bar \sigma}}c_{jm'{\bar \sigma}}^{+}c_{jm\sigma}
n_{jm'' \sigma}+
c_{in{\bar \sigma}}^{+}c_{in'\sigma}c_{jm'\sigma}^{+}c_{jm{\bar \sigma}}
n_{jm'' \sigma}){\bigg ]} \nonumber 
\end{eqnarray}
\begin{eqnarray}
-\frac{1}{3(U_2-J)}\sum_{ij\sigma}\sum_{nn'}\sum_{mm'm''}    
(1-\delta_{mm'})(1-\delta_{mm''})(1-\delta_{m'm''}) \nonumber
\end{eqnarray}
\begin{eqnarray}
\times {\bigg [}t_{ij}^{nm} t_{ji}^{mn'}
{\big (}c_{in\sigma}^{+}c_{in'\sigma}c_{jm'\sigma}^{+}c_{jm'{\bar \sigma}}
c_{jm''{\bar \sigma}}^{+}c_{jm''\sigma}
+c_{in\sigma}^{+}c_{in'{\bar \sigma}}
n_{jm'\sigma}c_{jm'' {\bar \sigma}}^{+}c_{jm''\sigma}+
c_{in{\bar \sigma}}^{+}c_{in'\sigma}c_{jm'\sigma}^{+}c_{jm'{\bar \sigma}}
n_{jm''\sigma}{\big )} \nonumber 
\end{eqnarray}
\begin{eqnarray}
-t_{ij}^{nm} t_{ji}^{m'n'}
{\big (}c_{in\sigma}^{+}c_{in'\sigma}c_{jm'{\bar\sigma}}^{+}
c_{jm \sigma}
c_{jm'' \sigma}^{+}c_{jm''{\bar \sigma}}
+c_{in\sigma}^{+}c_{in' \sigma}
c_{jm' {\bar \sigma}}^{+}c_{jm{\bar \sigma}}n_{jm''\sigma}
+c_{in\sigma}^{+}c_{in'\sigma}c_{jm'\sigma}^{+}c_{jm{\bar \sigma}}
c_{jm''{\bar \sigma}}c_{jm''\sigma} \nonumber 
\end{eqnarray}
\begin{eqnarray}
+c_{in\sigma}^{+}c_{in'{\bar \sigma}}c_{jm'\sigma}^{+}c_{jm \sigma}
c_{jm''{\bar \sigma}}^{+}c_{jm''\sigma}
+c_{in{\bar \sigma}}^{+}c_{in'\sigma}c_{jm'\sigma}^{+}c_{jm{ \sigma}}
c_{jm''\sigma}^{+}c_{jm''{\bar\sigma}}+
c_{in{\bar \sigma}}^{+}c_{in'{\bar \sigma}}c_{jm'\sigma}^{+}c_{jm \sigma}
n_{jm''\sigma}
{\big )}
{\bigg ]}
\label{h9}
\end{eqnarray}

\begin{eqnarray}
H_{\rm eff}^{(2)}=-\frac{1}{2(U_2+2J)}
\sum_{ij\sigma}\sum_{nn'}\sum_{mm'}
(1-\delta_{mm'})
{\bigg [}t_{ij}^{nm} t_{ji}^{mn'}
(c^{+}_{in\sigma}c_{in'\sigma}n_{jm{\bar\sigma}}n_{jm'\sigma}+
c^{+}_{in{\bar \sigma}}c_{in'{\bar\sigma}}n_{jm\sigma}n_{jm'\sigma} \nonumber 
\end{eqnarray}
\begin{eqnarray}
-c^{+}_{in\sigma}c_{in'{\bar\sigma}}c^{+}_{jm{\bar\sigma}}c_{jm\sigma}
n_{jm'\sigma}
-c^{+}_{in{\bar\sigma}}c_{in'\sigma}c^{+}_{jm\sigma}c_{jm{\bar\sigma}}
n_{jm'\sigma}
{\bigg ]}+
\frac{1}{2(U_2+2J)}\sum_{ij\sigma}\sum_{nn'}\sum_{mm'm''}    
(1-\delta_{mm'})(1-\delta_{mm''})(1-\delta_{m'm''}) \nonumber 
\end{eqnarray}
\begin{eqnarray}
\times {\bigg [}t_{ij}^{nm} t_{ji}^{m'n'}
(c^{+}_{in\sigma}c_{in'\sigma}c^{+}_{jm{\bar\sigma}}c_{jm'{\bar\sigma}}
n_{jm''\sigma}
-c^{+}_{in\sigma}c_{in'{\bar\sigma}}c^{+}_{jm{\bar\sigma}}c_{jm'\sigma}
n_{jm''\sigma} \nonumber 
\end{eqnarray}
\begin{eqnarray}
-c^{+}_{in{\bar\sigma}}c_{in'\sigma}c^{+}_{jm\sigma}c_{jm'{\bar\sigma}}
n_{jm''\sigma}
+c^{+}_{in{\bar\sigma}}c_{in'{\bar\sigma}}c^{+}_{jm\sigma}c_{jm'\sigma}
n_{jm''\sigma})
{\bigg ]} \nonumber 
\end{eqnarray}
\begin{eqnarray}
-\frac{2}{3(U_2+2J)}\sum_{ij\sigma}\sum_{nn'}\sum_{mm'\atop{m''}}    
(1-\delta_{mm'})(1-\delta_{mm''})(1-\delta_{m'm''})
{\bigg [}t_{ij}^{nm} t_{ji}^{mn'}
(c_{in\sigma}^{+}c_{in'\sigma}n_{jm'\sigma}n_{jm''{\bar \sigma}}+\frac{1}{2}c_{in{\bar \sigma}}^{+}c_{in'{\bar \sigma}}
n_{jm'\sigma}n_{jm'' \sigma}) \nonumber 
\end{eqnarray}
\begin{eqnarray}
-t_{ij}^{nm} t_{ji}^{m'n'}
(c_{in\sigma}^{+}c_{in'\sigma}c_{jm'\sigma}^{+}c_{jm\sigma}
n_{jm''{\bar \sigma}}+
c_{in\sigma}^{+}c_{in'{\bar \sigma}}c_{jm'{\bar \sigma}}^{+}c_{jm\sigma}
n_{jm'' \sigma}+
c_{in{\bar \sigma}}^{+}c_{in'\sigma}c_{jm'\sigma}^{+}c_{jm{\bar \sigma}}
n_{jm'' \sigma}){\bigg ]} \nonumber 
\end{eqnarray}
\begin{eqnarray}
+\frac{1}{3(U_2+2J)}\sum_{ij\sigma}\sum_{nn'}\sum_{mm'm''}    
(1-\delta_{mm'})(1-\delta_{mm''})(1-\delta_{m'm''})
{\bigg [}t_{ij}^{nm} t_{ji}^{mn'}
{\big (}c_{in\sigma}^{+}c_{in'\sigma}c_{jm'\sigma}^{+}c_{jm'{\bar \sigma}}
c_{jm''{\bar \sigma}}^{+}c_{jm''\sigma} \nonumber 
\end{eqnarray}
\begin{eqnarray}
+c_{in\sigma}^{+}c_{in'{\bar \sigma}}
n_{jm'\sigma}c_{jm'' {\bar \sigma}}^{+}c_{jm''\sigma}+
c_{in{\bar \sigma}}^{+}c_{in'\sigma}c_{jm'\sigma}^{+}c_{jm'{\bar \sigma}}
n_{jm''\sigma}{\big )}-
t_{ij}^{nm} t_{ji}^{m'n'}
{\big (}c_{in\sigma}^{+}c_{in'\sigma}c_{jm'{\bar\sigma}}^{+}
c_{jm \sigma}
c_{jm'' \sigma}^{+}c_{jm''{\bar \sigma}} \nonumber 
\end{eqnarray}
\begin{eqnarray}
+c_{in\sigma}^{+}c_{in' \sigma}
c_{jm' {\bar \sigma}}^{+}c_{jm{\bar \sigma}}n_{jm''\sigma}+
c_{in\sigma}^{+}c_{in'\sigma}c_{jm'\sigma}^{+}c_{jm{\bar \sigma}}
c_{jm''{\bar \sigma}}c_{jm''\sigma}+
c_{in\sigma}^{+}c_{in'{\bar \sigma}}c_{jm'\sigma}^{+}c_{jm \sigma}
c_{jm''{\bar \sigma}}c_{jm''\sigma} \nonumber 
\end{eqnarray}
\begin{eqnarray}
+c_{in{\bar \sigma}}^{+}c_{in'\sigma}c_{jm'\sigma}^{+}c_{jm{\bar \sigma}}
c_{jm''\sigma}c_{jm''{\bar\sigma}}+
c_{in\sigma}^{+}c_{in'{\bar \sigma}}c_{jm'\sigma}^{+}c_{jm \sigma}
n_{jm''\sigma}
{\big )}
{\bigg ]}
\label{h11}
\end{eqnarray}

\begin{eqnarray}
H_{\rm eff}^{(3)}=-\frac{1}{2(U_2+4J)}
\sum_{ij\sigma}\sum_{nn'}\sum_{mm'}
(1-\delta_{mm'})
{\bigg [}t_{ij}^{nm} t_{ji}^{mn'}
(c^{+}_{in\sigma}c_{in'\sigma}n_{jm{\bar\sigma}}n_{jm'\sigma}+
c^{+}_{in{\bar \sigma}}c_{in'{\bar\sigma}}n_{jm\sigma}n_{jm'\sigma} \nonumber 
\end{eqnarray}
\begin{eqnarray}
-c^{+}_{in\sigma}c_{in'{\bar\sigma}}c^{+}_{jm{\bar\sigma}}c_{jm\sigma}n_{jm'\sigma}
-c^{+}_{in{\bar\sigma}}c_{in'\sigma}c^{+}_{jm\sigma}c_{jm{\bar\sigma}}
n_{jm'\sigma}
{\bigg ]}-
\frac{1}{2(U_2+4J)}\sum_{ij\sigma}\sum_{nn'}\sum_{mm'm''}    
(1-\delta_{mm'})(1-\delta_{mm''})(1-\delta_{m'm''}) \nonumber 
\end{eqnarray}
\begin{eqnarray}
\times {\bigg [}t_{ij}^{nm} t_{ji}^{m'n'}
(c^{+}_{in\sigma}c_{in'\sigma}c^{+}_{jm{\bar\sigma}}c_{jm'{\bar\sigma}}
n_{jm''\sigma}
-c^{+}_{in\sigma}c_{in'{\bar\sigma}}c^{+}_{jm{\bar\sigma}}c_{jm'\sigma}
n_{jm''\sigma} \nonumber
\end{eqnarray}
\begin{eqnarray}
-c^{+}_{in{\bar\sigma}}c_{in'\sigma}c^{+}_{jm\sigma}c_{jm'{\bar\sigma}}
n_{jm''\sigma}
+c^{+}_{in{\bar\sigma}}c_{in'{\bar\sigma}}c^{+}_{jm\sigma}c_{jm'\sigma}
n_{jm''\sigma})
{\bigg ]}
\label{h11bis}
\end{eqnarray}

\section{Commutation relations.}

In the following we evaluate the commutators used to derive the effective Hamiltonian in section II.

As usual, given two generic operators A and B, we define $[A,B] \equiv AB-BA$ and  $\{ A,B \} \equiv AB+BA$

\begin{eqnarray}
\begin{array}{l}
[n_{m\sigma}n_{m'\sigma'}
n_{m''\sigma''}, c^{+}_{m'''\sigma'''} ] =
\delta_{mm'''}\delta_{\sigma\sigma'''}c^{+}_{m\sigma}n_{m'\sigma'}n_{m''\sigma''}\\
+\delta_{m'm'''}\delta_{\sigma'\sigma'''}n_{m\sigma}c^{+}_{m'\sigma'}
n_{m''\sigma''}+
\delta_{m''m'''}\delta_{\sigma''\sigma'''}n_{m\sigma}n_{m'\sigma'}
c^{+}_{m''\sigma''}
\end{array}
\label{c2}
\end{eqnarray}

\begin{eqnarray}
\begin{array}{l}
\{ c_{m''''\sigma''''}, [n_{m\sigma}n_{m'\sigma'}
n_{m''\sigma''}, c^{+}_{m'''\sigma'''} ] \}=
\delta_{mm''''}\delta_{\sigma\sigma''''}{\big (}
\delta_{mm'''}\delta_{\sigma\sigma'''}n_{m'\sigma'}n_{m''\sigma''}\\
+\delta_{m'm'''}\delta_{\sigma'\sigma'''}c_{m\sigma}c^{+}_{m'\sigma'}
n_{m''\sigma''}
+\delta_{m''m'''}\delta_{\sigma''\sigma'''}c_{m\sigma}n_{m'\sigma'}
c^{+}_{m''\sigma''}{\big )}\\
+\delta_{m'm''''}\delta_{\sigma'\sigma''''}{\big (}
\delta_{m''m'''}\delta_{\sigma''\sigma'''}
n_{m\sigma}c_{m'\sigma'}c^{+}_{m''\sigma''}
+\delta_{m'm'''}\delta_{\sigma'\sigma'''}
n_{m\sigma}n_{m''\sigma''}\\
-\delta_{mm'''}\delta_{\sigma\sigma'''}
c^{+}_{m\sigma}c_{m'\sigma'}n_{m''\sigma''}
{\big )}+
\delta_{m''m''''}\delta_{\sigma''\sigma''''}{\big (}
\delta_{m''m'''}\delta_{\sigma''\sigma'''}
n_{m\sigma}n_{m'\sigma'}\\
-\delta_{m'm'''}\delta_{\sigma'\sigma'''}
n_{m\sigma}c^{+}_{m'\sigma'}c_{m''\sigma''}
-\delta_{mm'''}\delta_{\sigma\sigma'''}
c^{+}_{m\sigma}n_{m'\sigma'}c_{m''\sigma''}
{\big )}
\label{c1}
\end{array}
\end{eqnarray}

\begin{eqnarray}
\begin{array}{l}
[n_{m''\sigma''}c^{+}_{m\sigma}c_{m'\sigma}c^{+}_{m'\bar{\sigma}} 
c_{m\bar{\sigma}}, c^{+}_{m'''\sigma'''}] =
\delta_{m''m'''}\delta_{\sigma''\sigma'''}c^{+}_{m''\sigma''}
c^{+}_{m\sigma}
c_{m'\sigma}c^{+}_{m'\bar{\sigma}}c_{m\bar{\sigma}}\\
+n_{m''\sigma''}c^{+}_{m\sigma}c^{+}_{m'\sigma'}
(\delta_{m'm'''}\delta_{\sigma\sigma'''}
c_{m\bar{\sigma}}-
\delta_{mm'''}\delta_{\bar{\sigma}\sigma'''}
c_{m'\sigma})
\end{array}
\label{c3}
\end{eqnarray}

\begin{eqnarray}
\begin{array}{l}
\{ c_{m''''\sigma''''}, 
[n_{m''\sigma''}c^{+}_{m\sigma}c^{+}_{m\bar{\sigma}}
c_{m'\bar{\sigma}}c_{m'\sigma}, c^{+}_{m'''\sigma'''} ] \}=
\delta_{m''m''''}\delta_{\sigma''\sigma''''}{\big (}
\delta_{m''m'''}\delta_{\sigma''\sigma'''}c^{+}_{m\sigma}
c^{+}_{m\bar{\sigma}}
c_{m'\bar{\sigma}}c_{m'\sigma}\\
+\delta_{m'm'''}c_{m''\sigma''}c^{+}_{m\sigma}
c^{+}_{m{\bar\sigma}}(\delta_{\sigma'''\sigma}c_{m'\bar{\sigma}}\\
-\delta_{\sigma'''\bar{\sigma}}c_{m'\sigma}){\big )}
+\delta_{mm''''}\delta_{\sigma\sigma''''}{\big (}
\delta_{m'm'''}n_{m''\sigma''}c^{+}_{m\bar{\sigma}}
(\delta_{\sigma\sigma'''}c_{m'\bar{\sigma}}-
\delta_{\bar{\sigma}\sigma'''}c_{m'\sigma})\\
-\delta_{m''m'''}\delta_{\sigma''\sigma'''}
c^{+}_{m''\sigma''}c^{+}_{m\bar{\sigma}}
c_{m'{\bar\sigma}}c_{m'\sigma}{\big )}
+\delta_{mm''''}\delta_{\bar{\sigma}\sigma''''}{\big (}
\delta_{m''m'''}\delta_{\sigma''\sigma'''}c^{+}_{m''\sigma''}
c^{+}_{m\sigma}c_{m'\bar{\sigma}}c_{m'\sigma}\\
-\delta_{m'm'''}n_{m''\sigma''}c^{+}_{m\sigma}(
\delta_{\sigma\sigma'''}c_{m'{\bar\sigma}}-
\delta_{\bar{\sigma}\sigma'''}c_{m'\sigma}){\big )} 
\end{array}
\label{c4}
\end{eqnarray}

\begin{eqnarray}
\begin{array}{l}
[n_{m''\sigma''}c^{+}_{m\sigma}c^{+}_{m\bar{\sigma}}
c_{m'\bar{\sigma}} 
c_{m'\sigma}, c^{+}_{m'''\sigma'''}] =
\delta_{m''m'''}\delta_{\sigma''\sigma'''}c^{+}_{m''\sigma''}
c^{+}_{m\sigma}
c^{+}_{m\bar{\sigma}}c_{m'{\bar\sigma}}c_{m'\sigma}\\
+n_{m''\sigma''}c^{+}_{m\sigma}c^{+}_{m\bar{\sigma}}
\delta_{m'm'''}(\delta_{\sigma\sigma'''}c_{m'\bar{\sigma}}-
\delta_{\bar{\sigma}\sigma'''}
c_{m'\sigma})
\end{array}
\label{c3a}
\end{eqnarray}

\begin{eqnarray}
\begin{array}{l}
\{ c_{m''''\sigma''''}, 
[n_{m''\sigma''}c^{+}_{m\sigma}c_{m'\sigma}
c^{+}_{m'\bar{\sigma}}c_{m\bar{\sigma}}, 
c^{+}_{m'''\sigma'''} ] \}=
\delta_{m''m''''}\delta_{\sigma''\sigma''''}{\big (}
\delta_{m''m'''}\delta_{\sigma''\sigma'''}c^{+}_{m\sigma}c_{m'\sigma}
c^{+}_{m'\bar{\sigma}}c_{m\bar{\sigma}}\\
+c_{m''\sigma''}c^{+}_{m\sigma}c^{+}_{m'{\bar\sigma}}
(\delta_{m'm'''}\delta_{\sigma\sigma'''}
c_{m\bar{\sigma}}-
\delta_{mm'''}\delta_{\sigma'''\bar{\sigma}}c_{m'\sigma}){\big )}+
\delta_{mm''''}\delta_{\sigma\sigma''''}{\big (}
n_{m''\sigma''}c^{+}_{m'\bar{\sigma}}\\
\times (\delta_{m'm'''}\delta_{\sigma\sigma'''}c_{m\bar{\sigma}}-
\delta_{mm'''}\delta_{{\bar\sigma}\sigma'''}c_{m'\sigma})-
\delta_{m''m'''}\delta_{\sigma''\sigma'''}c^{+}_{m''\sigma''}
c_{m'\sigma}c^{+}_{m'\bar{\sigma}}c_{m\bar{\sigma}}
{\big )}\\
-\delta_{m'm''''}\delta_{\bar{\sigma}\sigma''''}{\big (}
\delta_{m''m'''}\delta_{\sigma''\sigma'''}c^{+}_{m''\sigma''}
c_{m\sigma}^{+}c_{m'\sigma}c_{m\bar{\sigma}}+
n_{m''\sigma''}c^{+}_{m\sigma}(
\delta_{m'm'''}\delta_{\sigma\sigma'''}c_{m\bar{\sigma}}-
\delta_{mm'''}\delta_{\bar{\sigma}\sigma'''}c_{m'\sigma}){\big )} 
\end{array}
\label{c5}
\end{eqnarray}

\section{The effective spin-orbital Hamiltonian.}

As anticipated in section III, the spin orbital representation of the $H_{\rm eff}$ is 
quite cumbersome, so it was not reported in the main text. 
Nonetheless we think it could be useful to write down its complete form
 because, despite its apparent complications, it becomes very simple to
 handle for real materials, where, due to symmetry considerations, 
many of the terms could be zero.\\
Starting from the expression of Appendix A, we introduce the spin-pseudospin representation defined in section III and write down $H_{\rm eff}$ in the following form:

\begin{eqnarray}
H_{\rm eff}^{(1)}=-\frac{1}{3}\frac{1}{U_2-J}\sum_{ij}\sum_{nn'}
 c^{+}_{in}c_{in'}{\bigg [} \frac{1}{2}t_{ij}^{n1}t_{ji}^{1n'}
\tau_{jz}(1+\tau_{jz})+ \frac{1}{2}t_{ij}^{n2}t_{ji}^{2n'}
\tau_{jz}(\tau_{jz}-1)+t_{ij}^{n3}t_{ji}^{3n'}(1-\tau_{jz}^2)
-\frac{1}{2}t_{ij}^{n1}t_{ji}^{2n'}\tau_{j}^{+}\tau_{j}^{+}
\nonumber
\end{eqnarray}
\begin{eqnarray}
+\frac{1}{\sqrt{2}}t_{ij}^{n1}t_{ji}^{3n'}\tau_{zj}\tau_{j}^{+}
-\frac{1}{2}t_{ij}^{n2}t_{ji}^{1n'}\tau_{j}^-\tau_{j}^-
+\frac{1}{\sqrt{2}}t_{ij}^{n2}t_{ji}^{3n'}\tau_{zj}\tau_{j}^-
+\frac{1}{\sqrt{2}}t_{ij}^{n3}t_{ji}^{1n'}\tau_{j}^-\tau_{zj}
+\frac{1}{\sqrt{2}}t_{ij}^{n3}t_{ji}^{2n'}\tau_{j}^{+}\tau_{zj}
{\bigg ]}{\big [}
\vec S_i\cdot \vec S_j+2 {\big ]}
\label{h16}
\end{eqnarray}
\begin{eqnarray}
H_{\rm eff}^{(2)}=-\frac{1}{12}\frac{1}{U_2+2J}\sum_{ij}\sum_{nn'}
 c^{+}_{in}c_{in'}{\bigg [} \frac{1}{2}t_{ij}^{n1}t_{ji}^{1n'}
(\tau_{jz}^2+\tau_{jz}+6)+ \frac{1}{2}t_{ij}^{n2}t_{ji}^{2n'}
(\tau_{jz}^2-\tau_{jz}+6)+t_{ij}^{n3}t_{ji}^{3n'}(4-\tau_{jz}^2)
\nonumber
\end{eqnarray}
\begin{eqnarray}
+t_{ij}^{n1}t_{ji}^{2n'}\left(-\frac{3}{2}\tau_{j}^-\tau_{j}^-
+\tau_{j}^{+}\tau_{j}^{+}\right)
+t_{ij}^{n1}t_{ji}^{3n'}\left(\frac{3}{\sqrt{2}}\tau_{j}^-\tau_{jz}
-\sqrt{2}\tau_{jz}\tau_{j}^{+}\right)
+t_{ij}^{n2}t_{ji}^{1n'}\left(-\frac{3}{2}\tau_{j}^{+}\tau_{j}^{+}
+\tau_{j}^-\tau_{j}^-\right)
\label{h17}
\end{eqnarray}
\begin{eqnarray}
+t_{ij}^{n2}t_{ji}^{3n'}\left(\frac{3}{\sqrt{2}}\tau_{j}^{+}\tau_{jz}
-\sqrt{2}\tau_{jz}\tau_{j}^-\right)+
t_{ij}^{n3}t_{ji}^{1n'}\left(\frac{3}{\sqrt{2}}\tau_{jz}\tau_{j}^{+}
-\sqrt{2}\tau_{j}^-\tau_{zj}\right)+
t_{ij}^{n3}t_{ji}^{2n'}\left(\frac{3}{\sqrt{2}}\tau_{jz}\tau_{j}^-
-\sqrt{2}\tau_{j}^{+}\tau_{jz}\right){\bigg ]}{\big [}
1-\vec S_i\cdot \vec S_j {\big ]}
\nonumber
\end{eqnarray}
\begin{eqnarray}
H_{\rm eff}^{(3)}=-\frac{1}{4}\frac{1}{U_2+4J}\sum_{ij}\sum_{nn'}
 c^{+}_{in}c_{in'}{\bigg [} \frac{1}{2}t_{ij}^{n1}t_{ji}^{1n'}
(1-\tau_{jz})(2+\tau_{jz})+ \frac{1}{2}t_{ij}^{n2}t_{ji}^{2n'}
(2-\tau_{jz})(1+\tau_{jz})+t_{ij}^{n3}t_{ji}^{3n'}\tau_{jz}^2
\nonumber
\end{eqnarray}
\begin{eqnarray}
+\frac{1}{2}t_{ij}^{n1}t_{ji}^{2n'}\tau_{j}^-\tau_{j}^-
-\frac{1}{\sqrt{2}}t_{ij}^{n1}t_{ji}^{3n'}\tau_{j}^-\tau_{jz}
+\frac{1}{2}t_{ij}^{n2}t_{ji}^{1n'}\tau_{j}^{+}\tau_{j}^{+}
-\frac{1}{\sqrt{2}}t_{ij}^{n2}t_{ji}^{3n'}\tau_{j}^{+}\tau_{jz}
\label{h18}
\end{eqnarray}
\begin{eqnarray}
-\frac{1}{\sqrt{2}}t_{ij}^{n3}t_{ji}^{1n'}\tau_{jz}\tau_{j}^{+}
-\frac{1}{\sqrt{2}}t_{ij}^{n3}t_{ji}^{2n'}\tau_{jz}\tau_{j}^-
{\bigg ]}{\big [} 1-\vec S_i\cdot \vec S_j {\big ]}
\nonumber
\end{eqnarray}

The summation on $n$ and $n'$ must be taken according to the rules of the following table:
\begin{eqnarray}
\begin{array}{ccc}
n=1,~n'=1 & \Longrightarrow & c^{+}_{in}c_{in'}=\frac{1}{2}(1-\tau_{iz})(2+\tau_{iz}) \\
~ &~  &~  \\
n=1,~n'=2 & \Longrightarrow & c^{+}_{in}c_{in'}=\frac{1}{2}\tau_{i}^-\tau_{i}^- \\
~ &~  &~  \\
n=1,~n'=3 & \Longrightarrow & c^{+}_{in}c_{in'}=-\frac{1}{\sqrt{2}}\tau_{i}^-\tau_{iz} \\
~ & ~ &~  \\
n=2,~n'=1 & \Longrightarrow & c^{+}_{in}c_{in'}=\frac{1}{2}\tau_{i}^{+}\tau_{i}^{+} \\
~ &~  &~  \\
n=2,~n'=2 & \Longrightarrow & c^{+}_{in}c_{in'}=\frac{1}{2}(2-\tau_{iz})(1+\tau_{iz})  \\
~ &~  &~  \\
n=2,~n'=3 & \Longrightarrow & c^{+}_{in}c_{in'}=-\frac{1}{\sqrt{2}}\tau_{i}^{+}\tau_{iz}  \\
~ &~  &~  \\
n=3,~n'=1 & \Longrightarrow & c^{+}_{in}c_{in'}=-\frac{1}{\sqrt{2}}\tau_{iz}\tau_{i}^{+}  \\
~ & ~ &~  \\
n=3,~n'=2 & \Longrightarrow & c^{+}_{in}c_{in'}=-\frac{1}{\sqrt{2}}\tau_{iz}\tau_{i}^-  \\
~ &~  &~  \\
n=3,~n'=3 & \Longrightarrow & c^{+}_{in}c_{in'}=\tau_{iz}^2  
\end{array}
\nonumber
\end{eqnarray}

If we perform the summation over $n$ and $n'$ explicitly, we get:

\begin{eqnarray}
H_{\rm eff}^{(1)}=-\frac{1}{3}\frac{1}{U_2-J}\sum_{ij}{\big [}
\vec S_i\cdot \vec S_j+2 {\big ]}
{\bigg [} \frac{1}{4}(t_{ij}^{11})^2 (
2-\tau_{iz}-\tau_{iz}^2)\tau_{jz}(\tau_{jz}+1)
\nonumber
\end{eqnarray}
\begin{eqnarray}
+\frac{1}{4}(t_{ij}^{22})^2 
(2+\tau_{iz}-\tau_{iz}^2)\tau_{jz}(\tau_{jz}-1)
+\frac{1}{4}(t_{ij}^{33})^2 
\tau_{iz}^2(1-\tau_{jz}^2)
\nonumber
\end{eqnarray}
\begin{eqnarray}
-\frac{1}{2}t_{ij}^{11}t_{ji}^{12}
(\tau_i^-\tau_i^-+\tau_i^+\tau_i^+)(1-\tau_{jz}-\tau_{jz}^2)
+\frac{1}{\sqrt{2}}t_{ij}^{11}t_{ji}^{13}
(\tau_i^-\tau_{iz}+\tau_{iz}\tau_i^+)(1-\tau_{jz}-\tau_{jz}^2)
\nonumber
\end{eqnarray}
\begin{eqnarray}
-\frac{1}{2}t_{ij}^{11}t_{ji}^{22}
\tau_i^-\tau_i^-\tau_j^+\tau_j^+
+\frac{1}{\sqrt{2}}t_{ij}^{11}t_{ji}^{23}\left(
\tau_i^-\tau_i^-\tau_{jz}\tau_j^+
+\tau_i^+\tau_i^+\tau_j^-\tau_{jz}\right)
\nonumber
\end{eqnarray}
\begin{eqnarray}
-t_{ij}^{11}t_{ji}^{33}\left(
\tau_i^-\tau_{iz}\tau_{jz}\tau_j^+\right)
+\frac{1}{2}(t_{ij}^{12})^2 \left(
2\tau_{iz}^2+\tau_{iz}\tau_{jz}-\tau_{iz}^2\tau_{jz}^2
-\frac{1}{2}(\tau_i^-\tau_i^-\tau_j^-\tau_j^-+\tau_j^+\tau_j^+\tau_i^+\tau_i^+)\right)
\nonumber
\end{eqnarray}
\begin{eqnarray}
+\frac{1}{\sqrt{2}}t_{ij}^{12}t_{ji}^{13}\left(
\tau_i^-\tau_i^-\tau_j^-\tau_{jz}
+\tau_i^+\tau_i^+\tau_{jz}\tau_j^+
+(\tau_{iz}\tau_i^-+\tau_i^+\tau_{iz})(1-\tau_{jz}-\tau_{jz}^2)\right)
\nonumber
\end{eqnarray}
\begin{eqnarray}
+\frac{1}{\sqrt{2}}t_{ij}^{12}t_{ji}^{23}\left(
\tau_i^-\tau_i^-\tau_{jz}\tau_j^-+\tau_i^+\tau_i^+\tau_j^+\tau_{jz}
+(\tau_{iz}\tau_i^++\tau_i^-\tau_{iz})(1+\tau_{jz}-\tau_{jz}^2)\right)
\nonumber
\end{eqnarray}
\begin{eqnarray}
-\frac{1}{2}t_{ij}^{12}t_{ji}^{22}\left(
\tau_i^-\tau_i^-+\tau_i^+\tau_i^+\right)\left(1+\tau_{jz}-\tau_{jz}^2\right)
-t_{ij}^{12}t_{ji}^{33}\left(
\tau_i^-\tau_{iz}\tau_{jz}\tau_j^-
+\tau_{iz}\tau_i^+\tau_j^+\tau_{jz}\right)
\nonumber
\end{eqnarray}
\begin{eqnarray}
+\frac{1}{2}(t_{ij}^{13})^2 \left(
(2-3\tau_{iz}^2-\tau_{iz}+2\tau_{iz}\tau_{jz}^2+2\tau_{iz}^2\tau_{jz}^2)
-(\tau_i^-\tau_{iz}\tau_j^-\tau_{jz}+\tau_{jz}\tau_j^+\tau_{iz}\tau_i^+)\right)
\nonumber
\end{eqnarray}
\begin{eqnarray}
-t_{ij}^{13}t_{ji}^{23}\left(
\tau_i^-\tau_{iz}\tau_j^+\tau_{jz}+\tau_{iz}\tau_i^+\tau_{jz}\tau_j^-
+\frac{1}{2}(\tau_i^+\tau_i^++\tau_i^-\tau_i^-)(1-2\tau_{jz}^2)\right)
\nonumber
\end{eqnarray}
\begin{eqnarray}
+\frac{1}{\sqrt{2}}t_{ij}^{13}t_{ji}^{22}\left(
\tau_i^-\tau_i^-\tau_j^+\tau_{jz}+\tau_i^+\tau_i^+\tau_{jz}\tau_j^-\right)
-\frac{1}{\sqrt{2}}t_{ij}^{13}t_{ji}^{33}\left(\tau_{iz}\tau_i^++\tau_i^-\tau_{iz}\right)\left(1-2\tau_{jz}^2\right)
\nonumber
\end{eqnarray}
\begin{eqnarray}
+\frac{1}{\sqrt{2}}t_{ij}^{22}t_{ji}^{23}\left(\tau_{iz}\tau_i^-+\tau_i^+\tau_{iz}\right)\left(1+\tau_{jz}-\tau_{jz}^2\right)
-\frac{1}{\sqrt{2}}t_{ij}^{23}t_{ji}^{33}\left(\tau_{iz}\tau_i^-+\tau_i^+\tau_{iz}\right)\left(1-2\tau_{jz}^2\right){\big ]}
\nonumber
\end{eqnarray}
\begin{eqnarray}
-t_{ij}^{22}t_{ji}^{33}\left(
\tau_{iz}\tau_i^-\tau_j^+\tau_{jz}\right)
+\frac{1}{2}(t_{ij}^{23})^2 \left(
(2-3\tau_{iz}^2+\tau_{iz}-2\tau_{iz}\tau_{jz}^2+2\tau_{iz}^2\tau_{jz}^2)
-(\tau_i^+\tau_{iz}\tau_j^+\tau_{jz}+\tau_{jz}\tau_j^-\tau_{iz}\tau_i^-)\right) {\bigg ]}
\nonumber
\end{eqnarray}

\vspace{1.0cm}

\begin{eqnarray}
H_{\rm eff}^{(2)}=-\frac{1}{12}\frac{1}{U_2+2J}\sum_{ij}{\big [}
1- \vec S_i\cdot \vec S_j {\big ]}
{\bigg [} (t_{ij}^{11})^2 \left(
3-\tau_{iz}-\tau_{iz}^2-\frac{1}{4}\tau_{iz}\tau_{jz}(1+2\tau_{iz}+\tau_{iz}\tau_{jz})\right)
\nonumber
\end{eqnarray}
\begin{eqnarray}
+(t_{ij}^{22})^2 \left(3+\tau_{iz}-\tau_{iz}^2-\frac{1}{4}\tau_{iz}\tau_{jz}(1-2\tau_{iz}+\tau_{iz}\tau_{jz})\right)
+(t_{ij}^{33})^2 \left(
4\tau_{iz}^2-\tau_{iz}^2\tau_{jz}^2\right)
\nonumber
\end{eqnarray}
\begin{eqnarray}
+\frac{1}{2}t_{ij}^{11}t_{ji}^{12}\left(
\tau_i^-\tau_i^-+\tau_i^+\tau_i^+\right)\left(2+\tau_{jz}+\tau_{jz}^2\right)
-\frac{1}{\sqrt{2}}t_{ij}^{11}t_{ji}^{13}\left(
(\tau_i^-\tau_{iz}+\tau_{iz}\tau_i^+\right)\left(2+\tau_{jz}+\tau_{jz}^2)\right)
\nonumber
\end{eqnarray}
\begin{eqnarray}
-\frac{1}{4}t_{ij}^{11}t_{ji}^{22}\left(
\tau_i^-\tau_i^-\tau_j^-\tau_j^-+\tau_j^+\tau_j^+\tau_i^+\tau_i^++2(\tau_i^-\tau_i^--\tau_i^+\tau_i^+)(\tau_j^-\tau_j^--\tau_j^+\tau_j^+)\right)
\nonumber
\end{eqnarray}
\begin{eqnarray}
-\frac{1}{\sqrt{2}}t_{ij}^{11}t_{ji}^{23}\left(\tau_i^-\tau_{iz}(2\tau_j^+\tau_j^+-3\tau_j^-\tau_j^-)+\tau_{iz}\tau_i^+(2\tau_j^-\tau_j^--3\tau_j^+\tau_j^+)
\right)
\nonumber
\end{eqnarray}
\begin{eqnarray}
+\frac{1}{2}t_{ij}^{11}t_{ji}^{33}\left(\tau_i^-\tau_{iz}(2\tau_{jz}\tau_j^+-3\tau_j^-\tau_{jz})+\tau_{iz}\tau_i^+(2\tau_j^-\tau_{jz}-3\tau_{jz}\tau_j^+)
\right)
\nonumber
\end{eqnarray}
\begin{eqnarray}
+\frac{1}{2}(t_{ij}^{12})^2 \left[ \left(\tau_i^+\tau_i^+(\tau_j^+\tau_j^+-3\tau_j^-\tau_j^-)+\tau_i^-\tau_i^-\tau_j^-\tau_j^-\right)+(12-\tau_{iz}^2\tau_{jz}^2-4\tau_{iz}^2+\tau_{iz}\tau_{jz}) \right]
\nonumber
\end{eqnarray}
\begin{eqnarray}
-\frac{1}{\sqrt{2}}t_{ij}^{12}t_{ji}^{13}\left(\tau_i^-\tau_{iz}(2\tau_j^-\tau_j^--3\tau_j^+\tau_j^+)+\tau_{iz}\tau_i^+(2\tau_j^+\tau_j^+-3\tau_j^-\tau_j^-)
+(\tau_{iz}\tau_i^-+\tau_i^+\tau_{iz})(2+\tau_{jz}+\tau_{jz}^2)\right)
\nonumber
\end{eqnarray}
\begin{eqnarray}
+\frac{1}{2}t_{ij}^{12}t_{ji}^{22}\left(
(\tau_i^-\tau_i^-+\tau_i^+\tau_i^+)(2-\tau_{jz}+\tau_{jz}^2)\right)
\nonumber
\end{eqnarray}
\begin{eqnarray}
-\frac{1}{\sqrt{2}}t_{ij}^{12}t_{ji}^{23}\left(\tau_{iz}\tau_i^-(2\tau_j^-\tau_j^--3\tau_j^+\tau_j^+)+\tau_i^+\tau_{iz}(2\tau_j^+\tau_j^+-3\tau_j^-\tau_j^-)
+(\tau_{iz}\tau_i^++\tau_i^-\tau_{iz})(2-\tau_{jz}+\tau_{jz}^2)\right)
\nonumber
\end{eqnarray}
\begin{eqnarray}
+t_{ij}^{12}t_{ji}^{33}\left(\tau_{iz}\tau_i^-(2\tau_j^-\tau_{jz}-3\tau_{jz}\tau_j^+)+\tau_i^+\tau_{iz}(2\tau_{jz}\tau_j^+-3\tau_j^-\tau_{jz})\right)
\nonumber
\end{eqnarray}
\begin{eqnarray}
+(t_{ij}^{13})^2 \left(
(4-2\tau_{iz}+\tau_{iz}\tau_{jz}^2+\tau_{iz}^2\tau_{jz}^2)
+\tau_i^-\tau_{iz}(\tau_j^-\tau_{jz}-3\tau_{jz}\tau_j^+)+\tau_{iz}\tau_i^+\tau_{jz}\tau_j^+\right)
\nonumber
\end{eqnarray}
\begin{eqnarray}
-\frac{1}{\sqrt{2}}t_{ij}^{13}t_{ji}^{22}\left(
\tau_{iz}\tau_i^-(2\tau_j^+\tau_j^+-3\tau_j^-\tau_j^-)+\tau_i^+\tau_{iz}(2\tau_j^-\tau_j^--3\tau_j^+\tau_j^+)\right)
\nonumber
\end{eqnarray}
\begin{eqnarray}
+t_{ij}^{13}t_{ji}^{23}\left(
\tau_{iz}\tau_i^-(2\tau_{jz}\tau_j^+-3\tau_j^-\tau_{jz})+\tau_i^+\tau_{iz}(2\tau_j^-\tau_{jz}-3\tau_{jz}\tau_j^+)+(\tau_i^+\tau_i^++\tau_i^-\tau_i^-)(2-\tau_{jz}^2)\right)
\nonumber
\end{eqnarray}
\begin{eqnarray}
-{\sqrt{2}}t_{ij}^{13}t_{ji}^{33}\left((\tau_{iz}\tau_i^++\tau_i^-\tau_{iz})(2-\tau_{jz}^2)\right)
-\frac{1}{\sqrt{2}}t_{ij}^{22}t_{ji}^{23}\left((\tau_{iz}\tau_i^-+\tau_i^+\tau_{iz})(2-\tau_{jz}+\tau_{jz}^2)\right)
\nonumber
\end{eqnarray}
\begin{eqnarray}
+\frac{1}{2}t_{ij}^{22}t_{ji}^{33}\left(
\tau_{iz}\tau_i^-(2\tau_j^+\tau_{jz}-3\tau_{jz}\tau_j^-)+\tau_i^+\tau_{iz}(2\tau_{jz}\tau_j^--3\tau_j^+\tau_{jz})\right)
-{\sqrt{2}}t_{ij}^{23}t_{ji}^{33}\left((\tau_{iz}\tau_i^-+\tau_i^+\tau_{iz})(2-\tau_{jz}^2)\right){\big ]}
\nonumber
\end{eqnarray}
\begin{eqnarray}
+(t_{ij}^{23})^2 \left(
(4+2\tau_{iz}-\tau_{iz}\tau_{jz}^2+\tau_{iz}^2\tau_{jz}^2)
+\tau_i^+\tau_{iz}(\tau_j^+\tau_{jz}-3\tau_{jz}\tau_j^-)+\tau_{iz}\tau_i^-\tau_{jz}\tau_j^-\right) {\bigg ] }
\nonumber
\end{eqnarray}

\vspace{1.0cm}

\begin{eqnarray}
H_{\rm eff}^{(3)}=-\frac{1}{4}\frac{1}{U_2+4J}\sum_{ij}{\big [}
1-\vec S_i\cdot \vec S_j {\big ]}
{\bigg [} \frac{1}{4}(t_{ij}^{11})^2 
(2-\tau_{iz}-\tau_{iz}^2)(2-\tau_{jz}-\tau_{jz}^2)
\nonumber
\end{eqnarray}
\begin{eqnarray}
+\frac{1}{4}(t_{ij}^{22})^2 
(2+\tau_{iz}-\tau_{iz}^2)(2+\tau_{jz}-\tau_{jz}^2)
+(t_{ij}^{33})^2 \left(
\tau_{iz}^2\tau_{jz}^2\right)
\nonumber
\end{eqnarray}
\begin{eqnarray}
+\frac{1}{2}t_{ij}^{11}t_{ji}^{12}(\tau_i^-\tau_i^-+\tau_i^+\tau_i^+)(2-\tau_{jz}-\tau_{jz}^2)
-\frac{1}{\sqrt{2}}t_{ij}^{11}t_{ji}^{13}(\tau_i^-\tau_{iz}+\tau_{iz}\tau_i^+)(2-\tau_{jz}-\tau_{jz}^2)
\nonumber
\end{eqnarray}
\begin{eqnarray}
+\frac{1}{4}t_{ij}^{11}t_{ji}^{22}\left(
\tau_i^-\tau_i^-\tau_j^-\tau_j^-+\tau_j^+\tau_j^+\tau_i^+\tau_i^+\right)
-\frac{1}{\sqrt{2}}t_{ij}^{11}t_{ji}^{23}\left(
\tau_i^-\tau_i^-\tau_j^-\tau_{jz}+\tau_i^+\tau_i^+\tau_{jz}\tau_j^+\right)
\nonumber
\end{eqnarray}
\begin{eqnarray}
+\frac{1}{2}t_{ij}^{11}t_{ji}^{33}\left(\tau_i^-\tau_{iz}\tau_j^-\tau_{jz}+\tau_{jz}\tau_j^+\tau_{iz}\tau_i^+\right)
+\frac{1}{2}(t_{ij}^{12})^2 \left(
(2-\tau_{iz}-\tau_{iz}^2)(2+\tau_{jz}-\tau_{jz}^2)+\tau_i^-\tau_i^-\tau_j^+\tau_j^+\right)
\nonumber
\end{eqnarray}
\begin{eqnarray}
-\frac{1}{\sqrt{2}}t_{ij}^{12}t_{ji}^{13}\left(
\tau_i^-\tau_i^-\tau_{jz}\tau_j^++\tau_i^+\tau_i^+\tau_j^-\tau_{jz}+(2-\tau_{iz}-\tau_{iz}^2)(\tau_j^+\tau_{jz}+\tau_{jz}\tau_j^-)\right)
\nonumber
\end{eqnarray}
\begin{eqnarray}
+\frac{1}{2}t_{ij}^{12}t_{ji}^{22}(\tau_i^-\tau_i^-+\tau_i^+\tau_i^+)(2+\tau_{jz}-\tau_{jz}^2)
\nonumber
\end{eqnarray}
\begin{eqnarray}
-\frac{1}{\sqrt{2}}t_{ij}^{12}t_{ji}^{23}\left(
\tau_i^-\tau_i^-\tau_j^+\tau_{jz}+\tau_i^+\tau_i^+\tau_{jz}\tau_j^-+(2+\tau_{jz}-\tau_{jz}^2)(\tau_j^-\tau_{jz}+\tau_{jz}\tau_j^+)\right)
\nonumber
\end{eqnarray}
\begin{eqnarray}
+t_{ij}^{12}t_{ji}^{33}\left(
\tau_i^-\tau_{iz}\tau_j^+\tau_{jz}+\tau_{iz}\tau_i^+\tau_{jz}\tau_j^-\right)
+(t_{ij}^{13})^2 \left(
(2-\tau_{iz}-\tau_{iz}^2)\tau_{jz}^2+\tau_i^-\tau_{iz}\tau_{jz}\tau_j^+\right)
\nonumber
\end{eqnarray}
\begin{eqnarray}
-\frac{1}{\sqrt{2}}t_{ij}^{13}t_{ji}^{22}\left(
\tau_i^-\tau_i^-\tau_{jz}\tau_j^-+\tau_i^+\tau_i^+\tau_j^+\tau_{jz}\right)
+t_{ij}^{13}t_{ji}^{23}\left(
\tau_i^-\tau_{iz}\tau_{jz}\tau_j^-+\tau_{iz}\tau_i^+\tau_j^+\tau_{jz}+\tau_{jz}^2(\tau_i^-\tau_i^-+\tau_i^+\tau_i^+)\right)
\nonumber
\end{eqnarray}
\begin{eqnarray}
-\sqrt{2}t_{ij}^{13}t_{ji}^{33}(\tau_i^-\tau_{iz}+\tau_{iz}\tau_i^+)\tau_{jz}^2
-\frac{1}{\sqrt{2}}t_{ij}^{22}t_{ji}^{23}(\tau_i^+\tau_{iz}+\tau_{iz}\tau_i^-)(2+\tau_{jz}-\tau_{jz}^2)
\nonumber
\end{eqnarray}
\begin{eqnarray}
-\sqrt{2}t_{ij}^{23}t_{ji}^{33}(\tau_i^+\tau_{iz}+\tau_{iz}\tau_i^-)\tau_{jz}^2
\nonumber
\end{eqnarray}
\begin{eqnarray}
+\frac{1}{2}t_{ij}^{22}t_{ji}^{33}\left(
\tau_i^+\tau_{iz}\tau_j^+\tau_{jz}+\tau_{jz}\tau_j^-\tau_{iz}\tau_i^-\right)
+(t_{ij}^{23})^2 \left(
(2+\tau_{iz}-\tau_{iz}^2)\tau_{jz}^2+\tau_i^+\tau_{iz}\tau_{jz}\tau_j^-\right){\bigg ]}
\nonumber
\end{eqnarray}

\end{widetext}
\begin{multicols}{2}

For the sake of brevity, we have not symmetrized the above expressions over 
$i$ and $j$, so that we wrote, for example, the expression
$\tau_{1z}\tau_2^{+} + \tau_{2z}\tau_1^{+}+\tau_{1z}^2+\tau_{2z}^2$
 as
$$\sum_{ij=1,2}\tau_{iz}\tau_j^{+}+\tau_{iz}^2$$
instead of
$$\frac{1}{2}\sum_{ij=1,2}(\tau_{iz}\tau_j^{+} + \tau_{jz}\tau_i^{+}+\tau_{iz}^2+\tau_{jz}^2)$$

Note that even if in principle we should have 81 terms ($3^4$), we have 
assumed, as in CNR,\cite{cnr1,cnr2}
that $t_{ij}^{mm'}=t_{ij}^{m'm}$ so that some regrouping of the terms has
been possible (each of the three $H_{\rm eff}^{(i)}$ contains 
only 21 terms). Unfortunately, as anticipated in the text, the fact that there is no 
conservation law for the pseudospin quantum number $\tau_z$, during each 
hopping process, prevents from having more symmetrical expressions for 
the orbital part.

\section{Action of $H_{\rm eff}$ on molecular and atomic states.}

In this appendix we show how to derive Eqs. (\ref{san1}), (\ref{san2}). 
For a given horizontal bond (for example, $\delta_1$ in Fig. \ref{bonds}) 
nine two-site orbital states are involved in the product state (\ref{st1}). They are listed below:

\begin{eqnarray}
\begin{array}{lll}
\parallel  1\rangle_{ac} & = & |-1\rangle_a|-1\rangle_c \\[0.1cm] 
\parallel   2\rangle_{ac} & = & |-1\rangle_a|0\rangle_c \\[0.1cm]
\parallel   3\rangle_{ac} & = & |-1\rangle_a|1\rangle_c \\[0.1cm]
\parallel  4\rangle_{ac} & = & |0\rangle_a|-1\rangle_c \\[0.1cm]
\parallel  5\rangle_{ac} & = & |0\rangle_a|0\rangle_c \\[0.1cm]
\parallel  6\rangle_{ac} & = & |0\rangle_a|1\rangle_c \\[0.1cm]
\parallel  7\rangle_{ac} & = & |1\rangle_a|-1\rangle_c\\[0.1cm]
\parallel  8\rangle_{ac} & = & |1\rangle_a|0\rangle_c\\[0.1cm]
\parallel  9\rangle_{ac} & = & |1\rangle_a|1\rangle_c\\[0.1cm]
\end{array}
\label{st9}
\end{eqnarray}

If the spin configuration of the bond is ferromagnetic, then after 
some simple algebra we find:
\begin{eqnarray}
\begin{array}{l}
H_{\delta_1}^{F}\parallel  1\rangle_{ac}=-\frac{2\tau^2+2\chi^2}{U_2-J}
\parallel  1\rangle_{ac}+\frac{2\chi^2}{U_2-J}\parallel  9\rangle_{ac}
\\[0.2cm]
H_{\delta_1}^{F}\parallel  2\rangle_{ac}=
-\frac{\beta ^2+\sigma^2+\chi^2+\theta^2 }{U_2-J}
\parallel  2\rangle_{ac}
\\[0.2cm]
H_{\delta_1}^{F}\parallel  3\rangle_{ac}=-\frac{\alpha ^2+\beta ^2 
+\tau ^2+\theta^2}{U_2-J}
\parallel  3\rangle_{ac}-
\frac{2\alpha\beta}{U_2-J} \parallel  7\rangle_{ac}
\\[0.2cm]
H_{\delta_1}^{F}\parallel  4\rangle_{ac}=-\frac{\beta ^2+\sigma^2+\chi^2+\theta^2}{U_2-J}
\parallel  4\rangle_{ac}
\\[0.2cm]
H_{\delta_1}^{F}\parallel  5\rangle_{ac}=-\frac{2\tau^2+2\theta^2}{U_2-J}
\parallel  5\rangle_{ac}
\\[0.2cm]
H_{\delta_1}^{F}\parallel  6\rangle_{ac}=-\frac{\alpha ^2+\sigma ^2 +\tau ^2+\chi^2}{U_2-J}
\parallel  6\rangle_{ac}
\\[0.2cm]
H_{\delta_1}^{F}\parallel  7\rangle_{ac}=
-\frac{\alpha ^2+\beta ^2 +\tau ^2+\theta^2}{U_2-J}
\parallel  7\rangle_{ac}-
\frac{2\alpha\beta}{U_2-J} \parallel  3\rangle_{ac}
\\[0.2cm]
H_{\delta_1}^{F}\parallel  8\rangle_{ac}=-\frac{\alpha ^2+\sigma ^2 +
\tau ^2+\chi^2}{U_2-J}
\parallel  8\rangle_{ac}
\\[0.2cm]
H_{\delta_1}^{F}\parallel  9\rangle_{ac}=-\frac{2\chi^2 +2\theta^2}{U_2-J}
\parallel  9\rangle_{ac}+\frac{2\chi^2}{U_2-J}\parallel  1\rangle_{ac} 
\end{array}
\label{st10}
\end{eqnarray}

\noindent whereas for the antiferromagnetic configuration we obtain:

\begin{eqnarray}
\begin{array}{lll}
H_{\delta_1}^{A}\parallel  1\rangle_{ac} & = &
-(\frac{\alpha^2+\sigma^2+2\theta^2}{(U_2+4J)}\\[0.4cm]
 & & +\frac{3\alpha^2+3\sigma^2+4\tau^2+4\chi^2+6\theta^2}{3(U_2+2J)})
\parallel  1\rangle_{ac}\\[0.4cm]
 & & +(\frac{\alpha\beta }{U_2+4J}-\frac{\alpha\beta }{U_2+2J} )
\parallel  9\rangle_{ac}\\[0.4cm]
H_{\delta_1}^{A}\parallel  2\rangle_{ac} & = &
-( \frac{\alpha^2+\tau^2 +\chi^2+\theta^2 }{(U_2+4J)}\\[0.4cm]
 & &+\frac{3\alpha^2+2\beta^2+2\sigma^2+3\tau^2+5\chi^2+5\theta^2 }{3(U_2+2J)})
\parallel  2\rangle_{ac}\\[0.4cm]
H_{\delta_1}^{A}\parallel  3\rangle_{ac} & = &
-(\frac{\sigma^2 +\tau^2+\chi^2+\theta^2}{(U_2+4J)}\\[0.4cm]
 & &+\frac{2\alpha^2+2\beta^2+3\sigma^2+5\tau^2+3\chi^2+5\theta^2 }{3(U_2+2J)})
\parallel  3\rangle_{ac}\\[0.4cm]
 & &+\frac{2}{3(U_2+2J)}\alpha\beta
\parallel  7\rangle_{ac}\\[0.4cm]
H_{\delta_1}^{A}\parallel  4\rangle_{ac} & = &
-( \frac{\alpha^2+\tau^2+\chi^2+\theta^2 }{(U_2+4J)}\\[0.4cm]
 & &+\frac{3\alpha^2+2\beta^2+2\sigma^2+3\tau^2+5\chi^2+5\theta^2 }{3(U_2+2J)})
\parallel  4\rangle_{ac}\\[0.4cm]
\end{array}
\label{st11}
\end{eqnarray}
\begin{eqnarray}
\begin{array}{lll}
H_{\delta_1}^{A}\parallel  5\rangle_{ac} & = &
-(\frac{\alpha^2+\beta^2+2\chi^2}{(U_2+4J)}\\[0.4cm]
 & &+\frac{3\alpha^2+3\beta^2+4\tau^2+6\chi^2+4\theta^2 }{3(U_2+2J)}) 
\parallel  5\rangle_{ac}\\[0.4cm]
H_{\delta_1}^{A}\parallel  6\rangle_{ac} & = &
-( \frac{\beta^2+\tau^2+\chi^2+\theta^2 }{(U_2+4J)}\\[0.4cm]
 & &+\frac{2\alpha^2+3\beta^2+2\sigma^2+5\tau^2+5\chi^2+3\theta^2 }
{3(U_2+2J)} ) \parallel  6\rangle_{ac}\\[0.4cm]
H_{\delta_1}^{A}\parallel  7\rangle_{ac} & = &
-(\frac{\sigma^2+\tau^2+\chi^2+\theta^2 }{(U_2+4J)}\\[0.4cm]
 & &+\frac{2\alpha^2+2\beta^2+3\sigma^2+5\tau^2+3\chi^2+5\theta^2 }{3(U_2+2J)})
 \parallel  7\rangle_{ac}\\[0.4cm]
 & &+\frac{2\alpha\beta}{3(U_2+2J)}
\parallel  3\rangle_{ac}\\[0.4cm]
H_{\delta_1}^{A}\parallel  8\rangle_{ac} & = &
-( \frac{\beta^2+\tau^2+\chi^2+\theta^2 }{(U_2+4J)}\\[0.4cm]
 & &+\frac{2\alpha^2+3\beta^2+2\sigma^2+5\tau^2+5\chi^2+3\theta^2 }{3(U_2+2J)})
\parallel  8\rangle_{ac}\\[0.4cm]
H_{\delta_1}^{A}\parallel  9\rangle_{ac} & = &
-( \frac{\beta^2+\sigma^2+2\tau^2 }{(U_2+4J)}\\[0.4cm]
 & &+ \frac{3\beta^2+3\sigma^2+6\tau^2+4\chi^2+4\theta^2 }{3(U_2+2J)})
\parallel 9\rangle_{ac}\\[0.4cm]
 & &+(\frac{\alpha\beta}{U_2+4J}-\frac{\alpha\beta}{U_2+2J})
\parallel  1\rangle_{ac}
\end{array}
\label{st11b}
\end{eqnarray}

As in the text, we have retained the 
hopping integrals $\theta=t^{13}$ and $\chi=t^{12}$ which are zero in the 
corundum phase and can be different from zero in the monoclinic one.

Given the form of the molecular wave function in Eq. (\ref{psidim}), 
these are the only terms which survive when taking an average of the
kind shown in Eq. (\ref{st1}).

\end{multicols} 

\end{document}